\documentclass[preprints,article,accept,moreauthors,pdftex]{mdpi} 
\def\hb{H$\beta$\/}
\def\oiii{[{O}{\sc iii}]$\lambda\-\lambda$\-4959,\-5007\/}

\def\oiiiseven{[{O}{\sc iii}]$\lambda$5007\/}

\def\mgii{{Mg}{\sc ii}$\lambda$2800\/}

\def\feii{{Fe}{\sc ii}\/}
\def\kms{km\,s$^{-1}$\/}

\def\rfe{$R_\mathrm{{Fe}{\textsc{ii}}}$\/}
\def\mbh{$M_\mathrm{\rm BH}$\/}
\def\lledd{$L/L_\mathrm{\rm Edd}$\/}

\def\hbbc{H$\beta_\mathrm{\rm{BC}}$\/}
\def\c14{$c$(1/4)\/}
\def\ergs{erg s$^{-1}$\/}
\def\apjs{The Astrophysical Journal Supplements}
\def\aap{Astronomy \&\ Astrophysics}
\def\mnras{Monthly Notices Royal Astronomical Sociey\/}
\def\apjl{The Astrophysical Journal Letters\/}
\def\apj{The Astrophysical Journal\/}
\def\pasj{Publ. Astr. Soc. Japan\/}
\def\pasp{Publication Astronomical Society Pacific\/}
\def\aj{The Astronomical Journal\/}
\def\nat{Nature\/}
\def\apss{Astrophyscs Space  Science\/}

\def\araa{Annual Review Astronomy \& Astrophysics}
\def\ssr{Space Science Review}
\def\prd{Physical Review D}
\def\jcap{Journal  Cosmology Astroparticle Physics}
\def\aapr{Astronomy Astrophysics Review}
\firstpage{1} 
\pubvolume{1}
\issuenum{1}
\articlenumber{0}
\pubyear{2023}
\copyrightyear{2022}
\datereceived{} 
\daterevised{}
\dateaccepted{} 
\datepublished{} 
\hreflink{https://doi.org/} % If needed use \linebreak
\Title{Where to search for supermassive binary black holes}

\TitleCitation{Supermassive binary black holes}

\Author{Paola Marziani $^{1,*}$\orcidA{}, Edi Bon$^{2}\orcidC{}$, Natasa Bon$^{2}$\orcidD{},
Mauro D'Onofrio$^{3,1}$\orcidB{}}

\address{
$^{1}$ \quad National Institute for Astrophysics (INAF), Astronomical Observatory of Padova, IT-35122 Padova, Italy; paola.marziani@inaf.it\\ 
$^{2}$ \quad Astronomical Observatory Belgrade, Volgina 7, 11060 Belgrade; ebon@aob.rs, nbon@aob.rs\\ 
$^{3}$ \quad Department of Physics and Astronomy, University of Padua, IT-35122 Padova, Italy; mauro.donofrio@unipd.it
}
\corres{Correspondence: paola.marziani@inaf.it}

\nolinenumbers
\abstract{
Supermassive binary black holes (SMBBHs) are the anticipated byproducts of galaxy mergers and play a pivotal role in shaping galaxy evolution, gravitational wave emissions, and accretion physics. Despite their theoretical prevalence, direct observational evidence for SMBBHs remains elusive, with only a handful of candidates identified to date. This paper explores optimal strategies and key environments for locating SMBBHs, focusing on observational signatures in the broad Balmer lines.  { We present a preliminary analysis on a flux-limited sample of sources belonging to an evolved spectral type along the quasar main sequence, and  } we discuss the  spectroscopic clues indicative of binary activity and highlight the critical role of time-domain spectroscopic surveys in uncovering periodic variability linked to binary systems.
}


\begin{document}
\nolinenumbers
%%%%%%%%%%%%%%%%%%%%%%%%%%%%%%%%%%%%%%%%%%
%\setcounter{section}{-1} %% Remove this when starting to work on the template.

%\tableofcontents 

% The order of the section titles is different for some journals. Please refer to the "Instructions for Authors” on the journal homepage.
 % Recent advancements in observational techniques, particularly 
%pace-based telescopes and interferometric arrays, have provided unprecedented insights into the inner workings of quasars during periods of super-Eddington accretion. These observations reveal complex dynamics within the accretion disk, including rapid fluctuations in the X-ray luminosity, extreme ionization states of emitted gas (both low and high!), and occasionally, relativistic jets.

\section{Introduction}

% The  astonishing luminosity  of quasars (which may reach $\sim 10^{48}$ \ergs, 10,000 the luminosity of a giant galaxy) suggests a highly efficient energy release mechanism, much beyond the efficiency yielded by nuclear reactions in stars. Such high efficiency can only be achieved through infall into the deep gravitational well of a compact object.  Any lingering doubt regarding the fundamental tenet of what is now known as the standard model of active galactic nuclei (AGN) \citep[e.g.,][ and references therein]{peterson97,franketal02,donofrioetal12,netzer13}, namely accretion onto a supermassive black hole, has  definitively passed into history \citep{eht19}.  

Supermassive black hole binaries (SMBBHs) are an expected consequence of hierarchical galaxy formation. As galaxies merge, their central supermassive black holes (SMBHs), with masses ranging from 10$^{6} -10^{{9}}\ M_{\odot}$, are brought together by dynamical friction and may eventually form a bound binary system \cite{colpi14,komossaetal21}. SMBBHs are relevant for understanding a wide range of astrophysical processes \cite{komossagrupe24}, including galaxy evolution, the fueling of active galactic nuclei (AGN), and the generation of low-frequency gravitational waves detectable by pulsar timing arrays (PTAs, \cite{burke-spolaoretal11,thereauetal21}) and future space-based observatories like LISA \cite[][]{bertietal06,amaro-seoaneetal17,donofriomarziani18}. Despite their theoretical significance, the observational evidence for SMBBHs remains sparse, with only a handful of candidates  identified through periodicity in AGN light curves, dual-AGN systems, or distinct velocity offsets in emission lines \cite{sillanpaaetal88,valtonen2008,valtonenetal23}. Several claims of periodic behavior were nullified by monitoring on a large temporal basis \cite{eracleousetal97,gezarietal07,liuetal18}.  Current estimates of the population of SMBBHs vary depending on the assumptions about galaxy merger rates, SMBH masses, and the efficiency of binary coalescence \cite{volonteri2003}. Models predict that thousands of SMBBHs could exist within the detectable range of PTAs, assuming typical binary separations of 0.1–10 parsecs and orbital periods ranging from months to decades \cite{volonterietal09,burke-spolaoretal11}. However, the true population remains highly uncertain due to the complex interplay of gas dynamics, stellar interactions, and gravitational wave emission in the binary evolution \cite{merritt2005}. Future surveys and gravitational wave observations are expected to provide a more robust census.

The electromagnetic phenomenology of  SMBBHs is all but clear.  Previous attempts to identify SMBBHs have resorted to the detection of periodic phenomena involving photometric and spectroscopic properties.  Photometric  periodicities can arise from several mechanisms, including the modulation of accretion rates due to the gravitational interaction between the binary components, relativistic Doppler boosting of emission from the mini-disk around one of the black holes, or periodic disruptions of gas streams. Examples include the AGN PG 1302-102, which exhibits quasi-periodic oscillations with a period of approximately 5 years, interpreted as evidence of a binary with a sub-parsec separation  \cite{grahametal15}. Additional candidates have been identified in large surveys such as the Palomar Transient Factory (PTF) and Zwicky Transient Facility (ZTF) \cite{chambersetal16,bellmetal19,grahametal19,mascietal19}, where systematic searches for periodic variability in AGN light curves have revealed populations consistent with binary models \cite{charisietal16,chen2022}. Furthermore, irregular photometric dips or flares, potentially linked to gas dynamics in the circum-binary disk or mini-disks, provide complementary evidence for the presence of SMBBHs \cite{valtonen2008}. 

The interpretation of quasi-periodic variability as evidence for supermassive black hole binaries (SMBBHs) faces significant challenges, particularly from red noise in AGN light curves. AGNs exhibit stochastic variability across a broad range of timescales, which can produce apparent periodic signals through random fluctuations. This variability is characterized by a power-law power spectral density (PSD), where lower-frequency (longer timescale) variations dominate, often described as red noise. Simulated light curves with red noise properties have been shown to produce spurious periodicities that mimic the signals expected from SMBBHs \cite{vaughanetal16,edelman2023}. Criticism of candidates like PG 1302-102, originally proposed as a binary due to its periodic variability \cite{grahametal15}, highlights the difficulty of distinguishing true periodic signals from statistical artifacts \cite{liuetal18}. Rigorous tests, such as the Lomb-Scargle periodogram and wavelet analysis, must account for the PSD of AGN variability to avoid false positives \cite{hornebaliunas86}. Long monitoring timescales are essential to verify periodicity over multiple cycles and rule out red noise as the source of variability. Systematic studies, such as those using Pan-STARRS and ZTF, increasingly incorporate red noise models to refine binary candidates and improve the reliability of photometric searches for SMBBHs \cite{sesar2007}. Applying the same criterion customarily applied for binary stars (at least three full orbits monitoring \cite{hilditch01}), would imply a monitoring time of several decades, and suitable data are not yet available for SMBBHs. 

Broad emission line profile variability offers a promising avenue for detecting SMBBHs. In a binary system, the orbital motion of the black holes, coupled with their interaction with surrounding gas, can produce periodic changes in the broad-line profiles, such as centroid shifts, line asymmetries, or double peaks. Variability in the broad-line profiles of AGN such as NGC 4151 suggests that gravitational interactions between binary components might influence the emission region \cite{bonetal12}. \citet{shenetal13} conducted a systematic search for binary SMBHs by examining velocity shifts in broad-line centroids using multi-epoch spectroscopy from the Sloan Digital Sky Survey (SDSS), providing statistical evidence for potential binaries. Simulations  further demonstrated that periodic line profile changes \citep[e.g., ][]{lietal16}, including double peaks and shifts, can arise from mini-disks around individual SMBHs in a binary system.  However, explored long-term variability of quasars with double-peaked and peculiar profiles disfavored the idea that periodic changes in the broad \hb\  line could be attributed to a  SMBBH system \cite{gezarietal07}. 

Additional evidence for  SMBBHs is arguably based on peculiar broad line profiles.  The broad line region (BLR) of 1E 1821+643 exhibits significant velocity offsets of its broad emission line profiles, with the \hb\ line centroid shifted by thousands of kilometers per second relative to the host galaxy's systemic velocity \cite{robinsonetal10}. Such offsets could be interpreted as evidence of a gravitationally bound binary system, where the orbital motion of one black hole contributes to the Doppler shift of the observed emission lines. Alternatively, the velocity shifts could indicate a  recoiling black hole, where a gravitational wave ``kick" following a binary black hole merger ejects the remnant black hole at high velocity from the center of the galaxy \cite{robinsonetal10,jadhavetal21}. Both scenarios are considered compelling but remain speculative due to the complex dynamics of the BLR and alternative explanations involving anisotropic gas motion or asymmetric accretion flows. Future multi-wavelength monitoring, including precision radial velocity measurements and time-domain studies, may help distinguish between these interpretations and provide deeper insight into the nature of 1E 1821+643.  

While observational evidence remains sparse due to the complexity of BLR dynamics and the long orbital timescales of SMBBHs, high-resolution spectroscopic monitoring and advanced time-domain surveys are paving the way for more robust detections. However, interpretation is made more complex by the lack of consensus over the expected phenomenology of emission line profiles. Also, in the case of extreme mass ratio inspirals, a second black hole may not provide a detectable electromagnetic signal, while still perturbing line emission or circum-primary accretion disk.  

The previous considerations are meant to stress that an efficient search for SMBBHs should be highly focused. A blind search over the full AGN population will be probably overwhelmed by the red noise characteristics of the light curves and by the lack of contextualization of emission line profile properties. The aim of this note is to point out that there is a well defined segment of the quasar main sequence where  SMBBHs  can be more easily detected. We begin by summarizing the concept of the quasar main sequence (MS), a framework that has proven to be an effective tool for organizing the diversity of type-1 AGN (\S \ref{ms}). From the MS, we focus on a specific spectral type and conduct a detailed analysis of the emission line profiles of the Balmer HI line \hb\ and the UV resonance line \mgii\ (\S \ref{b1pp}).  This particular spectral type exhibits several sources with emission line profiles that can be interpreted as being perturbed by the presence of a secondary compact object. Based on this interpretation, we identify a set of promising candidates for further monitoring (\S \ref{bin} and \S \ref{resu}) and outline the fundamental multi-messenger characteristics expected of such binary systems (\S \ref{discu}).

 \begin{figure}
 \centering
 \includegraphics[width=0.95\linewidth]{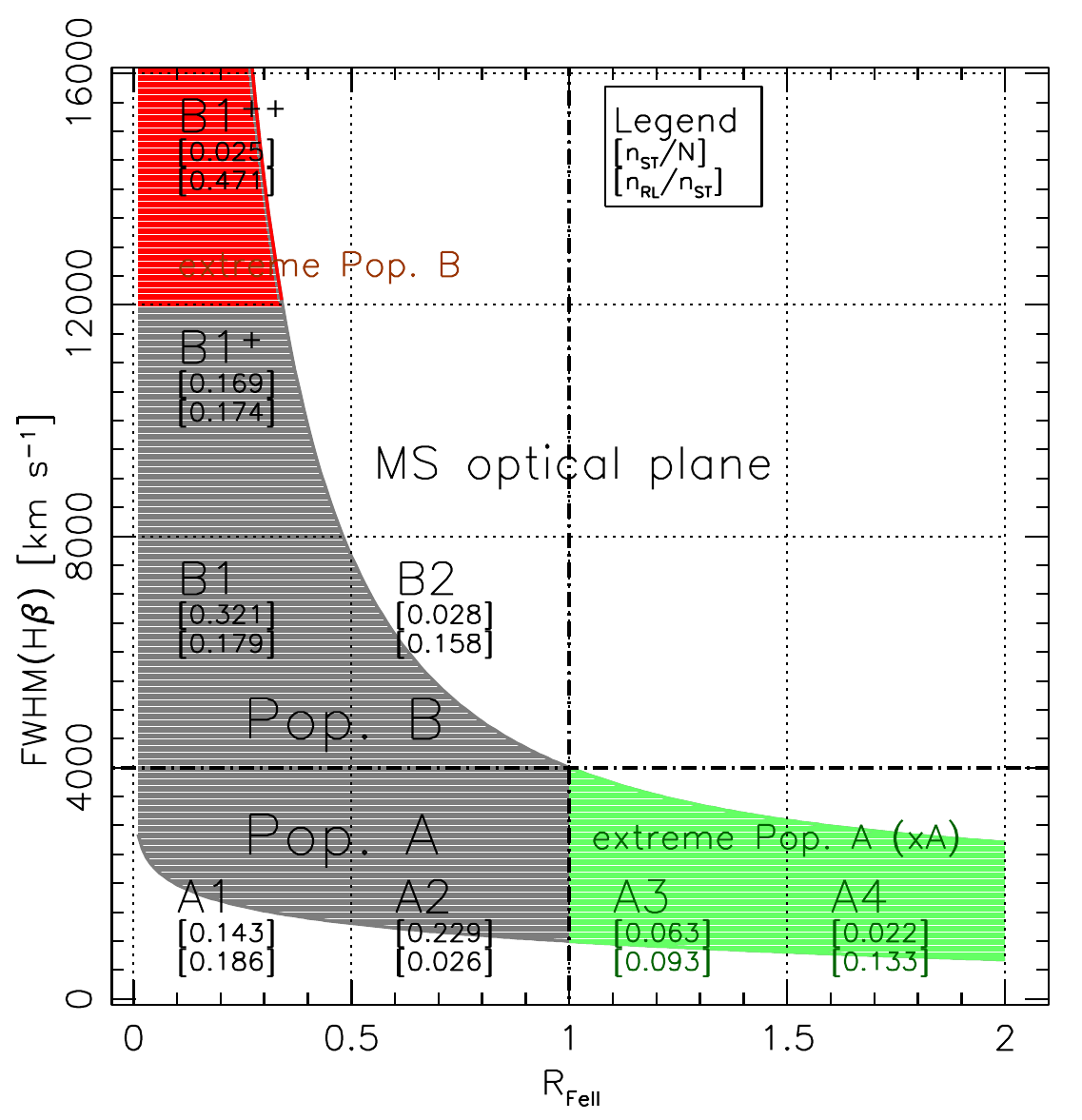}
 \caption{A schematic representation of the optical plane of the quasar main sequence, with the subdivisions into spectral bins clearly marked. The numbers in square brackets beneath each spectral type label indicate the prevalence of the spectral type ({ number $n_\mathrm{ST}$\ of sources in each spectral type normalized by the number of sources in the full sample}), and the fraction of RL (jetted) sources { $n_\mathrm{RL}/n_\mathrm{ST}$} within each type. The data are from the sample of \citet{marzianietal13a}.}
 \label{fig:e1bin}
\end{figure}

\section{The quasar main sequence: a brief synopsis}
\label{ms}

The concept of the main sequence   in quasar studies originated from a Principal Component Analysis (PCA) applied to the spectra of Palomar Green quasars, which identified a dominant trend known as Eigenvector 1 (E1; \citep{borosongreen92}). The E1 revealed an anti-correlation between the strength of the singly ionized iron emission blend centered at  $\lambda$4570, and the full width at half maximum (FWHM) of H$\beta$, or alternatively, the peak intensity of \oiiiseven. The MS can be represented in a plane where FWHM(H$\beta$) is plotted against the parameter \rfe, defined as the flux ratio  F(\feii$\lambda$4570)/F(\hbbc). Figure \ref{fig:e1bin} illustrates the quasar MS in this parameter space. Over time, the MS has been validated and extended from the original sample of 80 objects \citep{borosongreen92}, encompassing now hundreds \citep{marzianietal03b} and tens of thousands sources \citep{shenho14}. The trends identified in the MS have been linked to physical properties of the accretion process and outflows, confirmed by multi-frequency data \citep{sulenticetal00a, sulenticetal00b,pandaetal19} and advanced techniques like locally linear embedding in manifold learning \citep{jankovetal21,ghojoghetal20}. The MS captures fundamental correlations related to quasar dynamics, orientation, and the accretion structure. 

The MS encompasses only type-1 AGN, where broad emission lines are visible, and orientation constraints (angle between the line of sight and the AGN symmetry axis $\theta \lesssim$ 45) affect the observed line widths. The parameters FWHM(H$\beta$) and \rfe\ are particularly revealing. FeII emission, which spans from the UV to the IR, is largely self-similar in type-1 AGN but varies in relative strength to H$\beta$, with \rfe\ values ranging from undetectable to 
$\gtrsim$2 \citep{sulenticetal00a,marinelloetal16} which can be directly linked to Eddington ratio \cite{marzianietal01,marzianietal03b,sunshen15,duetal16a}. The FWHM of H$\beta$ reflects the virialized velocity field of the low-ionization  BLR  and is influenced by orientation, with the angle  $\theta$  playing a key role \citep{antonucci93,urrypadovani95,marinantonucci16}. 

A defining feature of the MS is its ``elbow" shape, dividing sources into Population A (Pop. A) and Population B (Pop. B) \citep{sulenticetal00a}. Pop. A quasars, characterized by FWHM(H$\beta$) 
$\lesssim$4000 km/s, exhibit sharp H$\beta$ profiles, prominent \feii\  emission, and weak [O{\sc iii}], while Pop. B quasars, with broader H$\beta$ profiles, display weaker \feii\ and stronger [O{\sc iii}] \citep{sulenticetal00a}. Additionally, high-ionization lines such as C{\sc iv}
$\lambda$1549 show systematic blueshifts relative to narrow low-ionization lines like [O{\sc ii}]
$\lambda$3727  or \hb\ narrow component \citep{bonetal20}. These distinctions suggest a heuristic subdivision of the BLR into a virialized low-ionization region, likely associated with the accretion disk, and a wind-dominated high-ionization region, particularly pronounced in extreme Population A (xA) sources where \rfe\ 
$>1$ \citep{collinsouffrinetal88,marzianietal96,elvis00}.

\subsection{A special section of the main sequence}
\label{b1pp}

Extreme Population B sources are characterized by extremely broad Balmer line profiles, with conventional FWHM limits of  $\gtrsim 12,000$ \kms\ or even $\gtrsim 16,000$ \kms. They  include only the most extreme cases, and occupy the spectral type B1$^{++}$ in Figure \ref{fig:e1bin}). A visual inspection shows symmetric very broad profiles (the red wing of \hb\ is invariably extended beyond the \oiii\ emission).   These sources also exhibit very low, often undetectable \rfe, in contrast to Population A, where the majority of sources have measurable \rfe. A hallmark of Population B is the most-pronounced redward asymmetry in low-ionization lines, such as H$\beta$, a feature well-documented in the literature \citep{netzer77,marzianietal03b,wangetal17,wolfetal20,baoetal22,marziani23,zastrockyetal24}.  However, the situation for spectral  type B1$^{++}$\ is less clear: the \mgii\ and \hb\ median spectra show an asymmetry index consistent with a symmetric profile (AI $\approx 0.02 \pm 0.06$ and $\approx 0.04 \pm 0.06$\ for the two lines, \cite{marzianietal13a}). 

The \rfe\ parameter, central to the definition of the MS, depends not only on the gas physical conditions but also on chemical abundances. This makes it crucial to understand how metallicity (Z) influences a source location in the optical MS plane \citep{punslyetal18a,pandaetal18,pandaetal19}. Another key insight from MS studies is the distinct differences in the prevalence of radio-loud (RL) sources across populations. Extreme Population B hosts the largest fraction of ``jetted" sources ($\approx 50$\%\ as reported in  Fig. \ref{fig:e1bin}), distinguished from radio-quiet (RQ) ones by their distribution of blueshifts in high-ionization lines. Interestingly, this radio-loudness has a minor effect on the low-ionization lines \citep{sulenticetal07,richardsetal11}. While extreme Population A also includes powerful radio emitters, these objects are likely fundamentally different in nature from the powerful, jetted sources in Population B. The radio-emitting sources  of extreme Population A might be dominated by star formation, with the possible contribution of other mechanisms \cite{bonzinietal15,gancietal19,panessaetal19,chenetal24}.

\section{Candidate SMBBHs along the quasar main sequence}
\label{bin}

\subsection{Samples and measurements}

The sample of \citet{marzianietal13,marzianietal13a} is flux-limited and encompasses 680 objects from the DR7 of the SDSS, in the redshift range $0.4 \le z \le 0.7$\ to make it possible to simultaneously cover the \hb\ and the \mgii\ line. The flux limit ensures that the S/N is high enough for a reliable analysis of the \hb\ and \mgii\ profiles that are contaminated especially by extended \feii\ emission \citep{marzianietal03a,kovacevicdojcinovicpopopvic15}. The sample is large enough to ensure that all spectral bins identified in Fig. \ref{fig:e1bin}\ are sufficiently populated. In addition, even if \mgii\ lines are known to be narrower than \hb\ in Population B \cite{wangetal09,trakhtenbrotnetzer12,marzianietal13a,popovicetal19}, suggesting a larger emissivity-weighted distance from the continuum source for the \mgii\ line \cite{shenetal16,leetal20,yueetal24}, \hb\ and \mgii\ are both low-ionization lines expected to be predominantly emitted in a low-ionization, virialized part of the BLR \cite{collinsouffrin87,marzianietal96,marzianietal10}, and therefore to show consistent profiles.  

Our selection criterion is based on the presence of peculiarities in the emission line profiles. The correlations along the main sequence MS provide a robust framework to identify what is peculiar or, at the very least, rare. For example, the prominent red wing observed in the \hb\ profile is a characteristic feature encompassing most of Population B, which constitutes approximately half of optically selected quasar samples \citep{marzianietal03b,zamfiretal10}. While the origin of this feature remains unclear, it is not considered a peculiarity. In contrast, double- or multi-peaked profiles, as well as strongly shifted emission lines, are rare. A systematic line profile analysis along the MS reveals that they tend to be concentrated among quasars with the broadest \hb\ profiles \citep{sulenticetal00a}. These cases also frequently exhibit rapid profile variability, occasionally leading to the ``changing look" phenomenon \cite{pandasniegowska24,komossaetal24}. { In Population A, large inter-line shifts are easily understood in the framework of a system accretion disk + wind \cite{marzianietal96}.}

Seventeen sources belong to the spectral type B1$^{++}$, about 2.5 \%\ of the full sample. Their identification is reported in Table \ref{tab:sample}, along with the SDSS redshift, radio classification, qualitative \hb\ profile classification following \citet{sulentic89}, centroid shift at 0.9 fractional intensity. Measurements were obtained through non-linear, multi-components fits of the SDSS spectra within {\tt IRAF v.2.18} \cite{tody86,tody93,fitzpatricketal24}, using the task {\tt specfit} \cite{kriss94}. The same multi-component decomposition has an heuristic base and has been applied in tens of works in the past 25 years \cite{marzianietal10}. For a previous application on median spectra of the same sample considered here, see  \citet{marzianietal13a}.

The classification of \citet{sulentic89} assigns an asymmetry code to the broad line profile (asymmetric to the red (AR) or to the blue (AB), or symmetric) and a code for the line peak shift (R or B).  The centroid at different fractional intensity are as originally defined by \citet{marzianietal96} and applied in several more recent works \cite[{ for example,}][]{marzianietal22}.

\begin{table}[]
% \centering 
  \caption{Basic properties of the B1$^{++}$\ sample. }
\begin{tabular}{lcccccc}\hline\hline
 Identification & Jcode & $z_\mathrm{SDSS}$ & Radio & \hb\ & $c(0.9)$ \hb & Bin.? \\ 
 \multicolumn{1}{c}{(1)}& \multicolumn{1}{c}{(2)}& \multicolumn{1}{c}{(3)}&
 \multicolumn{1}{c}{(4)}& \multicolumn{1}{c}{(5)}&  \multicolumn{1}{c}{(6)} & \multicolumn{1}{c}{(7)} 
\\
 \hline
WISEA J093642.95+551119.2 &J093642&0.4971& RQ&AR,R&1309&Y\\
SDSS J101230.78+182021.1 &J101230&0.4623&RL&AR,B&$-480$&Y\\
FBQS J110001.0+231412 &J110001&0.5567&RL&AR,R&490&--\\
3C 254 &J111438&0.7359&RL&AR,R&2225&Y\\
FBQS J111903.2+385852 &J111903&0.7344&RL&AR,B&$-$2149&Y  \\
HB89 1156+631 &J115839&0.5924& RQ&AR,B&$-$349&Y\\
PG 1201+436 &J120424&0.6617& RQ&AR,B&-1374&Y$^\mathrm{a}$\\
FBQS J1300+2830 &J130028&0.6467&RL&AB,R&1781&Y\\
WISEA J130704.39+091004.1 &J130704&0.5247& RQ&AR,B&3086$^\mathrm{b}$&Y\\
SDSS J133051.90+184932.9 &J133051&0.5141&RL&AR,B&$-885$&Y\\
WISE J133655.49+654115.9 &J133655 &0.4378&RQ&AR,R&172&Y$^\mathrm{a}$\\
FBQS J140012.6+353930 &J140012&0.5184&RQ&AR,R&2341&Y\\
WISEA J141312.59+564113.3 &J141312&0.6686&RL&AR,0&53&—\\
WISEA J150249.02+081305.9 &J150249&0.5186& RQ&AR,B&7&Y\\
FBQS J153159.1+242047 &J153159&0.6321&RL&AR,R&597&—\\
WISEA J155330.23+223010.3 &J155330&0.6404&RQ&AR,B&$-$1809&Y\\
WISEA J163206.04+441659.5 &J163206&0.5304&RQ&AB,B&$-$411&Y\\ \hline
 \end{tabular}
{ Notes:} { Col. (1): common name recognizable by the NASA extragalactic ddatabase (NED); (2) code in the form Jhhmmss; (3) redshift   reported in the header of the original SDSS spectra; (4) radio classification following \citet{zamfiretal08}; (6) centroid at 0.9 fractional intensity of \hb, in \kms; (7) answer to the question: is the target a potential SMBBH candidate? } $^\mathrm{a}$: boxy $^\mathrm{b}$: Peak shifted to the blue $v \sim -3550$ \kms.\\
 \label{tab:sample}
\end{table}

\section{Results}
\label{resu}

The benchmark classification for non-peculiar sources in Population B is AR,R or AR,0, where peak shifts at peak are usually modest, $\lesssim$ a few hundreds \kms\ \cite{marzianietal03a,marzianietal03b,marzianietal09}. Large shifts ($\gtrsim 1000$ \kms) are rare and should be considered peculiar. Table \ref{tab:sample} reports the amplitude of the centroid at 0.9 fractional intensity $c(0.9)$\ that is considered a proxy of the line peak shift.  Only three sources in Table 1 show an \hb\ profile consistent with the reference AR,R profile for Population B. All the remaining sources show some peculiarities, with 8 sources showing large shifts with absolute amplitude $\gtrsim 1000$ \kms, suggesting that the origin of the perturbation is located in correspondence of the BLR. In one case (J130704) the actual peak is { even } displaced to the blue while the $c(0.9)$ is shifted to the red.\ This case highlights the large prevalence discordant asymmetries and peak shifts. %Lower shifts ($\sim \pm 500$ \kms) might indicate a perturber in the outer BLR. 

\begin{figure}
 \centering
 \includegraphics[width=0.95\linewidth]{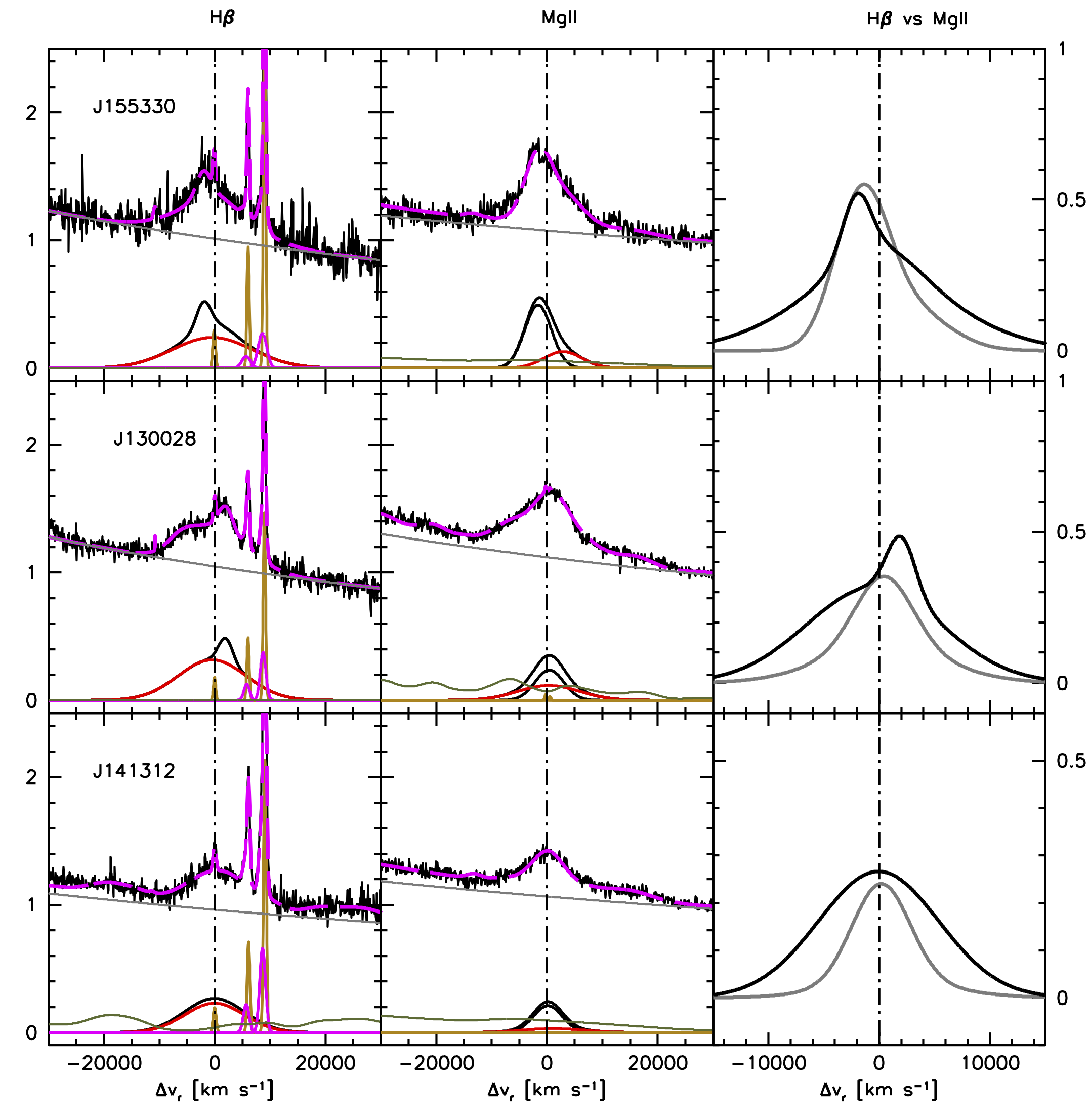}
 \caption{Examples of emission line profile analysis for \hb\ (left column) and \mgii\ (middle column). The top plot of each panel shows the original spectrum (thin black line) and the full model (dashed magenta line) along with the adopted continuum (gray) The bottom plots show the multiple components employed in the non-linear fit: broad and very broad \hb\ component (black and red Gaussians)  \cite{marzianisulentic93,sneddengaskell07}, \hb\ and \oiii\ narrow line emission (gold), semi-broad emission of \oiii\ (magenta) and \feii\ (dark green). \feii\ emission is very weak in most of the sample, both in the optical and in the UV.   The right column shows a comparison between the \hb\ and \mgii\ (gray) broad profiles.  }
 \label{fig:hbmgf}
\end{figure}

\begin{figure}
 \centering
 \includegraphics[width=1.\linewidth]{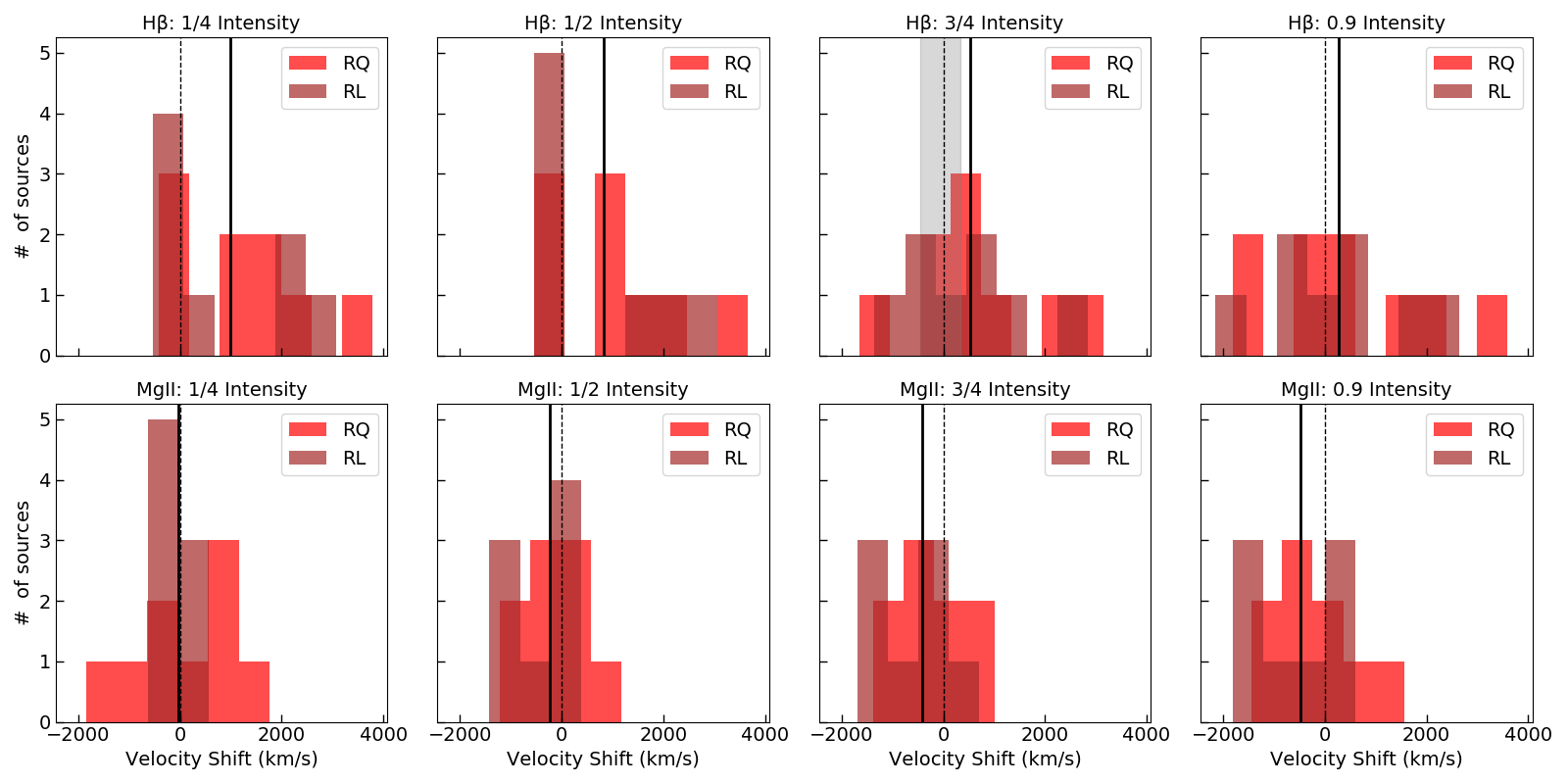}
 \caption{  Distribution of centroids at four different fractional intensity levels, from left to right $\frac{1}{4}$, $\frac{1}{2}$, $\frac{3}{4}$\ and $0.9$. Top: \hb, bottom: \mgii. Radio-loud sources are shaded brown. The rest frame radial velocity is represented by the dashed line, while the thick black line is the average of the sample. The shaded gray box shows the $\pm 1 \sigma$\ range for the $c(\frac{3}{4})$\ measurement of the B1 spectral type from the sample of \citet{zamfiretal10}. }
 \label{fig:centro}
\end{figure}

Fig. \ref{fig:hbmgf} shows three cases two of which depict the profiles of a binary system with a perturber with different, extreme projections of its velocity along the line of sight. The profile of \mgii\ is narrower and more symmetric, as found in previous studies \cite{wangetal09,trakhtenbrotnetzer12,shenetal16} and shown in the third column of Fig. \ref{fig:hbmgf}.  The classifications reported in Table \ref{tab:sample} suggest that most sources have a peak shift that might be associated with excess emission over an AR profile. This is confirmed by the distribution of centroids for \hb\ and \mgii. Fig. \ref{fig:centro} shows the distribution of the centroids at $\frac{1}{4}$, $\frac{1}{2}$, $\frac{3}{4}$\ and $0.9$\ fractional intensity. The $c(0.9)$ of \hb\ (bottom rightmost panel of Fig. \ref{fig:centro}) shows a uniform distribution over a range $\sim -2000 - +3000$ \kms. A consistent distribution is observed also for $c(\frac{3}{4})$. At lower fractional intensity, the shifts to the red associated with the prominent \hb\ red wing typical of population B dominates. In other words, apart from a few cases, the impression is the one of a perturbation affecting the top of the profile.

Fig. \ref{fig:hbmgf} and Fig. \ref{fig:centro} suggest two possibilities. The first is that the profile is affected by an outflow. The second is that there is a perturbation yielding a shifted peak. In the case of B1$^{++}$\ the outflow cannot be ruled out, but is not favored by the large shift of the profiles.  An outflow component is present in the \hb\ profile of Pop. B; however, it is typically associated with the semi-broad component of \oiii\ \citep{marzianietal22b} and with shifts of a few hundred \kms. 
The centroid at $\frac{3}{4}$\ measurements are available only for the sample of \citet{zamfiretal10}. Out of 169 B1 sources, only 3 show |$c(\frac{3}{4})$|$\gtrsim 1000$ \kms. The distribution of B1 is much more centrally concentrated with a modest dispersion (gray box in Fig. \ref{fig:centro}).  A Kolmogorov-Sminrnov test carried out on the 169 B1 and on the 13 B1$^{++}$\ confirms that the two distributions are significantly different at a confidence level $\gtrsim 4 \sigma$.

Prominent outflows affecting the  profiles up to $\gtrsim 1000$ \kms\ are seen only in Population A and extreme Population A at the opposite end of the sequence, where the ratio radiative  to gravitational forces is highest \cite{sulenticetal07,richardsetal11,temple24}. 

\begin{figure}
 \centering
 \includegraphics[width=0.95\linewidth]{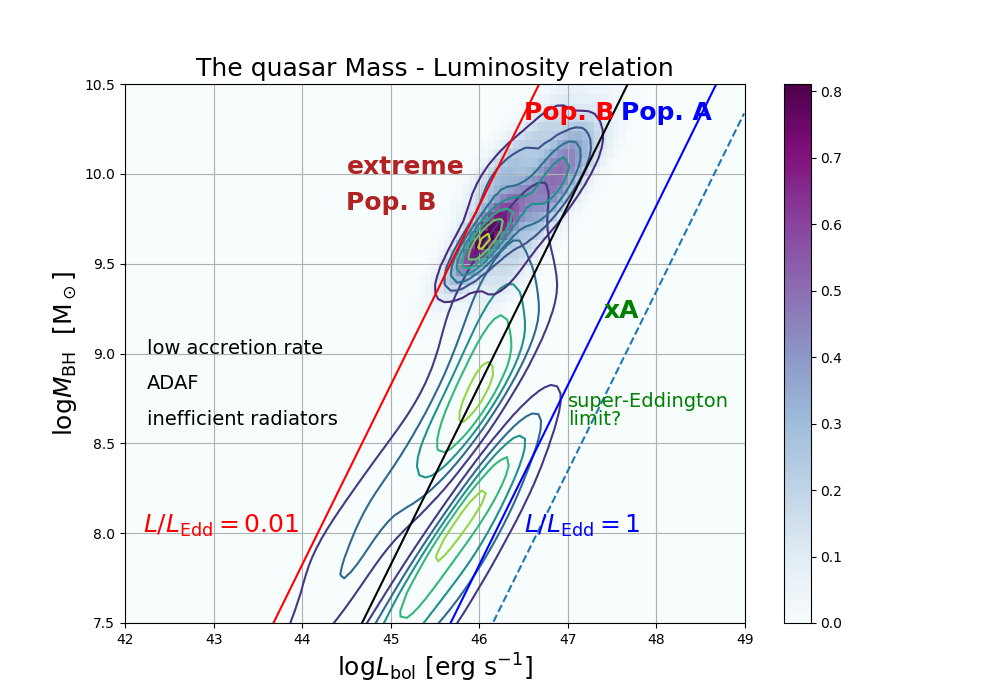}
 \caption{The black hole mass -- luminosity diagram for a low-$z$\ type quasar sample. The quasar location in the \mbh -- $L$\ diagram corresponds to the contour in the middle upper part of the diagram. The wide majority are located close  to the limit \lledd $ \sim 10^{-2}$, close to the domain expected for inefficient radiators. The black diagonal line at \lledd $= 0.1$\ separates Population A and B. On the high Eddington ratio side, the possible super-Eddington accretors are labeled as xA (extreme Population A), with Eddington ratios not exceeding 1 by a large factor, as found observationally \cite{marzianietal06} and as derived from theory of super-Eddington accretion disks as well \cite{mineshigeetal00,sadowski09,dotanshaviv11,abramowiczstaub14}.
 \label{fig:ml}}
\end{figure}

The interpretation of the red wing is not fully clear. However, the consensus is that it is a \mbh\ effect, either because of infall \cite{wangetal17} or because of gravitational and transverse redshift \cite{netzer77,corbin95,gavrilovicetal07,jonicetal16,bonetal15,marziani23}. The very broad, red wing becomes more prominent for very massive black holes powering the most luminous quasars \cite{marzianietal09,wolfetal20}, supporting the idea that the \hb\ line profile is mostly broadened in the velocity field of a very massive black hole  (as are the ones of extreme Population B). Fig. \ref{fig:ml} presents the luminosity-to-mass diagram for a sample of type-1 AGN, with the distribution of quasars from the considered sample overlaid for comparison. The data points cluster in correspondence of a fairly well defined limit in Eddington ratio \cite{marzianietal03b}, below which the black holes are expected to enter in an inefficient radiative domain   \cite[e.g.,][]{heckman80,narayanyi94,soriaetal06,giustiniproga19}. They involve very massive sources, if black hole mass is estimated using standard relations \cite{vestergaardpeterson06}. This property is consistent with the prominent redward asymmetry and large shifts to the red  toward the \hb\ line base that are observed in the B1$^{++}$\ sample, if they are due to gravitational and transverse redshift. The main body of the line profiles is attributed to the velocity field of virialized gas within the gravitational field of a supermassive black hole, with gravitational redshift becoming increasingly significant as the emitting gas approaches the center of gravity. In contrast, the peaks of the profiles appear to be influenced by perturbations that account for only a small fraction of the line flux. The last column of Table \ref{tab:sample} identifies the profiles that can be interpreted along this framework i.e., the sources that could be considered candidates for eventual monitoring.

%\subsection{}

\section{Discussion}
\label{discu}

\subsection{A SMBBH system}

The simplest interpretation in terms of a SMBBH system is of a very massive primary (leading to the \hb\ profile with redshift increasing toward the line base) with a secondary of small mass ratio $q \sim 0.1$, capable of inducing a measurable effect on the BLR region velocity field and accretion disk, and even jet precession \cite[e.g.,][]{romerozulema00,krauseetal11}. There are several factors that may increase the detectability of SMBBHs  in the spectral type B1$^{++}$. The first one is due to orientation. The current interpretation of the BLR of Pop. B is the one of a highly flattened system, coplanar with the accretion disk \cite{decarlietal11,afanasievetal19,marzianietal22b}.  The extreme line width, and the high prevalence of radio sources whose radio axis is seen at a large angle \cite{willsbrowne86,marin16,gancietal19}, indicates that the velocity field projection along the line of sight is maximized for B1$^{++}$\ sources. B1$^{++}$ can be connected to blazars (sources of low Eddington ratio and minimal \feii\ emission \cite{punslyetal18} by an exclusive orientation effect, { as blazars are oriented with the jet closely aligned to the line-of-sight (Fig. 3 by \citet{marzianietal22a}), but are otherwise believed to be low accretors}.  However, orientation alone cannot explain the properties of the B1$^{++}$.

The low FeII emission, and the extreme line width with extreme redward asymmetries point toward very extremely high \mbh\ and extremely low \lledd.  Indeed, the B1$^{++}$\ sources show similar \mbh $\sim 10^{10}$ solar masses, and moderate luminosity due to low Eddington ratio $\sim 10^{-2}$.  
 In other words, they are evolved system close to (but not yet in) the condition of ``spent" quasars \cite{lynden-bell69}.  The presence of a second black hole with $q \gtrsim 0.04$\ may induce a truncation in the BLR emission, as  the   mass flow in circum-binary Keplerian disk is perturbed by tidal forces which carve annular gaps in the second black hole orbital path \cite{linpapaloizou86, artymowiczlubow94}, hampering to various extent the accretion process \cite{tiedeetal22}. This might be the case of the classical double ``peakers" with broad line profile consistent with the emission from an accretion disk truncated at $\sim 1000$\ gravitational radii \cite{chenetal89,eracleoushalpern03,stratevaetal03,terefemengistueetal23}.

The average optical luminosity of the B1$^{++}$\ quasars    $\sim 5 \cdot 10^{45}$ \ergs, implies a BLR radius { $\sim 5 \cdot  10^{17}$cm} $ \lesssim 1 $ pc, identifying these  systems as  bona-fide sub-parsec  binary candidates.  The time to coalescence because of gravitational wave emission is provided by the relation  
$
T_\mathrm{c} = \frac{15}{304} \frac{c^5 a_0^4}{G^3 M_1 M_2 (M_1 + M_2)} \cdot (1 - e_0^2)^{7/2} \cdot \frac{1}{g(e_0)}
$,
 where $ g(e_0) = 1 + \frac{73}{24} e_0^2 + \frac{37}{96} e_0^4$ \cite{peters64,deng22}, where $e_0$\ is the initial eccentricity. In our case, assuming that $M_2 = 0.1 M_1$, with $M_1 = 10^{10}$ solar masses, we obtain a coalescence time $\sim   10^7$ yr for a circular orbit for $e_0 = 0$. The $
T_\mathrm{c}$\ can be significantly shortened only if the eccentricity is rather high: by a factor $10$ if $e_0 = 0.6$, and by a factor 100 if  $e_0 = 0.8$. With an orbital { period} $P \gtrsim 10^2$ yr, the immediate prediction is that the peculiar, displaced peaks { should} not show any short term variation ($\lesssim 1$ yr), { while a systematic shift over 10 -- 20 yr might occur}. At least, the  B1$^{++}$ spectral type can be considered an appropriate testbed for the SMBBH model { \cite[see also][]{nguyenbogdanovic16}.}  

The inspiral motion of a less massive black hole in the gravitational field of a supermassive black hole produces a distinctive gravitational wave signal characterized by long-duration, low-frequency waveforms. These signals result from the gradual inspiral of a compact object toward the SMBH. { For the sample considered in this study, the black hole masses are extremely large and inspiral motions may produces waves of frequencies in the domain of the Pulsar Timing Array (\cite{hobbsetal10,mingarellietal17}, $\nu \sim 10^{-9}$). }

\subsection{The latest phases of the evolution along the quasar main sequence}

{ We might expect a rich phenomenology   in the frequency domains covered by planned space- and ground-based detectors  due to the remnants of a nuclear stellar system \cite{antoniniperets12}.}
Nuclear (i.e., {\em within the molecular torus}) { and circumnuclear}  star formation occurs in most AGN with moderate or high accretion rates \citep{wangetal09sf,wangetal10sf,wangetal11sf,wangetal12sf}. The outer self-gravitating disk and torus provide gas for nuclear fueling, with star formation naturally associated with high or super-Eddington accretion \citep{artymowiczetal93,lin97,collinzahn99,wangetal23,dittmanncantiello24,fabjetal24,liuetal24}. As quasars evolve, feedback processes deplete surrounding material, leading to lower \lledd\ ratios and a transition into Population B, characterized by weaker outflows and stronger { core} \oiii\ emission \citep{sulenticetal00a}. \citet{fraix-burnetetal17} list several multi-frequency parameters that are systematically different between Population A and B.  Extreme population B sources can be thought as simply reaching the most extreme values in most observational parameters of Population B.  The torus may be absent in such low-luminosity AGN \citep{elitzurshlosman06}, leaving the black hole without a large reservoir of accreting matter. The point here is what may happen in the nuclear regions of such evolved { stellar } system.

The fate of  such a system remains unclear.  { The best prospect for detecting it lies in } gravitational radiation, since the electromagnetic signal is overwhelmed by the luminosity of the AGN.    The scenario envisaged for super-Eddington accretors will ultimately lead to a population of stellar mass compact objects, and  to intermediate mass black holes (IMBHs)  \cite{lin97,portegieszwartetal02,portegiesetal04,gaeteetal24,chenlin24}.

\subsection{GW from IMBH and stellar mass black hole coalescence}

{ Observed nuclear star clusters have extremely high stellar density, in excess of  $10^6$\ M$_\odot$\ \cite{neumayeretal20,boker09}, and runaway collisions of massive stars   \cite{portegieszwartetal02,portegiesetal04} could be a  pathway to generate IMBHs  of $10^4 - 10^5$ M$_{\odot}$. A second pathway is via the enhancement of stellar black hole merger rates in the dense environment of the accretion disk due to   the increased probability of BH–BH encounters, binary formation via orbital migration and migration  traps \cite{tagawaetal20}. }  An IMBH     could in turn become a seed black hole in a binary system, and lead to an extreme mass ratio inspirals (EMRI) phenomenology.   The gravitational wave frequencies from EMRIs  typically fall in the millihertz range, which makes them ideal targets for future space-based gravitational wave detectors like the Laser Interferometer Space Antenna \cite[LISA,][]{amaro-seoaneetal17,donofriomarziani18}.  LISA will be sensitive to the low-frequency gravitational waves emitted by EMRIs with primary masses covering the domain of IMBHs up to moderately SMBBH ($\sim 10^8$ M$_\odot$, \cite{amaro-seoaneetal17,baracketal19}), which cannot be detected by ground-based observatories due to their low frequency range. 

{ The first IMBH  gravitational-wave detections, such as GW150914,    involved black holes of about 29 and 36 solar masses  \cite{abbottetal15} which   are thought to form  by direct collapse of massive stars in low-metallicity environments \cite{ebisuzakietal01,mapelli16}.}  If the progenitor of the remnant black holes were high metallicity stars, it { might} difficult that the { initial} mass after collapse could be as large as 30 M$_\odot$
{ \cite{belczynskietal10,giacobbomapelli18}}. In AGN disks, { stellar-mass black hole} mergers are expected to occur at a higher rate than in isolated environments \cite{grobneretal20}, { and produce gravitational wave bursts \cite{wangetal21a}}. The high frequency of mergers is expected in turn to produce   a post-AGN population of $\sim 10$ M$_\odot$\ black holes \cite{jermynetal22},  along with a population of IMBHs \cite{jermynetal22,chenlin24}.  The signal amplitude of gravitational waves from a  binary BH  merger is typically expressed in terms of the strain amplitude $h$, which depends on the masses of the black holes, the distance to the merger, and the orientation of the system relative to the observer.  For a stellar-mass black hole merger, let's assume both black holes have masses of $ 5 M_\odot $, typical of a  { general } population of remnant black holes  \cite{siciliaetal22}. The chirp mass    is $  M_\mathrm{c} = \frac{(M_1 M_2)^{3/5}}{(M_1 + M_2)^{1/5}} \approx 4.35 M_\odot$. The GW strain amplitude $ h $ for a circular binary inspiral of two masses $ M_1 $ and $ M_2 $ is approximated by  $h \approx \frac{4 (G M_c)^{5/3}}{c^4 D_L} ( \pi f )^{2/3}$, 
where $ G $ is the gravitational constant,    $ f $ is the gravitational wave frequency,  $ D_L $ is the luminosity distance to the source, and $ c $ is the speed of light. Frequencies in the range of 10 Hz to 1000 Hz are typically considered in a binary black hole inspiral. Ground-based interferometers like the Laser Interferometer Gravitational-Wave Observatory     \cite[LIGO][]{LIGOcollaboration15} and Virgo \cite{acerneseetal15} are optimized for detecting the mergers of stellar-mass black holes.   A frequency $ f = 100 \, \textrm{Hz} $  is in the sensitivity range for both LIGO and the Einstein Telescope \cite[ET, ][]{punturoetal10,maggioreetal20}. The strain $ h $ for a merger at $ z \approx 0.55 $ is $h \sim  10^{-23} $. 
%LIGO is sensitive to gravitational waves with frequencies between 10 Hz and 1000 Hz, with the best sensitivity at frequencies around 100 Hz. 
The strain sensitivity of LIGO's O3 run is about $ h \sim 10^{-23} $ to $ 10^{-24} $ at its peak sensitivity. This means that LIGO   detects BBH mergers up to typical redshift of the B1$^{++}$ sample. The ET will be 10 to 100 times more sensitive than LIGO, extending the detection horizon to much higher redshifts.  The ET will be capable of detecting strain amplitudes as low as $ h \sim 10^{-25} $ or even smaller, and sample comfortably the stellar-mass BH remnant population expected in evolved  AGN.

\section{Conclusions}

This study has highlighted the potential of the quasar main sequence as a framework to identify promising candidates for supermassive binary black holes (SMBBHs). By focusing on a specific spectral type along the quasar main sequence, B1$^{++}$, we identified emission line peculiarities, such as large centroid shifts, which may indicate the presence of secondary compact objects perturbing the accretion dynamics. These findings suggest that high-mass black holes with low Eddington ratios, typical of B1$^{++}$ sources, are favorable environments for detecting SMBBHs.

Future multi-wavelength campaigns and gravitational wave observations will be essential to confirm these systems. Sub-parsec binaries within this spectral class may also provide crucial insights into the role of tidal forces and binary-induced disk perturbations in shaping AGN emission profiles and evolutionary pathways.
 
%%%%%%%%%%%%%%%%%%%%%%%%%%%%%%%%%%%%%%%%%%
\authorcontributions{Writing---original draft preparation, P.M.; Conceptualization, reading and correcting draft, E.B., N.B., and M.D.O.}

\funding{ }

%\institutionalreview{ }

%\informedconsent{

\dataavailability{This paper is based  entirely on published data (SDSS). A table of measurement has been uploaded along with the paper.  } 

\acknowledgments{E.B. and N.B. acknowledge the support of the Ministry of Science, Technological Development and 
Innovation of the Republic of Serbia, contract No. 451-03-66/2024-03/200002.

Funding for the SDSS and SDSS-II has been provided by the Alfred P. Sloan Foundation, the Participating Institutions, the National Science Foundation, the U.S. Department of Energy, the National Aeronautics and Space Administration, the Japanese Monbukagakusho, the Max Planck Society, and the Higher Education Funding Council for England. The SDSS Web Site is http://www.sdss.org/.

The SDSS is managed by the Astrophysical Research Consortium for the Participating Institutions. The Participating Institutions are the American Museum of Natural History, Astrophysical Institute Potsdam, University of Basel, University of Cambridge, Case Western Reserve University, University of Chicago, Drexel University, Fermilab, the Institute for Advanced Study, the Japan Participation Group, Johns Hopkins University, the Joint Institute for Nuclear Astrophysics, the Kavli Institute for Particle Astrophysics and Cosmology, the Korean Scientist Group, the Chinese Academy of Sciences (LAMOST), Los Alamos National Laboratory, the Max-Planck-Institute for Astronomy (MPIA), the Max-Planck-Institute for Astrophysics (MPA), New Mexico State University, Ohio State University, University of Pittsburgh, University of Portsmouth, Princeton University, the United States Naval Observatory, and the University of Washington.
}

%\conflictsofinterest{ } 

%%%%%%%%%%%%%%%%%%%%%%%%%%%%%%%%%%%%%%%%%%
%% Optional
%\sampleavailability{Samples of the compounds are available from the authors.}

%% Only for journal Encyclopedia
%\entrylink{The Link to this entry published on the encyclopedia platform.}

\abbreviations{Abbreviations}{
The following abbreviations are used in this manuscript:\\

\noindent 
\begin{tabular}{@{}ll}
AGN & Active Galactic Nucleus/i \\
BH & Black Hole \\
BLR & Broad Line Region \\
E1 & Eigenvector 1\\
ET & Einstein Telescope \\
EMRI & Extreme mass ratio inspiral \\
FWHM & Full Width Half Maximum \\
GW & Gravitational Wave \\
LIGO & Laser Interferometer Gravitational-Wave Observatory\\
LISA & Laser Interferometer Space Antenna \\ 
LIL & Low Ionization Line\\
MDPI & Multidisciplinary Digital Publishing Institute\\
MS & Main Sequence\\
NLSy1 &  Narrow  Line Seyfert 1 \\
PCA & Principal Component Analysis \\
PTF & Palomar Transient Factory \\ 
RL & Radio-Loud\\
RQ & Radio-Quiet\\

SDSS & Sloan Digital Sky Survey\\
SMBH & Super-Massive Black Hole \\
SMBBH & Super-Massive Binary Black Hole \\
xA & extreme (Population) A \\
ZTF & Zwicky Transient Facility \\
\end{tabular}
}

%%%%%%%%%%%%%%%%%%%%%%%%%%%%%%%%%%%%%%%%%%
%% Optional
%\appendixtitles{no} % Leave argument "no" if all appendix headings stay EMPTY (then no dot is printed after "Appendix A"). If the appendix sections contain a heading then change the argument to "yes".
%\appendixstart
%\appendix
%\section[\appendixname~\thesection]{}
%\subsection[\appendixname~\thesubsection]{}

%dix is an optional section that can contain details and data supplemental to the main text---for example, explanations of experimental details that would disrupt the flow of the main text but nonetheless remain crucial to understanding and reproducing the research shown; figures of replicates for experiments of which representative data are shown in the main text can be added here if brief, or as Supplementary Data. Mathematical proofs of results not central to the paper can be added as an appendix.

%%%%%%%%%%%%%%%%%%%%%%%%%%%%%%%%%%%%%%%%%%
\begin{adjustwidth}{-\extralength}{0cm}

\reftitle{References}


%\bibliography{biblioletter2}

\begin{thebibliography}{999}

\bibitem[{Colpi}(2014)]{colpi14}
{Colpi}, M.
\newblock {Massive Binary Black Holes in Galactic Nuclei and Their Path to
  Coalescence}.
\newblock {\em \ssr} {\bf 2014}, {\em 183},~189--221,
  \href{http://xxx.lanl.gov/abs/1407.3102}{{\normalfont
  [arXiv:astro-ph.GA/1407.3102]}}.
\newblock {\url{https://doi.org/10.1007/s11214-014-0067-1}}.

\bibitem[{Komossa} \em{et~al.}(2021){Komossa}, {Ciprini}, {Dey}, {Gallo},
  {Gomez}, {Gonzalez}, {Grupe}, {Kraus}, {Laine}, {Parker}, {Valtonen},
  {Chandra}, {Gopakumar}, {Haggard}, and {Nowak}]{komossaetal21}
{Komossa}, S.; {Ciprini}, S.; {Dey}, L.; {Gallo}, L.C.; {Gomez}, J.L.;
  {Gonzalez}, A.; {Grupe}, D.; {Kraus}, A.; {Laine}, S.J.; {Parker}, M.L.;
  et~al.
\newblock {Supermassive Binary Black Holes and the Case of OJ 287}.
\newblock In Proceedings of the XIX Serbian Astronomical Conference,
  Publications of the Astronomical Observatory of Belgrade;
  {Kova{\v{c}}evi{\'c}}, A.; {Kova{\v{c}}evi{\'c} Doj{\v{c}}inovi{\'c}}, J.;
  {Mar{\v{c}}eta}, D.; {Oni{\'c}}, D., Eds.,  2021, Vol. 100, pp. 29--42,
  \href{http://xxx.lanl.gov/abs/2104.12901}{{\normalfont
  [arXiv:astro-ph.HE/2104.12901]}}.
\newblock {\url{https://doi.org/10.48550/arXiv.2104.12901}}.

\bibitem[{Komossa} and {Grupe}(2024)]{komossagrupe24}
{Komossa}, S.; {Grupe}, D.
\newblock {The Extremes of Continuum and Emission-Line Variability of AGN:
  Changing-Look Events and Binary SMBHS}.
\newblock {\em Serbian Astronomical Journal} {\bf 2024}, {\em 209},~1--24.
\newblock {\url{https://doi.org/10.2298/SAJ2409001K}}.

\bibitem[{Burke-Spolaor} \em{et~al.}(2011){Burke-Spolaor}, {Bailes},
  {Johnston}, {Bates}, {Bhat}, {Burgay}, {D'Amico}, {Jameson}, {Keith},
  {Kramer}, {Levin}, {Milia}, {Possenti}, {Stappers}, and {van
  Straten}]{burke-spolaoretal11}
{Burke-Spolaor}, S.; {Bailes}, M.; {Johnston}, S.; {Bates}, S.D.; {Bhat},
  N.D.R.; {Burgay}, M.; {D'Amico}, N.; {Jameson}, A.; {Keith}, M.J.; {Kramer},
  M.;  et~al.
\newblock {The High Time Resolution Universe Pulsar Survey - III. Single-pulse
  searches and preliminary analysis}.
\newblock {\em \mnras} {\bf 2011}, {\em 416},~2465--2476,
  \href{http://xxx.lanl.gov/abs/1102.4111}{{\normalfont
  [arXiv:astro-ph.SR/1102.4111]}}.
\newblock {\url{https://doi.org/10.1111/j.1365-2966.2011.18521.x}}.

\bibitem[{Theureau} \em{et~al.}(2021){Theureau}, {Babak}, {Berthereau},
  {Chalumeau}, {Chen}, {Cognard}, {Falxa}, {Guillemot}, and
  {Petiteau}]{thereauetal21}
{Theureau}, G.; {Babak}, S.; {Berthereau}, A.; {Chalumeau}, A.; {Chen}, S.;
  {Cognard}, I.; {Falxa}, M.; {Guillemot}, L.; {Petiteau}, A.
\newblock {Pulsar Timing Arrays and gravitational waves : the first steps
  towards detection?}
\newblock In Proceedings of the SF2A-2021: Proceedings of the Annual meeting of
  the French Society of Astronomy and Astrophysics; {Siebert}, A.;
  {Bailli{\'e}}, K.; {Lagadec}, E.; {Lagarde}, N.; {Malzac}, J.; {Marquette},
  J.B.; {N'Diaye}, M.; {Richard}, J.; {Venot}, O., Eds.,  2021, pp. 23--28.

\bibitem[{Berti} \em{et~al.}(2006){Berti}, {Cardoso}, and {Will}]{bertietal06}
{Berti}, E.; {Cardoso}, V.; {Will}, C.M.
\newblock {Gravitational-wave spectroscopy of massive black holes with the
  space interferometer LISA}.
\newblock {\em \prd} {\bf 2006}, {\em 73},~064030,
  \href{http://xxx.lanl.gov/abs/gr-qc/0512160}{{\normalfont
  [arXiv:gr-qc/gr-qc/0512160]}}.
\newblock {\url{https://doi.org/10.1103/PhysRevD.73.064030}}.

\bibitem[{Amaro-Seoane} \em{et~al.}(2017){Amaro-Seoane}, {Audley}, {Babak},
  {Baker}, {Barausse}, {Bender}, {Berti}, {Binetruy}, {Born}, {Bortoluzzi},
  {Camp}, {Caprini}, {Cardoso}, {Colpi}, {Conklin}, {Cornish}, {Cutler},
  {Danzmann}, {Dolesi}, {Ferraioli}, {Ferroni}, {Fitzsimons}, {Gair}, {Gesa
  Bote}, {Giardini}, {Gibert}, {Grimani}, {Halloin}, {Heinzel}, {Hertog},
  {Hewitson}, {Holley-Bockelmann}, {Hollington}, {Hueller}, {Inchauspe},
  {Jetzer}, {Karnesis}, {Killow}, {Klein}, {Klipstein}, {Korsakova}, {Larson},
  {Livas}, {Lloro}, {Man}, {Mance}, {Martino}, {Mateos}, {McKenzie},
  {McWilliams}, {Miller}, {Mueller}, {Nardini}, {Nelemans}, {Nofrarias},
  {Petiteau}, {Pivato}, {Plagnol}, {Porter}, {Reiche}, {Robertson},
  {Robertson}, {Rossi}, {Russano}, {Schutz}, {Sesana}, {Shoemaker}, {Slutsky},
  {Sopuerta}, {Sumner}, {Tamanini}, {Thorpe}, {Troebs}, {Vallisneri},
  {Vecchio}, {Vetrugno}, {Vitale}, {Volonteri}, {Wanner}, {Ward}, {Wass},
  {Weber}, {Ziemer}, and {Zweifel}]{amaro-seoaneetal17}
{Amaro-Seoane}, P.; {Audley}, H.; {Babak}, S.; {Baker}, J.; {Barausse}, E.;
  {Bender}, P.; {Berti}, E.; {Binetruy}, P.; {Born}, M.; {Bortoluzzi}, D.;
  et~al.
\newblock {Laser Interferometer Space Antenna}.
\newblock {\em arXiv e-prints} {\bf 2017}, p. arXiv:1702.00786,
  \href{http://xxx.lanl.gov/abs/1702.00786}{{\normalfont
  [arXiv:astro-ph.IM/1702.00786]}}.
\newblock {\url{https://doi.org/10.48550/arXiv.1702.00786}}.

\bibitem[{D'Onofrio} and {Marziani}(2018)]{donofriomarziani18}
{D'Onofrio}, M.; {Marziani}, P.
\newblock {A multimessenger view of galaxies and quasars from now to
  mid-century}.
\newblock {\em Frontiers in Astronomy and Space Sciences} {\bf 2018}, {\em
  5},~31,  \href{http://xxx.lanl.gov/abs/1807.07435}{{\normalfont
  [arXiv:astro-ph.GA/1807.07435]}}.
\newblock {\url{https://doi.org/10.3389/fspas.2018.00031}}.

\bibitem[{Sillanpaa} \em{et~al.}(1988){Sillanpaa}, {Haarala}, {Valtonen},
  {Sundelius}, and {Byrd}]{sillanpaaetal88}
{Sillanpaa}, A.; {Haarala}, S.; {Valtonen}, M.J.; {Sundelius}, B.; {Byrd}, G.G.
\newblock {OJ 287 - Binary pair of supermassive black holes}.
\newblock {\em \apj} {\bf 1988}, {\em 325},~628--634.
\newblock {\url{https://doi.org/10.1086/166033}}.

\bibitem[Valtonen \em{et~al.}(2008)Valtonen, Lehto, et~al.]{valtonen2008}
Valtonen, M.J.; Lehto, H.J.;  et~al.
\newblock A massive binary black-hole system in OJ 287 and a test of general
  relativity.
\newblock {\em Nature} {\bf 2008}, {\em 452},~851--853.
\newblock {\url{https://doi.org/10.1038/nature06896}}.

\bibitem[{Valtonen} \em{et~al.}(2023){Valtonen}, {Dey}, {Gopakumar}, {Zola},
  {L{\"a}hteenm{\"a}ki}, {Tornikoski}, {Gupta}, {Pursimo}, {Knudstrup},
  {Gomez}, {Hudec}, {Jel{\'\i}nek}, {{\v{S}}trobl}, {Berdyugin}, {Ciprini},
  {Reichart}, {Kouprianov}, {Matsumoto}, {Drozdz}, {Mugrauer}, {Sadun},
  {Zejmo}, {Sillanp{\"a}{\"a}}, {Lehto}, {Nilsson}, {Imazawa}, and
  {Uemura}]{valtonenetal23}
{Valtonen}, M.J.; {Dey}, L.; {Gopakumar}, A.; {Zola}, S.;
  {L{\"a}hteenm{\"a}ki}, A.; {Tornikoski}, M.; {Gupta}, A.C.; {Pursimo}, T.;
  {Knudstrup}, E.; {Gomez}, J.L.;  et~al.
\newblock {Observational Implications of OJ 287's Predicted 2022 Disk Impact in
  the Black Hole Binary Model}.
\newblock {\em Galaxies} {\bf 2023}, {\em 11},~82,
  \href{http://xxx.lanl.gov/abs/2308.01878}{{\normalfont
  [arXiv:astro-ph.HE/2308.01878]}}.
\newblock {\url{https://doi.org/10.3390/galaxies11040082}}.

\bibitem[{Eracleous} \em{et~al.}(1997){Eracleous}, {Halpern}, {Gilbert},
  {Newman}, and {Filippenko}]{eracleousetal97}
{Eracleous}, M.; {Halpern}, J.P.; {Gilbert}, A.M.; {Newman}, J.A.;
  {Filippenko}, A.V.
\newblock {Rejection of the Binary Broad-Line Region Interpretation of
  Double-peaked Emission Lines in Three Active Galactic Nuclei}.
\newblock {\em \apj} {\bf 1997}, {\em 490},~216--+,
  \href{http://xxx.lanl.gov/abs/arXiv:astro-ph/9706222}{{\normalfont
  [arXiv:astro-ph/9706222]}}.
\newblock {\url{https://doi.org/10.1086/304859}}.

\bibitem[{Gezari} \em{et~al.}(2007){Gezari}, {Halpern}, and
  {Eracleous}]{gezarietal07}
{Gezari}, S.; {Halpern}, J.P.; {Eracleous}, M.
\newblock {Long-Term Profile Variability of Double-peaked Emission Lines in
  Active Galactic Nuclei}.
\newblock {\em \apjs} {\bf 2007}, {\em 169},~167--212,
  \href{http://xxx.lanl.gov/abs/arXiv:astro-ph/0702594}{{\normalfont
  [arXiv:astro-ph/0702594]}}.
\newblock {\url{https://doi.org/10.1086/511032}}.

\bibitem[{Liu} \em{et~al.}(2018){Liu}, {Gezari}, and {Miller}]{liuetal18}
{Liu}, T.; {Gezari}, S.; {Miller}, M.C.
\newblock {Did ASAS-SN Kill the Supermassive Black Hole Binary Candidate
  PG1302-102?}
\newblock {\em \apjl} {\bf 2018}, {\em 859},~L12,
  \href{http://xxx.lanl.gov/abs/1803.05448}{{\normalfont
  [arXiv:astro-ph.HE/1803.05448]}}.
\newblock {\url{https://doi.org/10.3847/2041-8213/aac2ed}}.

\bibitem[Volonteri \em{et~al.}(2003)Volonteri, Haardt, and
  Madau]{volonteri2003}
Volonteri, M.; Haardt, F.; Madau, P.
\newblock The Assembly and Merging History of Supermassive Black Holes in
  Hierarchical Models of Galaxy Formation.
\newblock {\em The Astrophysical Journal} {\bf 2003}, {\em 582},~559--573.
\newblock {\url{https://doi.org/10.1086/344675}}.

\bibitem[{Volonteri} \em{et~al.}(2009){Volonteri}, {Miller}, and
  {Dotti}]{volonterietal09}
{Volonteri}, M.; {Miller}, J.M.; {Dotti}, M.
\newblock {Sub-Parsec Supermassive Binary Quasars: Expectations at z {$<$} 1}.
\newblock {\em \apjl} {\bf 2009}, {\em 703},~L86--L89,
  \href{http://xxx.lanl.gov/abs/0903.3947}{{\normalfont
  [arXiv:astro-ph.CO/0903.3947]}}.
\newblock {\url{https://doi.org/10.1088/0004-637X/703/1/L86}}.

\bibitem[Merritt and Milosavljevi{\'c}(2005)]{merritt2005}
Merritt, D.; Milosavljevi{\'c}, M.
\newblock Massive black hole binary evolution in stellar environments:
  Implications for gravitational wave detection.
\newblock {\em Living Reviews in Relativity} {\bf 2005}, {\em 8},~8.
\newblock {\url{https://doi.org/10.12942/lrr-2005-8}}.

\bibitem[{Graham} \em{et~al.}(2015){Graham}, {Djorgovski}, {Stern}, {Drake},
  {Mahabal}, {Donalek}, {Glikman}, {Larson}, and {Christensen}]{grahametal15}
{Graham}, M.J.; {Djorgovski}, S.G.; {Stern}, D.; {Drake}, A.J.; {Mahabal},
  A.A.; {Donalek}, C.; {Glikman}, E.; {Larson}, S.; {Christensen}, E.
\newblock {A systematic search for close supermassive black hole binaries in
  the Catalina Real-time Transient Survey}.
\newblock {\em \mnras} {\bf 2015}, {\em 453},~1562--1576,
  \href{http://xxx.lanl.gov/abs/1507.07603}{{\normalfont [1507.07603]}}.
\newblock {\url{https://doi.org/10.1093/mnras/stv1726}}.

\bibitem[{Chambers} \em{et~al.}(2016){Chambers}, {Magnier}, {Metcalfe},
  {Flewelling}, {Huber}, {Waters}, {Denneau}, {Draper}, {Farrow}, {Finkbeiner},
  {Holmberg}, {Koppenhoefer}, {Price}, {Rest}, {Saglia}, {Schlafly}, {Smartt},
  {Sweeney}, {Wainscoat}, {Burgett}, {Chastel}, {Grav}, {Heasley}, {Hodapp},
  {Jedicke}, {Kaiser}, {Kudritzki}, {Luppino}, {Lupton}, {Monet}, {Morgan},
  {Onaka}, {Shiao}, {Stubbs}, {Tonry}, {White}, {Ba{\~n}ados}, {Bell},
  {Bender}, {Bernard}, {Boegner}, {Boffi}, {Botticella}, {Calamida},
  {Casertano}, {Chen}, {Chen}, {Cole}, {Deacon}, {Frenk}, {Fitzsimmons},
  {Gezari}, {Gibbs}, {Goessl}, {Goggia}, {Gourgue}, {Goldman}, {Grant},
  {Grebel}, {Hambly}, {Hasinger}, {Heavens}, {Heckman}, {Henderson}, {Henning},
  {Holman}, {Hopp}, {Ip}, {Isani}, {Jackson}, {Keyes}, {Koekemoer}, {Kotak},
  {Le}, {Liska}, {Long}, {Lucey}, {Liu}, {Martin}, {Masci}, {McLean}, {Mindel},
  {Misra}, {Morganson}, {Murphy}, {Obaika}, {Narayan}, {Nieto-Santisteban},
  {Norberg}, {Peacock}, {Pier}, {Postman}, {Primak}, {Rae}, {Rai}, {Riess},
  {Riffeser}, {Rix}, {R{\"o}ser}, {Russel}, {Rutz}, {Schilbach}, {Schultz},
  {Scolnic}, {Strolger}, {Szalay}, {Seitz}, {Small}, {Smith}, {Soderblom},
  {Taylor}, {Thomson}, {Taylor}, {Thakar}, {Thiel}, {Thilker}, {Unger},
  {Urata}, {Valenti}, {Wagner}, {Walder}, {Walter}, {Watters}, {Werner},
  {Wood-Vasey}, and {Wyse}]{chambersetal16}
{Chambers}, K.C.; {Magnier}, E.A.; {Metcalfe}, N.; {Flewelling}, H.A.; {Huber},
  M.E.; {Waters}, C.Z.; {Denneau}, L.; {Draper}, P.W.; {Farrow}, D.;
  {Finkbeiner}, D.P.;  et~al.
\newblock {The Pan-STARRS1 Surveys}.
\newblock {\em arXiv e-prints} {\bf 2016}, p. arXiv:1612.05560,
  \href{http://xxx.lanl.gov/abs/1612.05560}{{\normalfont
  [arXiv:astro-ph.IM/1612.05560]}}.
\newblock {\url{https://doi.org/10.48550/arXiv.1612.05560}}.

\bibitem[{Bellm} \em{et~al.}(2019){Bellm}, {Kulkarni}, {Graham}, {Dekany},
  {Smith}, {Riddle}, {Masci}, {Helou}, {Prince}, {Adams}, {Barbarino},
  {Barlow}, {Bauer}, {Beck}, {Belicki}, {Biswas}, {Blagorodnova}, {Bodewits},
  {Bolin}, {Brinnel}, {Brooke}, {Bue}, {Bulla}, {Burruss}, {Cenko}, {Chang},
  {Connolly}, {Coughlin}, {Cromer}, {Cunningham}, {De}, {Delacroix}, {Desai},
  {Duev}, {Eadie}, {Farnham}, {Feeney}, {Feindt}, {Flynn}, {Franckowiak},
  {Frederick}, {Fremling}, {Gal-Yam}, {Gezari}, {Giomi}, {Goldstein},
  {Golkhou}, {Goobar}, {Groom}, {Hacopians}, {Hale}, {Henning}, {Ho}, {Hover},
  {Howell}, {Hung}, {Huppenkothen}, {Imel}, {Ip}, {Ivezi{\'c}}, {Jackson},
  {Jones}, {Juric}, {Kasliwal}, {Kaspi}, {Kaye}, {Kelley}, {Kowalski},
  {Kramer}, {Kupfer}, {Landry}, {Laher}, {Lee}, {Lin}, {Lin}, {Lunnan},
  {Giomi}, {Mahabal}, {Mao}, {Miller}, {Monkewitz}, {Murphy}, {Ngeow},
  {Nordin}, {Nugent}, {Ofek}, {Patterson}, {Penprase}, {Porter}, {Rauch},
  {Rebbapragada}, {Reiley}, {Rigault}, {Rodriguez}, {van Roestel}, {Rusholme},
  {van Santen}, {Schulze}, {Shupe}, {Singer}, {Soumagnac}, {Stein}, {Surace},
  {Sollerman}, {Szkody}, {Taddia}, {Terek}, {Van Sistine}, {van Velzen},
  {Vestrand}, {Walters}, {Ward}, {Ye}, {Yu}, {Yan}, and
  {Zolkower}]{bellmetal19}
{Bellm}, E.C.; {Kulkarni}, S.R.; {Graham}, M.J.; {Dekany}, R.; {Smith}, R.M.;
  {Riddle}, R.; {Masci}, F.J.; {Helou}, G.; {Prince}, T.A.; {Adams}, S.M.;
  et~al.
\newblock {The Zwicky Transient Facility: System Overview, Performance, and
  First Results}.
\newblock {\em \pasp} {\bf 2019}, {\em 131},~018002,
  \href{http://xxx.lanl.gov/abs/1902.01932}{{\normalfont
  [arXiv:astro-ph.IM/1902.01932]}}.
\newblock {\url{https://doi.org/10.1088/1538-3873/aaecbe}}.

\bibitem[{Graham} \em{et~al.}(2019){Graham}, {Kulkarni}, {Bellm}, {Adams},
  {Barbarino}, {Blagorodnova}, {Bodewits}, {Bolin}, {Brady}, {Cenko}, {Chang},
  {Coughlin}, {De}, {Eadie}, {Farnham}, {Feindt}, {Franckowiak}, {Fremling},
  {Gezari}, {Ghosh}, {Goldstein}, {Golkhou}, {Goobar}, {Ho}, {Huppenkothen},
  {Ivezi{\'c}}, {Jones}, {Juric}, {Kaplan}, {Kasliwal}, {Kelley}, {Kupfer},
  {Lee}, {Lin}, {Lunnan}, {Mahabal}, {Miller}, {Ngeow}, {Nugent}, {Ofek},
  {Prince}, {Rauch}, {van Roestel}, {Schulze}, {Singer}, {Sollerman}, {Taddia},
  {Yan}, {Ye}, {Yu}, {Barlow}, {Bauer}, {Beck}, {Belicki}, {Biswas}, {Brinnel},
  {Brooke}, {Bue}, {Bulla}, {Burruss}, {Connolly}, {Cromer}, {Cunningham},
  {Dekany}, {Delacroix}, {Desai}, {Duev}, {Feeney}, {Flynn}, {Frederick},
  {Gal-Yam}, {Giomi}, {Groom}, {Hacopians}, {Hale}, {Helou}, {Henning},
  {Hover}, {Hillenbrand}, {Howell}, {Hung}, {Imel}, {Ip}, {Jackson}, {Kaspi},
  {Kaye}, {Kowalski}, {Kramer}, {Kuhn}, {Landry}, {Laher}, {Mao}, {Masci},
  {Monkewitz}, {Murphy}, {Nordin}, {Patterson}, {Penprase}, {Porter},
  {Rebbapragada}, {Reiley}, {Riddle}, {Rigault}, {Rodriguez}, {Rusholme}, {van
  Santen}, {Shupe}, {Smith}, {Soumagnac}, {Stein}, {Surace}, {Szkody}, {Terek},
  {Van Sistine}, {van Velzen}, {Vestrand}, {Walters}, {Ward}, {Zhang}, and
  {Zolkower}]{grahametal19}
{Graham}, M.J.; {Kulkarni}, S.R.; {Bellm}, E.C.; {Adams}, S.M.; {Barbarino},
  C.; {Blagorodnova}, N.; {Bodewits}, D.; {Bolin}, B.; {Brady}, P.R.; {Cenko},
  S.B.;  et~al.
\newblock {The Zwicky Transient Facility: Science Objectives}.
\newblock {\em \pasp} {\bf 2019}, {\em 131},~078001,
  \href{http://xxx.lanl.gov/abs/1902.01945}{{\normalfont
  [arXiv:astro-ph.IM/1902.01945]}}.
\newblock {\url{https://doi.org/10.1088/1538-3873/ab006c}}.

\bibitem[{Masci} \em{et~al.}(2019){Masci}, {Laher}, {Rusholme}, {Shupe},
  {Groom}, {Surace}, {Jackson}, {Monkewitz}, {Beck}, {Flynn}, {Terek},
  {Landry}, {Hacopians}, {Desai}, {Howell}, {Brooke}, {Imel}, {Wachter}, {Ye},
  {Lin}, {Cenko}, {Cunningham}, {Rebbapragada}, {Bue}, {Miller}, {Mahabal},
  {Bellm}, {Patterson}, {Juri{\'c}}, {Golkhou}, {Ofek}, {Walters}, {Graham},
  {Kasliwal}, {Dekany}, {Kupfer}, {Burdge}, {Cannella}, {Barlow}, {Van
  Sistine}, {Giomi}, {Fremling}, {Blagorodnova}, {Levitan}, {Riddle}, {Smith},
  {Helou}, {Prince}, and {Kulkarni}]{mascietal19}
{Masci}, F.J.; {Laher}, R.R.; {Rusholme}, B.; {Shupe}, D.L.; {Groom}, S.;
  {Surace}, J.; {Jackson}, E.; {Monkewitz}, S.; {Beck}, R.; {Flynn}, D.;
  et~al.
\newblock {The Zwicky Transient Facility: Data Processing, Products, and
  Archive}.
\newblock {\em \pasp} {\bf 2019}, {\em 131},~018003,
  \href{http://xxx.lanl.gov/abs/1902.01872}{{\normalfont
  [arXiv:astro-ph.IM/1902.01872]}}.
\newblock {\url{https://doi.org/10.1088/1538-3873/aae8ac}}.

\bibitem[{Charisi} \em{et~al.}(2016){Charisi}, {Bartos}, {Haiman},
  {Price-Whelan}, {Graham}, {Bellm}, {Laher}, and {M{\'a}rka}]{charisietal16}
{Charisi}, M.; {Bartos}, I.; {Haiman}, Z.; {Price-Whelan}, A.M.; {Graham},
  M.J.; {Bellm}, E.C.; {Laher}, R.R.; {M{\'a}rka}, S.
\newblock {A population of short-period variable quasars from PTF as
  supermassive black hole binary candidates}.
\newblock {\em \mnras} {\bf 2016}, {\em 463},~2145--2171,
  \href{http://xxx.lanl.gov/abs/1604.01020}{{\normalfont [1604.01020]}}.
\newblock {\url{https://doi.org/10.1093/mnras/stw1838}}.

\bibitem[Chen \em{et~al.}(2022)Chen, Liu, et~al.]{chen2022}
Chen, X.; Liu, X.;  et~al.
\newblock Periodic optical variability in active galactic nuclei identified
  from Zwicky Transient Facility light curves.
\newblock {\em The Astrophysical Journal} {\bf 2022}, {\em 924},~7.
\newblock {\url{https://doi.org/10.3847/1538-4357/ac3299}}.

\bibitem[{Vaughan} \em{et~al.}(2016){Vaughan}, {Uttley}, {Markowitz},
  {Huppenkothen}, {Middleton}, {Alston}, {Scargle}, and {Farr}]{vaughanetal16}
{Vaughan}, S.; {Uttley}, P.; {Markowitz}, A.G.; {Huppenkothen}, D.;
  {Middleton}, M.J.; {Alston}, W.N.; {Scargle}, J.D.; {Farr}, W.M.
\newblock {False periodicities in quasar time-domain surveys}.
\newblock {\em \mnras} {\bf 2016}, {\em 461},~3145--3152,
  \href{http://xxx.lanl.gov/abs/1606.02620}{{\normalfont
  [arXiv:astro-ph.IM/1606.02620]}}.
\newblock {\url{https://doi.org/10.1093/mnras/stw1412}}.

\bibitem[Edelman and Liu(2023)]{edelman2023}
Edelman, A.; Liu, X.
\newblock The role of red noise in false periodicity detection in AGN light
  curves.
\newblock {\em The Astrophysical Journal} {\bf 2023}, {\em 949},~56.
\newblock {\url{https://doi.org/10.3847/1538-4357/acd387}}.

\bibitem[{Horne} and {Baliunas}(1986)]{hornebaliunas86}
{Horne}, J.H.; {Baliunas}, S.L.
\newblock {A Prescription for Period Analysis of Unevenly Sampled Time Series}.
\newblock {\em \apj} {\bf 1986}, {\em 302},~757.
\newblock {\url{https://doi.org/10.1086/164037}}.

\bibitem[Sesar \em{et~al.}(2007)Sesar et~al.]{sesar2007}
Sesar, B.;  et~al.
\newblock Exploring the nature of quasar variability: AGN or accretion disk
  instabilities?
\newblock {\em The Astronomical Journal} {\bf 2007}, {\em 134},~2236--2251.
\newblock {\url{https://doi.org/10.1086/522911}}.

\bibitem[{Hilditch}(2001)]{hilditch01}
{Hilditch}, R.W.
\newblock {\em {An Introduction to Close Binary Stars}}; Cambridge, UK:
  Cambridge University Press,  2001.

\bibitem[{Bon} \em{et~al.}(2012){Bon}, {Jovanovi{\'c}}, {Marziani},
  {Shapovalova}, {Bon}, {Borka Jovanovi{\'c}}, {Borka}, {Sulentic}, and
  {Popovi{\'c}}]{bonetal12}
{Bon}, E.; {Jovanovi{\'c}}, P.; {Marziani}, P.; {Shapovalova}, A.I.; {Bon}, N.;
  {Borka Jovanovi{\'c}}, V.; {Borka}, D.; {Sulentic}, J.; {Popovi{\'c}}, L.{\v
  C}.
\newblock {The First Spectroscopically Resolved Sub-parsec Orbit of a
  Supermassive Binary Black Hole}.
\newblock {\em \apj} {\bf 2012}, {\em 759},~118,
  \href{http://xxx.lanl.gov/abs/1209.4524}{{\normalfont
  [arXiv:astro-ph.HE/1209.4524]}}.
\newblock {\url{https://doi.org/10.1088/0004-637X/759/2/118}}.

\bibitem[{Shen} \em{et~al.}(2013){Shen}, {Liu}, {Loeb}, and
  {Tremaine}]{shenetal13}
{Shen}, Y.; {Liu}, X.; {Loeb}, A.; {Tremaine}, S.
\newblock {Constraining Sub-parsec Binary Supermassive Black Holes in Quasars
  with Multi-epoch Spectroscopy. I. The General Quasar Population}.
\newblock {\em \apj} {\bf 2013}, {\em 775},~49,
  \href{http://xxx.lanl.gov/abs/1306.4330}{{\normalfont
  [arXiv:astro-ph.CO/1306.4330]}}.
\newblock {\url{https://doi.org/10.1088/0004-637X/775/1/49}}.

\bibitem[{Li} \em{et~al.}(2016){Li}, {Wang}, {Ho}, {Lu}, {Qiu}, {Du}, {Hu},
  {Huang}, {Zhang}, {Wang}, and {Bai}]{lietal16}
{Li}, Y.R.; {Wang}, J.M.; {Ho}, L.C.; {Lu}, K.X.; {Qiu}, J.; {Du}, P.; {Hu},
  C.; {Huang}, Y.K.; {Zhang}, Z.X.; {Wang}, K.;  et~al.
\newblock {Spectroscopic Indication of a Centi-parsec Supermassive Black Hole
  Binary in the Galactic Center of NGC 5548}.
\newblock {\em \apj} {\bf 2016}, {\em 822},~4,
  \href{http://xxx.lanl.gov/abs/1602.05005}{{\normalfont [1602.05005]}}.
\newblock {\url{https://doi.org/10.3847/0004-637X/822/1/4}}.

\bibitem[{Robinson} \em{et~al.}(2010){Robinson}, {Young}, {Axon}, {Kharb}, and
  {Smith}]{robinsonetal10}
{Robinson}, A.; {Young}, S.; {Axon}, D.J.; {Kharb}, P.; {Smith}, J.E.
\newblock {Spectropolarimetric Evidence for a Kicked Supermassive Black Hole in
  the Quasar E1821+643}.
\newblock {\em \apjl} {\bf 2010}, {\em 717},~L122--L126,
  \href{http://xxx.lanl.gov/abs/1006.0993}{{\normalfont
  [arXiv:astro-ph.CO/1006.0993]}}.
\newblock {\url{https://doi.org/10.1088/2041-8205/717/2/L122}}.

\bibitem[{Jadhav} \em{et~al.}(2021){Jadhav}, {Robinson}, {Almeyda}, {Curran},
  and {Marconi}]{jadhavetal21}
{Jadhav}, Y.; {Robinson}, A.; {Almeyda}, T.; {Curran}, R.; {Marconi}, A.
\newblock {The spatially offset quasar E1821+643: new evidence for
  gravitational recoil}.
\newblock {\em \mnras} {\bf 2021}, {\em 507},~484--495,
  \href{http://xxx.lanl.gov/abs/2107.14711}{{\normalfont
  [arXiv:astro-ph.GA/2107.14711]}}.
\newblock {\url{https://doi.org/10.1093/mnras/stab2176}}.

\bibitem[{Marziani} \em{et~al.}(2013){Marziani}, {Sulentic}, {Plauchu-Frayn},
  and {del Olmo}]{marzianietal13a}
{Marziani}, P.; {Sulentic}, J.W.; {Plauchu-Frayn}, I.; {del Olmo}, A.
\newblock {Is Mg II 2800 a Reliable Virial Broadening Estimator for Quasars?}
\newblock {\em AAp} {\bf 2013}, {\em 555},~89, 16pp,
  \href{http://xxx.lanl.gov/abs/1305.1096}{{\normalfont
  [arXiv:astro-ph.CO/1305.1096]}}.

\bibitem[{Boroson} and {Green}(1992)]{borosongreen92}
{Boroson}, T.A.; {Green}, R.F.
\newblock {The Emission-Line Properties of Low-Redshift Quasi-stellar Objects}.
\newblock {\em \apjs} {\bf 1992}, {\em 80},~109.
\newblock {\url{https://doi.org/10.1086/191661}}.

\bibitem[{Marziani} \em{et~al.}(2003){Marziani}, {Zamanov}, {Sulentic}, and
  {Calvani}]{marzianietal03b}
{Marziani}, P.; {Zamanov}, R.K.; {Sulentic}, J.W.; {Calvani}, M.
\newblock {Searching for the physical drivers of eigenvector 1: influence of
  black hole mass and Eddington ratio}.
\newblock {\em MNRAS} {\bf 2003}, {\em 345},~1133--1144,
  \href{http://xxx.lanl.gov/abs/arXiv:astro-ph/0307367}{{\normalfont
  [arXiv:astro-ph/0307367]}}.
\newblock {\url{https://doi.org/10.1046/j.1365-2966.2003.07033.x}}.

\bibitem[{Shen} and {Ho}(2014)]{shenho14}
{Shen}, Y.; {Ho}, L.C.
\newblock {The diversity of quasars unified by accretion and orientation}.
\newblock {\em \nat} {\bf 2014}, {\em 513},~210--213,
  \href{http://xxx.lanl.gov/abs/1409.2887}{{\normalfont [1409.2887]}}.
\newblock {\url{https://doi.org/10.1038/nature13712}}.

\bibitem[{Sulentic} \em{et~al.}(2000{\natexlab{a}}){Sulentic}, {Marziani}, and
  {Dultzin-Hacyan}]{sulenticetal00a}
{Sulentic}, J.W.; {Marziani}, P.; {Dultzin-Hacyan}, D.
\newblock {Phenomenology of Broad Emission Lines in Active Galactic Nuclei}.
\newblock {\em ARA\&A} {\bf 2000}, {\em 38},~521--571.
\newblock {\url{https://doi.org/10.1146/annurev.astro.38.1.521}}.

\bibitem[{Sulentic} \em{et~al.}(2000{\natexlab{b}}){Sulentic}, {Marziani},
  {Zwitter}, {Dultzin-Hacyan}, and {Calvani}]{sulenticetal00b}
{Sulentic}, J.W.; {Marziani}, P.; {Zwitter}, T.; {Dultzin-Hacyan}, D.;
  {Calvani}, M.
\newblock {The Demise of the Classical Broad-Line Region in the Luminous Quasar
  PG 1416-129}.
\newblock {\em \apjl} {\bf 2000}, {\em 545},~L15--L18,
  \href{http://xxx.lanl.gov/abs/arXiv:astro-ph/0009326}{{\normalfont
  [arXiv:astro-ph/0009326]}}.
\newblock {\url{https://doi.org/10.1086/317330}}.

\bibitem[{Panda} \em{et~al.}(2019){Panda}, {Marziani}, and
  {Czerny}]{pandaetal19}
{Panda}, S.; {Marziani}, P.; {Czerny}, B.
\newblock {The Quasar Main Sequence Explained by the Combination of Eddington
  Ratio, Metallicity, and Orientation}.
\newblock {\em \apj} {\bf 2019}, {\em 882},~79,
  \href{http://xxx.lanl.gov/abs/1905.01729}{{\normalfont
  [arXiv:astro-ph.HE/1905.01729]}}.
\newblock {\url{https://doi.org/10.3847/1538-4357/ab3292}}.

\bibitem[{Jankov} \em{et~al.}(2021){Jankov}, {Ili{\'c}}, and
  {Kova{\v{c}}evi{\'c}}]{jankovetal21}
{Jankov}, I.; {Ili{\'c}}, D.; {Kova{\v{c}}evi{\'c}}, A.
\newblock {Manifold Learning in the Context of Quasar Spectral Diversity}.
\newblock In Proceedings of the XIX Serbian Astronomical Conference;
  {Kova{\v{c}}evi{\'c}}, A.; {Kova{\v{c}}evi{\'c} Doj{\v{c}}inovi{\'c}}, J.;
  {Mar{\v{c}}eta}, D.; {Oni{\'c}}, D., Eds.,  2021, Vol. 100, pp. 241--246.

\bibitem[{Ghojogh} \em{et~al.}(2020){Ghojogh}, {Ghodsi}, {Karray}, and
  {Crowley}]{ghojoghetal20}
{Ghojogh}, B.; {Ghodsi}, A.; {Karray}, F.; {Crowley}, M.
\newblock {Locally Linear Embedding and its Variants: Tutorial and Survey}.
\newblock {\em arXiv e-prints} {\bf 2020}, p. arXiv:2011.10925,
  \href{http://xxx.lanl.gov/abs/2011.10925}{{\normalfont
  [arXiv:stat.ML/2011.10925]}}.
\newblock {\url{https://doi.org/10.48550/arXiv.2011.10925}}.

\bibitem[{Marinello} \em{et~al.}(2016){Marinello}, {Rodriguez-Ardila},
  {Garcia-Rissmann}, {Sigut}, and {Pradhan}]{marinelloetal16}
{Marinello}, A.O.M.; {Rodriguez-Ardila}, A.; {Garcia-Rissmann}, A.; {Sigut},
  T.A.A.; {Pradhan}, A.K.
\newblock {The FeII emission in active galactic nuclei: excitation mechanisms
  and location of the emitting region}.
\newblock {\em \apj} {\bf 2016}, {\em 820},~116,
  \href{http://xxx.lanl.gov/abs/1602.05159}{{\normalfont [1602.05159]}}.

\bibitem[{Marziani} \em{et~al.}(2001){Marziani}, {Sulentic}, {Zwitter},
  {Dultzin-Hacyan}, and {Calvani}]{marzianietal01}
{Marziani}, P.; {Sulentic}, J.W.; {Zwitter}, T.; {Dultzin-Hacyan}, D.;
  {Calvani}, M.
\newblock {Searching for the Physical Drivers of the Eigenvector 1 Correlation
  Space}.
\newblock {\em \apj} {\bf 2001}, {\em 558},~553--560,
  \href{http://xxx.lanl.gov/abs/arXiv:astro-ph/0105343}{{\normalfont
  [arXiv:astro-ph/0105343]}}.
\newblock {\url{https://doi.org/10.1086/322286}}.

\bibitem[{Sun} and {Shen}(2015)]{sunshen15}
{Sun}, J.; {Shen}, Y.
\newblock {Dissecting the Quasar Main Sequence: Insight from Host Galaxy
  Properties}.
\newblock {\em \apjl} {\bf 2015}, {\em 804},~L15,
  \href{http://xxx.lanl.gov/abs/1503.08364}{{\normalfont [1503.08364]}}.
\newblock {\url{https://doi.org/10.1088/2041-8205/804/1/L15}}.

\bibitem[{Du} \em{et~al.}(2016){Du}, {Wang}, {Hu}, {Ho}, {Li}, and
  {Bai}]{duetal16a}
{Du}, P.; {Wang}, J.M.; {Hu}, C.; {Ho}, L.C.; {Li}, Y.R.; {Bai}, J.M.
\newblock {The Fundamental Plane of the Broad-line Region in Active Galactic
  Nuclei}.
\newblock {\em \apjl} {\bf 2016}, {\em 818},~L14,
  \href{http://xxx.lanl.gov/abs/1601.01391}{{\normalfont [1601.01391]}}.
\newblock {\url{https://doi.org/10.3847/2041-8205/818/1/L14}}.

\bibitem[{Antonucci}(1993)]{antonucci93}
{Antonucci}, R.
\newblock {Unified models for active galactic nuclei and quasars}.
\newblock {\em \araa} {\bf 1993}, {\em 31},~473--521.
\newblock {\url{https://doi.org/10.1146/annurev.aa.31.090193.002353}}.

\bibitem[{Urry} and {Padovani}(1995)]{urrypadovani95}
{Urry}, C.M.; {Padovani}, P.
\newblock {Unified Schemes for Radio-Loud Active Galactic Nuclei}.
\newblock {\em PASP} {\bf 1995}, {\em 107},~803,
  \href{http://xxx.lanl.gov/abs/arXiv:astro-ph/9506063}{{\normalfont
  [arXiv:astro-ph/9506063]}}.
\newblock {\url{https://doi.org/10.1086/133630}}.

\bibitem[{Marin} and {Antonucci}(2016)]{marinantonucci16}
{Marin}, F.; {Antonucci}, R.
\newblock {A Robust Derivation of the Tight Relationship of Radio Core
  Dominance to Inclination Angle in High Redshift 3CRR Sources}.
\newblock {\em \apj} {\bf 2016}, {\em 830},~82,
  \href{http://xxx.lanl.gov/abs/1607.04997}{{\normalfont
  [arXiv:astro-ph.GA/1607.04997]}}.
\newblock {\url{https://doi.org/10.3847/0004-637X/830/2/82}}.

\bibitem[{Bon} \em{et~al.}(2020){Bon}, {Marziani}, {Bon}, {Negrete}, {Dultzin},
  {del Olmo}, {D'Onofrio}, and {Mart{\'\i}nez-Aldama}]{bonetal20}
{Bon}, N.; {Marziani}, P.; {Bon}, E.; {Negrete}, C.A.; {Dultzin}, D.; {del
  Olmo}, A.; {D'Onofrio}, M.; {Mart{\'\i}nez-Aldama}, M.L.
\newblock {Selection of highly-accreting quasars. Spectral properties of Fe
  II$_{opt}$ emitters not belonging to extreme Population A}.
\newblock {\em \aap} {\bf 2020}, {\em 635},~A151,
  \href{http://xxx.lanl.gov/abs/2001.08765}{{\normalfont
  [arXiv:astro-ph.GA/2001.08765]}}.
\newblock {\url{https://doi.org/10.1051/0004-6361/201936773}}.

\bibitem[{Collin-Souffrin} \em{et~al.}(1988){Collin-Souffrin}, {Dyson},
  {McDowell}, and {Perry}]{collinsouffrinetal88}
{Collin-Souffrin}, S.; {Dyson}, J.E.; {McDowell}, J.C.; {Perry}, J.J.
\newblock {The environment of active galactic nuclei. I - A two-component broad
  emission line model}.
\newblock {\em MNRAS} {\bf 1988}, {\em 232},~539--550.

\bibitem[{Marziani} \em{et~al.}(1996){Marziani}, {Sulentic}, {Dultzin-Hacyan},
  {Calvani}, and {Moles}]{marzianietal96}
{Marziani}, P.; {Sulentic}, J.W.; {Dultzin-Hacyan}, D.; {Calvani}, M.; {Moles},
  M.
\newblock {Comparative Analysis of the High- and Low-Ionization Lines in the
  Broad-Line Region of Active Galactic Nuclei}.
\newblock {\em \apjs} {\bf 1996}, {\em 104},~37--+.
\newblock {\url{https://doi.org/10.1086/192291}}.

\bibitem[{Elvis}(2000)]{elvis00}
{Elvis}, M.
\newblock {A Structure for Quasars}.
\newblock {\em \apj} {\bf 2000}, {\em 545},~63--76,
  \href{http://xxx.lanl.gov/abs/arXiv:astro-ph/0008064}{{\normalfont
  [arXiv:astro-ph/0008064]}}.
\newblock {\url{https://doi.org/10.1086/317778}}.

\bibitem[{Netzer}(1977)]{netzer77}
{Netzer}, H.
\newblock {On the profiles of the broad lines in the spectra of QSOs and
  Seyfert galaxies}.
\newblock {\em \mnras} {\bf 1977}, {\em 181},~89P--92P.

\bibitem[{Wang} \em{et~al.}(2017){Wang}, {Du}, {Brotherton}, {Hu}, {Songsheng},
  {Li}, {Shi}, and {Zhang}]{wangetal17}
{Wang}, J.M.; {Du}, P.; {Brotherton}, M.S.; {Hu}, C.; {Songsheng}, Y.Y.; {Li},
  Y.R.; {Shi}, Y.; {Zhang}, Z.X.
\newblock {Tidally disrupted dusty clumps as the origin of broad emission lines
  in active galactic nuclei}.
\newblock {\em Nature Astronomy} {\bf 2017}, {\em 1},~775--783,
  \href{http://xxx.lanl.gov/abs/1710.03419}{{\normalfont [1710.03419]}}.
\newblock {\url{https://doi.org/10.1038/s41550-017-0264-4}}.

\bibitem[{Wolf} \em{et~al.}(2020){Wolf}, {Salvato}, {Coffey}, {Merloni},
  {Buchner}, {Arcodia}, {Baron}, {Carrera}, {Comparat}, {Schneider}, and
  {Nandra}]{wolfetal20}
{Wolf}, J.; {Salvato}, M.; {Coffey}, D.; {Merloni}, A.; {Buchner}, J.;
  {Arcodia}, R.; {Baron}, D.; {Carrera}, F.J.; {Comparat}, J.; {Schneider},
  D.P.;  et~al.
\newblock {Exploring the diversity of Type 1 active galactic nuclei identified
  in SDSS-IV/SPIDERS}.
\newblock {\em \mnras} {\bf 2020}, {\em 492},~3580--3601,
  \href{http://xxx.lanl.gov/abs/1911.01947}{{\normalfont
  [arXiv:astro-ph.HE/1911.01947]}}.
\newblock {\url{https://doi.org/10.1093/mnras/staa018}}.

\bibitem[{Bao} \em{et~al.}(2022){Bao}, {Brotherton}, {Du}, {McLane},
  {Zastrocky}, {Olson}, {Fang}, {Zhai}, {Huang}, {Wang}, {Zhao}, {Li}, {Yang},
  {Chen}, {Liu}, {Yao}, {Peng}, {Guo}, {Songsheng}, {Li}, {Jiang}, {Kasper},
  {Chick}, {Nguyen}, {Maithil}, {Kobulnicky}, {Dale}, {Hand}, {Adelman},
  {Carter}, {Murphree}, {Oeur}, {Schonsberg}, {Roth}, {Winkler}, {Marziani},
  {D'Onofrio}, {Hu}, {Xiao}, {Xue}, {Czerny}, {Aceituno}, {Ho}, {Bai}, {Wang},
  and {MAHA Collaboration}]{baoetal22}
{Bao}, D.W.; {Brotherton}, M.S.; {Du}, P.; {McLane}, J.N.; {Zastrocky}, T.E.;
  {Olson}, K.A.; {Fang}, F.N.; {Zhai}, S.; {Huang}, Z.P.; {Wang}, K.;  et~al.
\newblock {Monitoring AGNs with H{\ensuremath{\beta}} Asymmetry. III. Long-term
  Reverberation Mapping Results of 15 Palomar-Green Quasars}.
\newblock {\em \apjs} {\bf 2022}, {\em 262},~14,
  \href{http://xxx.lanl.gov/abs/2207.00297}{{\normalfont
  [arXiv:astro-ph.GA/2207.00297]}}.
\newblock {\url{https://doi.org/10.3847/1538-4365/ac7beb}}.

\bibitem[{Marziani}(2023)]{marziani23}
{Marziani}, P.
\newblock {Accretion/Ejection Phenomena and Emission-Line Profile (A)symmetries
  in Type-1 Active Galactic Nuclei}.
\newblock {\em Symmetry} {\bf 2023}, {\em 15},~1859.
\newblock {\url{https://doi.org/10.3390/sym15101859}}.

\bibitem[{Zastrocky} \em{et~al.}(2024){Zastrocky}, {Brotherton}, {Du},
  {McLane}, {Olson}, {Dale}, {Kobulnicky}, {Maithil}, {Nguyen}, {Chick},
  {Kasper}, {Hand}, {Adelman}, {Carter}, {Murphree}, {Oeur}, {Roth},
  {Schonsberg}, {Caradonna}, {Favro}, {Ferguson}, {Gonzalez}, {Hadding},
  {Hagler}, {Rogers}, {Stack}, {Chapman}, {Bao}, {Fang}, {Zhai}, {Yang},
  {Chen}, {Bai}, {Fu}, {Liu}, {Yao}, {Peng}, {Songsheng}, {Li}, {Bai}, {Hu},
  {Xiao}, {Ho}, and {Wang}]{zastrockyetal24}
{Zastrocky}, T.E.; {Brotherton}, M.S.; {Du}, P.; {McLane}, J.N.; {Olson}, K.A.;
  {Dale}, D.A.; {Kobulnicky}, H.A.; {Maithil}, J.; {Nguyen}, M.L.; {Chick},
  W.T.;  et~al.
\newblock {Monitoring AGNs with H{\ensuremath{\beta}} Asymmetry. IV. First
  Reverberation Mapping Results of 14 Active Galactic Nuclei}.
\newblock {\em \apjs} {\bf 2024}, {\em 272},~29,
  \href{http://xxx.lanl.gov/abs/2404.07343}{{\normalfont
  [arXiv:astro-ph.GA/2404.07343]}}.
\newblock {\url{https://doi.org/10.3847/1538-4365/ad3bad}}.

\bibitem[{Punsly} \em{et~al.}(2018){Punsly}, {Marziani}, {Bennert}, {Nagai},
  and {Gurwell}]{punslyetal18a}
{Punsly}, B.; {Marziani}, P.; {Bennert}, V.N.; {Nagai}, H.; {Gurwell}, M.A.
\newblock {Revealing the Broad Line Region of NGC 1275: The Relationship to Jet
  Power}.
\newblock {\em \apj} {\bf 2018}, {\em 869},~143,
  \href{http://xxx.lanl.gov/abs/1810.11716}{{\normalfont [1810.11716]}}.
\newblock {\url{https://doi.org/10.3847/1538-4357/aaec75}}.

\bibitem[Panda \em{et~al.}(2018)Panda, Czerny, Adhikari, Hryniewicz, Wildy,
  Kuraszkiewicz, and {\'{S}}niegowska]{pandaetal18}
Panda, S.; Czerny, B.; Adhikari, T.P.; Hryniewicz, K.; Wildy, C.;
  Kuraszkiewicz, J.; {\'{S}}niegowska, M.
\newblock Modeling of the Quasar Main Sequence in the Optical Plane.
\newblock {\em The Astrophysical Journal} {\bf 2018}, {\em 866},~115.
\newblock {\url{https://doi.org/10.3847/1538-4357/aae209}}.

\bibitem[{Sulentic} \em{et~al.}(2007){Sulentic}, {Bachev}, {Marziani},
  {Negrete}, and {Dultzin}]{sulenticetal07}
{Sulentic}, J.W.; {Bachev}, R.; {Marziani}, P.; {Negrete}, C.A.; {Dultzin}, D.
\newblock {C IV {\ensuremath{\lambda}}1549 as an Eigenvector 1 Parameter for
  Active Galactic Nuclei}.
\newblock {\em \apj} {\bf 2007}, {\em 666},~757--777,
  \href{http://xxx.lanl.gov/abs/0705.1895}{{\normalfont
  [arXiv:astro-ph/0705.1895]}}.
\newblock {\url{https://doi.org/10.1086/519916}}.

\bibitem[{Richards} \em{et~al.}(2011){Richards}, {Kruczek}, {Gallagher},
  {Hall}, {Hewett}, {Leighly}, {Deo}, {Kratzer}, and {Shen}]{richardsetal11}
{Richards}, G.T.; {Kruczek}, N.E.; {Gallagher}, S.C.; {Hall}, P.B.; {Hewett},
  P.C.; {Leighly}, K.M.; {Deo}, R.P.; {Kratzer}, R.M.; {Shen}, Y.
\newblock {Unification of Luminous Type 1 Quasars through C IV Emission}.
\newblock {\em \aj} {\bf 2011}, {\em 141},~167--+,
  \href{http://xxx.lanl.gov/abs/1011.2282}{{\normalfont
  [arXiv:astro-ph.GA/1011.2282]}}.
\newblock {\url{https://doi.org/10.1088/0004-6256/141/5/167}}.

\bibitem[{Bonzini} \em{et~al.}(2015){Bonzini}, {Mainieri}, {Padovani},
  {Andreani}, {Berta}, {Bethermin}, {Lutz}, {Rodighiero}, {Rosario}, {Tozzi},
  and {Vattakunnel}]{bonzinietal15}
{Bonzini}, M.; {Mainieri}, V.; {Padovani}, P.; {Andreani}, P.; {Berta}, S.;
  {Bethermin}, M.; {Lutz}, D.; {Rodighiero}, G.; {Rosario}, D.; {Tozzi}, P.;
  et~al.
\newblock {Star formation properties of sub-mJy radio sources}.
\newblock {\em \mnras} {\bf 2015}, {\em 453},~1079--1094,
  \href{http://xxx.lanl.gov/abs/1508.01905}{{\normalfont [1508.01905]}}.
\newblock {\url{https://doi.org/10.1093/mnras/stv1675}}.

\bibitem[{Ganci} \em{et~al.}(2019){Ganci}, {Marziani}, {D'Onofrio}, {del Olmo},
  {Bon}, {Bon}, and {Negrete}]{gancietal19}
{Ganci}, V.; {Marziani}, P.; {D'Onofrio}, M.; {del Olmo}, A.; {Bon}, E.; {Bon},
  N.; {Negrete}, C.A.
\newblock {Radio loudness along the quasar main sequence}.
\newblock {\em \aap} {\bf 2019}, {\em 630},~A110,
  \href{http://xxx.lanl.gov/abs/1908.07308}{{\normalfont
  [arXiv:astro-ph.GA/1908.07308]}}.
\newblock {\url{https://doi.org/10.1051/0004-6361/201936270}}.

\bibitem[{Panessa} \em{et~al.}(2019){Panessa}, {Baldi}, {Laor}, {Padovani},
  {Behar}, and {McHardy}]{panessaetal19}
{Panessa}, F.; {Baldi}, R.D.; {Laor}, A.; {Padovani}, P.; {Behar}, E.;
  {McHardy}, I.
\newblock {The origin of radio emission from radio-quiet active galactic
  nuclei}.
\newblock {\em Nature Astronomy} {\bf 2019}, {\em 3},~387--396,
  \href{http://xxx.lanl.gov/abs/1902.05917}{{\normalfont
  [arXiv:astro-ph.GA/1902.05917]}}.
\newblock {\url{https://doi.org/10.1038/s41550-019-0765-4}}.

\bibitem[{Chen} \em{et~al.}(2024){Chen}, {Laor}, {Behar}, {Baldi}, {Gelfand},
  {Kimball}, {McHardy}, {Orosz}, and {Paragi}]{chenetal24}
{Chen}, S.; {Laor}, A.; {Behar}, E.; {Baldi}, R.D.; {Gelfand}, J.D.; {Kimball},
  A.E.; {McHardy}, I.M.; {Orosz}, G.; {Paragi}, Z.
\newblock {Windy or Not: Radio Parsec-scale Evidence for a Broad-line Region
  Wind in Radio-quiet Quasars}.
\newblock {\em \apj} {\bf 2024}, {\em 975},~35,
  \href{http://xxx.lanl.gov/abs/2408.15934}{{\normalfont
  [arXiv:astro-ph.GA/2408.15934]}}.
\newblock {\url{https://doi.org/10.3847/1538-4357/ad74fc}}.

\bibitem[{Marziani} \em{et~al.}(2013){Marziani}, {Sulentic}, {Plauchu-Frayn},
  and {del Olmo}]{marzianietal13}
{Marziani}, P.; {Sulentic}, J.W.; {Plauchu-Frayn}, I.; {del Olmo}, A.
\newblock {Low-Ionization Outflows in High Eddington Ratio Quasars}.
\newblock {\em \apj} {\bf 2013}, {\em 764},
  \href{http://xxx.lanl.gov/abs/1301.0520}{{\normalfont
  [arXiv:astro-ph.CO/1301.0520]}}.

\bibitem[{Marziani} \em{et~al.}(2003){Marziani}, {Sulentic}, {Zamanov},
  {Calvani}, {Dultzin-Hacyan}, {Bachev}, and {Zwitter}]{marzianietal03a}
{Marziani}, P.; {Sulentic}, J.W.; {Zamanov}, R.; {Calvani}, M.;
  {Dultzin-Hacyan}, D.; {Bachev}, R.; {Zwitter}, T.
\newblock {An Optical Spectroscopic Atlas of Low-Redshift Active Galactic
  Nuclei}.
\newblock {\em \apjs} {\bf 2003}, {\em 145},~199--211.
\newblock {\url{https://doi.org/10.1086/346025}}.

\bibitem[{Kova{\v c}evi{\'c}-Doj{\v c}inovi{\'c}} and
  {Popovi{\'c}}(2015)]{kovacevicdojcinovicpopopvic15}
{Kova{\v c}evi{\'c}-Doj{\v c}inovi{\'c}}, J.; {Popovi{\'c}}, L.{\v C}.
\newblock {The Connections Between the UV and Optical Fe ii Emission Lines in
  Type 1 AGNs}.
\newblock {\em \apjs} {\bf 2015}, {\em 221},~35,
  \href{http://xxx.lanl.gov/abs/1509.03679}{{\normalfont [1509.03679]}}.
\newblock {\url{https://doi.org/10.1088/0067-0049/221/2/35}}.

\bibitem[{Wang} \em{et~al.}(2009){Wang}, {Dong}, {Wang}, {Ho}, {Yuan}, {Wang},
  {Zhang}, {Zhang}, and {Zhou}]{wangetal09}
{Wang}, J.; {Dong}, X.; {Wang}, T.; {Ho}, L.C.; {Yuan}, W.; {Wang}, H.;
  {Zhang}, K.; {Zhang}, S.; {Zhou}, H.
\newblock {Estimating Black Hole Masses in Active Galactic Nuclei Using the Mg
  II {$\lambda$}2800 Emission Line}.
\newblock {\em \apj} {\bf 2009}, {\em 707},~1334--1346,
  \href{http://xxx.lanl.gov/abs/0910.2848}{{\normalfont [0910.2848]}}.
\newblock {\url{https://doi.org/10.1088/0004-637X/707/2/1334}}.

\bibitem[{Trakhtenbrot} and {Netzer}(2012)]{trakhtenbrotnetzer12}
{Trakhtenbrot}, B.; {Netzer}, H.
\newblock {Black hole growth to z = 2 - I. Improved virial methods for
  measuring M$_{BH}$ and L/L$_{Edd}$}.
\newblock {\em \mnras} {\bf 2012}, {\em 427},~3081--3102,
  \href{http://xxx.lanl.gov/abs/1209.1096}{{\normalfont
  [arXiv:astro-ph.CO/1209.1096]}}.
\newblock {\url{https://doi.org/10.1111/j.1365-2966.2012.22056.x}}.

\bibitem[{Popovi{\'c}} \em{et~al.}(2019){Popovi{\'c}},
  {Kova{\v{c}}evi{\'c}-Doj{\v{c}}inovi{\'c}}, and
  {Mar{\v{c}}eta-Mandi{\'c}}]{popovicetal19}
{Popovi{\'c}}, L.{\v{C}}.; {Kova{\v{c}}evi{\'c}-Doj{\v{c}}inovi{\'c}}, J.;
  {Mar{\v{c}}eta-Mandi{\'c}}, S.
\newblock {The structure of the Mg II broad line emitting region in Type 1
  AGNs}.
\newblock {\em \mnras} {\bf 2019}, {\em 484},~3180--3197,
  \href{http://xxx.lanl.gov/abs/1901.03681}{{\normalfont
  [arXiv:astro-ph.GA/1901.03681]}}.
\newblock {\url{https://doi.org/10.1093/mnras/stz157}}.

\bibitem[{Shen} \em{et~al.}(2016){Shen}, {Horne}, {Grier}, {Peterson},
  {Denney}, {Trump}, {Sun}, {Brandt}, {Kochanek}, {Dawson}, {Green}, {Greene},
  {Hall}, {Ho}, {Jiang}, {Kinemuchi}, {McGreer}, {Petitjean}, {Richards},
  {Schneider}, {Strauss}, {Tao}, {Wood-Vasey}, {Zu}, {Pan}, {Bizyaev}, {Ge},
  {Oravetz}, and {Simmons}]{shenetal16}
{Shen}, Y.; {Horne}, K.; {Grier}, C.J.; {Peterson}, B.M.; {Denney}, K.D.;
  {Trump}, J.R.; {Sun}, M.; {Brandt}, W.N.; {Kochanek}, C.S.; {Dawson}, K.S.;
  et~al.
\newblock {The Sloan Digital Sky Survey Reverberation Mapping Project: First
  Broad-line H{\ensuremath{\beta}} and Mg II Lags at z {\ensuremath{\gtrsim}}
  0.3 from Six-month Spectroscopy}.
\newblock {\em \apj} {\bf 2016}, {\em 818},~30,
  \href{http://xxx.lanl.gov/abs/1510.02802}{{\normalfont
  [arXiv:astro-ph.GA/1510.02802]}}.
\newblock {\url{https://doi.org/10.3847/0004-637X/818/1/30}}.

\bibitem[{Le} \em{et~al.}(2020){Le}, {Woo}, and {Xue}]{leetal20}
{Le}, H.A.N.; {Woo}, J.H.; {Xue}, Y.
\newblock {Calibrating Mg II-based Black Hole Mass Estimators Using
  Low-to-high-luminosity Active Galactic Nuclei}.
\newblock {\em \apj} {\bf 2020}, {\em 901},~35,
  \href{http://xxx.lanl.gov/abs/2008.02990}{{\normalfont
  [arXiv:astro-ph.GA/2008.02990]}}.
\newblock {\url{https://doi.org/10.3847/1538-4357/abada0}}.

\bibitem[{Yue} \em{et~al.}(2024){Yue}, {Eilers}, {Simcoe}, {Mackenzie},
  {Matthee}, {Kashino}, {Bordoloi}, {Lilly}, and {Naidu}]{yueetal24}
{Yue}, M.; {Eilers}, A.C.; {Simcoe}, R.A.; {Mackenzie}, R.; {Matthee}, J.;
  {Kashino}, D.; {Bordoloi}, R.; {Lilly}, S.J.; {Naidu}, R.P.
\newblock {EIGER. V. Characterizing the Host Galaxies of Luminous Quasars at z
  {\ensuremath{\gtrsim}} 6}.
\newblock {\em \apj} {\bf 2024}, {\em 966},~176,
  \href{http://xxx.lanl.gov/abs/2309.04614}{{\normalfont
  [arXiv:astro-ph.GA/2309.04614]}}.
\newblock {\url{https://doi.org/10.3847/1538-4357/ad3914}}.

\bibitem[{Collin-Souffrin}(1987)]{collinsouffrin87}
{Collin-Souffrin}, S.
\newblock {Line and continuum radiation from the outer region of accretion
  discs in active galactic nuclei. I - Preliminary considerations}.
\newblock {\em \aap} {\bf 1987}, {\em 179},~60--70.

\bibitem[{Marziani} \em{et~al.}(2010){Marziani}, {Sulentic}, {Negrete},
  {Dultzin}, {Zamfir}, and {Bachev}]{marzianietal10}
{Marziani}, P.; {Sulentic}, J.W.; {Negrete}, C.A.; {Dultzin}, D.; {Zamfir}, S.;
  {Bachev}, R.
\newblock {Broad-line region physical conditions along the quasar eigenvector 1
  sequence}.
\newblock {\em \mnras} {\bf 2010}, {\em 409},~1033--1048,
  \href{http://xxx.lanl.gov/abs/1007.3187}{{\normalfont
  [arXiv:astro-ph.CO/1007.3187]}}.
\newblock {\url{https://doi.org/10.1111/j.1365-2966.2010.17357.x}}.

\bibitem[{Zamfir} \em{et~al.}(2010){Zamfir}, {Sulentic}, {Marziani}, and
  {Dultzin}]{zamfiretal10}
{Zamfir}, S.; {Sulentic}, J.W.; {Marziani}, P.; {Dultzin}, D.
\newblock {Detailed characterization of H{$\beta$} emission line profile in
  low-z SDSS quasars}.
\newblock {\em \mnras} {\bf 2010}, {\em 403},~1759,
  \href{http://xxx.lanl.gov/abs/0912.4306}{{\normalfont [0912.4306]}}.
\newblock {\url{https://doi.org/10.1111/j.1365-2966.2009.16236.x}}.

\bibitem[{Panda} and {{\'S}niegowska}(2024)]{pandasniegowska24}
{Panda}, S.; {{\'S}niegowska}, M.
\newblock {Changing-look Active Galactic Nuclei. I. Tracking the Transition on
  the Main Sequence of Quasars}.
\newblock {\em \apjs} {\bf 2024}, {\em 272},~13,
  \href{http://xxx.lanl.gov/abs/2206.10056}{{\normalfont
  [arXiv:astro-ph.HE/2206.10056]}}.
\newblock {\url{https://doi.org/10.3847/1538-4365/ad344f}}.

\bibitem[{Komossa} \em{et~al.}(2024){Komossa}, {Grupe}, {Marziani}, {Popovic},
  {Marceta-Mandic}, {Bon}, {Ilic}, {Kovacevic}, {Kraus}, {Haiman}, {Petrecca},
  {De Cicco}, {Dimitrijevic}, {Sreckovic}, {Kovacevic Dojcinovic},
  {Pannikkote}, {Bon}, {Gupta}, and {Iacob}]{komossaetal24}
{Komossa}, S.; {Grupe}, D.; {Marziani}, P.; {Popovic}, L.C.; {Marceta-Mandic},
  S.; {Bon}, E.; {Ilic}, D.; {Kovacevic}, A.B.; {Kraus}, A.; {Haiman}, Z.;
  et~al.
\newblock {The extremes of AGN variability: outbursts, deep fades, changing
  looks, exceptional spectral states, and semi-periodicities}.
\newblock {\em arXiv e-prints} {\bf 2024}, p. arXiv:2408.00089,
  \href{http://xxx.lanl.gov/abs/2408.00089}{{\normalfont
  [arXiv:astro-ph.HE/2408.00089]}}.
\newblock {\url{https://doi.org/10.48550/arXiv.2408.00089}}.

\bibitem[{Sulentic}(1989)]{sulentic89}
{Sulentic}, J.W.
\newblock {Toward a classification scheme for broad-line profiles in active
  galactic nuclei}.
\newblock {\em \apj} {\bf 1989}, {\em 343},~54--65.
\newblock {\url{https://doi.org/10.1086/167684}}.

\bibitem[{Tody}(1986)]{tody86}
{Tody}, D.
\newblock {The IRAF Data Reduction and Analysis System}.
\newblock In Proceedings of the Instrumentation in astronomy VI; {Crawford},
  D.L., Ed.,  1986, Vol. 627, {\em Society of Photo-Optical Instrumentation
  Engineers (SPIE) Conference Series}, p. 733.
\newblock {\url{https://doi.org/10.1117/12.968154}}.

\bibitem[{Tody}(1993)]{tody93}
{Tody}, D.
\newblock {IRAF in the Nineties}.
\newblock In Proceedings of the Astronomical Data Analysis Software and Systems
  II; {Hanisch}, R.J.; {Brissenden}, R.J.V.; {Barnes}, J., Eds.,  1993,
  Vol.~52, {\em Astronomical Society of the Pacific Conference Series}, p. 173.

\bibitem[{Fitzpatrick} \em{et~al.}(2024){Fitzpatrick}, {Placco}, {Bolton},
  {Merino}, {Ridgway}, and {Stanghellini}]{fitzpatricketal24}
{Fitzpatrick}, M.; {Placco}, V.; {Bolton}, A.; {Merino}, B.; {Ridgway}, S.;
  {Stanghellini}, L.
\newblock {Modernizing IRAF to Support Gemini Data Reduction}.
\newblock {\em arXiv e-prints} {\bf 2024}, p. arXiv:2401.01982,
  \href{http://xxx.lanl.gov/abs/2401.01982}{{\normalfont
  [arXiv:astro-ph.IM/2401.01982]}}.
\newblock {\url{https://doi.org/10.48550/arXiv.2401.01982}}.

\bibitem[{Kriss}(1994)]{kriss94}
{Kriss}, G.
\newblock {Fitting Models to UV and Optical Spectral Data}.
\newblock {\em Astronomical Data Analysis Software and Systems III, A.S.P.
  Conference Series} {\bf 1994}, {\em 61},~437.

\bibitem[{Marziani} \em{et~al.}(2022){Marziani}, {Olmo}, {Negrete}, {Dultzin},
  {Piconcelli}, {Vietri}, {Mart{\'\i}nez-Aldama}, {D'Onofrio}, {Bon}, {Bon},
  {Deconto Machado}, {Stirpe}, and {Buendia Rios}]{marzianietal22}
{Marziani}, P.; {Olmo}, A.d.; {Negrete}, C.A.; {Dultzin}, D.; {Piconcelli}, E.;
  {Vietri}, G.; {Mart{\'\i}nez-Aldama}, M.L.; {D'Onofrio}, M.; {Bon}, E.;
  {Bon}, N.;  et~al.
\newblock {The Intermediate-ionization Lines as Virial Broadening Estimators
  for Population A Quasars}.
\newblock {\em \apjs} {\bf 2022}, {\em 261},~30,
  \href{http://xxx.lanl.gov/abs/2205.07034}{{\normalfont
  [arXiv:astro-ph.GA/2205.07034]}}.
\newblock {\url{https://doi.org/10.3847/1538-4365/ac6fd6}}.

\bibitem[{Zamfir} \em{et~al.}(2008){Zamfir}, {Sulentic}, and
  {Marziani}]{zamfiretal08}
{Zamfir}, S.; {Sulentic}, J.W.; {Marziani}, P.
\newblock {New insights on the QSO radio-loud/radio-quiet dichotomy: SDSS
  spectra in the context of the 4D eigenvector1 parameter space}.
\newblock {\em MNRAS} {\bf 2008}, {\em 387},~856--870,
  \href{http://xxx.lanl.gov/abs/0804.0788}{{\normalfont [0804.0788]}}.
\newblock {\url{https://doi.org/10.1111/j.1365-2966.2008.13290.x}}.

\bibitem[{Marziani} \em{et~al.}(2009){Marziani}, {Sulentic}, {Stirpe},
  {Zamfir}, and {Calvani}]{marzianietal09}
{Marziani}, P.; {Sulentic}, J.W.; {Stirpe}, G.M.; {Zamfir}, S.; {Calvani}, M.
\newblock {VLT/ISAAC spectra of the H{$\beta$} region in intermediate-redshift
  quasars. III. H{$\beta$} broad-line profile analysis and inferences about BLR
  structure}.
\newblock {\em A\&Ap} {\bf 2009}, {\em 495},~83--112,
  \href{http://xxx.lanl.gov/abs/0812.0251}{{\normalfont [0812.0251]}}.
\newblock {\url{https://doi.org/10.1051/0004-6361:200810764}}.

\bibitem[{Marziani} \em{et~al.}(2022){Marziani}, {Deconto-Machado}, and {Del
  Olmo}]{marzianietal22b}
{Marziani}, P.; {Deconto-Machado}, A.; {Del Olmo}, A.
\newblock {Isolating an Outflow Component in Single-Epoch Spectra of Quasars}.
\newblock {\em Galaxies} {\bf 2022}, {\em 10},~54,
  \href{http://xxx.lanl.gov/abs/2203.09196}{{\normalfont
  [arXiv:astro-ph.GA/2203.09196]}}.
\newblock {\url{https://doi.org/10.3390/galaxies10020054}}.

\bibitem[{Temple}(2024)]{temple24}
{Temple}, M.J.
\newblock {Testing AGN outflow and accretion models with SDSS quasar
  demographics}.
\newblock In Proceedings of the IAU Symposium; {Bruni}, G.; {Diaz Trigo}, M.;
  {Laha}, S.; {Fukumura}, K., Eds.,  2024, Vol. 378, {\em IAU Symposium}, pp.
  27--29.
\newblock {\url{https://doi.org/10.1017/S1743921323003277}}.

\bibitem[{Corbin}(1995)]{corbin95}
{Corbin}, M.R.
\newblock {QSO Broad Emission Line Asymmetries: Evidence of Gravitational
  Redshift?}
\newblock {\em \apj} {\bf 1995}, {\em 447},~496--+.
\newblock {\url{https://doi.org/10.1086/175894}}.

\bibitem[{Gavrilovi{\'c}} \em{et~al.}(2007){Gavrilovi{\'c}}, {Popovi{\'c}}, and
  {Kollatschny}]{gavrilovicetal07}
{Gavrilovi{\'c}}, N.; {Popovi{\'c}}, L.{\v C}.; {Kollatschny}, W.
\newblock {The gravitational redshift in the broad line region of the active
  galactic nucleus Mrk 110}.
\newblock In Proceedings of the IAU Symposium; {Karas}, V.; {Matt}, G., Eds.,
  2007, Vol. 238, {\em IAU Symposium}, pp. 369--370.
\newblock {\url{https://doi.org/10.1017/S1743921307005492}}.

\bibitem[{Joni{\'c}} \em{et~al.}(2016){Joni{\'c}}, {Kova{\v c}evi{\'c}-Doj{\v
  c}inovi{\'c}}, {Ili{\'c}}, and {Popovi{\'c}}]{jonicetal16}
{Joni{\'c}}, S.; {Kova{\v c}evi{\'c}-Doj{\v c}inovi{\'c}}, J.; {Ili{\'c}}, D.;
  {Popovi{\'c}}, L.{\v C}.
\newblock {Virialization of the Broad Line Region in Active Galactic Nuclei -
  connection between shifts and widths of broad emission lines}.
\newblock {\em \apss} {\bf 2016}, {\em 361},~101,
  \href{http://xxx.lanl.gov/abs/1602.03668}{{\normalfont [1602.03668]}}.
\newblock {\url{https://doi.org/10.1007/s10509-016-2680-9}}.

\bibitem[{Bon} \em{et~al.}(2015){Bon}, {Bon}, {Marziani}, and
  {Jovanovi{\'c}}]{bonetal15}
{Bon}, N.; {Bon}, E.; {Marziani}, P.; {Jovanovi{\'c}}, P.
\newblock {Gravitational redshift of emission lines in the AGN spectra}.
\newblock {\em \apss} {\bf 2015}, {\em 360},~7,
  \href{http://xxx.lanl.gov/abs/1602.03688}{{\normalfont [1602.03688]}}.
\newblock {\url{https://doi.org/10.1007/s10509-015-2555-5}}.

\bibitem[{Heckman}(1980)]{heckman80}
{Heckman}, T.M.
\newblock {An optical and radio survey of the nuclei of bright galaxies -
  Activity in normal galactic nuclei}.
\newblock {\em \aap} {\bf 1980}, {\em 87},~152--164.

\bibitem[{Narayan} and {Yi}(1994)]{narayanyi94}
{Narayan}, R.; {Yi}, I.
\newblock {Advection-dominated Accretion: A Self-similar Solution}.
\newblock {\em \apjl} {\bf 1994}, {\em 428},~L13,
  \href{http://xxx.lanl.gov/abs/astro-ph/9403052}{{\normalfont
  [arXiv:astro-ph/astro-ph/9403052]}}.
\newblock {\url{https://doi.org/10.1086/187381}}.

\bibitem[{Soria} \em{et~al.}(2006){Soria}, {Graham}, {Fabbiano}, {Baldi},
  {Elvis}, {Jerjen}, {Pellegrini}, and {Siemiginowska}]{soriaetal06}
{Soria}, R.; {Graham}, A.W.; {Fabbiano}, G.; {Baldi}, A.; {Elvis}, M.;
  {Jerjen}, H.; {Pellegrini}, S.; {Siemiginowska}, A.
\newblock {Accretion and Nuclear Activity of Quiescent Supermassive Black
  Holes. II. Optical Study and Interpretation}.
\newblock {\em \apj} {\bf 2006}, {\em 640},~143--155,
  \href{http://xxx.lanl.gov/abs/astro-ph/0511341}{{\normalfont
  [arXiv:astro-ph/astro-ph/0511341]}}.
\newblock {\url{https://doi.org/10.1086/499935}}.

\bibitem[{Giustini} and {Proga}(2019)]{giustiniproga19}
{Giustini}, M.; {Proga}, D.
\newblock {A global view of the inner accretion and ejection flow around super
  massive black holes. Radiation-driven accretion disk winds in a physical
  context}.
\newblock {\em \aap} {\bf 2019}, {\em 630},~A94,
  \href{http://xxx.lanl.gov/abs/1904.07341}{{\normalfont
  [arXiv:astro-ph.GA/1904.07341]}}.
\newblock {\url{https://doi.org/10.1051/0004-6361/201833810}}.

\bibitem[{Vestergaard} and {Peterson}(2006)]{vestergaardpeterson06}
{Vestergaard}, M.; {Peterson}, B.M.
\newblock {Determining Central Black Hole Masses in Distant Active Galaxies and
  Quasars. II. Improved Optical and UV Scaling Relationships}.
\newblock {\em \apj} {\bf 2006}, {\em 641},~689--709,
  \href{http://xxx.lanl.gov/abs/arXiv:astro-ph/0601303}{{\normalfont
  [arXiv:astro-ph/0601303]}}.
\newblock {\url{https://doi.org/10.1086/500572}}.

\bibitem[{Marziani} and {Sulentic}(1993)]{marzianisulentic93}
{Marziani}, P.; {Sulentic}, J.W.
\newblock {Evidence for a very broad line region in PG 1138+222}.
\newblock {\em \apj} {\bf 1993}, {\em 409},~612--616,
  \href{http://xxx.lanl.gov/abs/arXiv:astro-ph/9210005}{{\normalfont
  [arXiv:astro-ph/9210005]}}.
\newblock {\url{https://doi.org/10.1086/172692}}.

\bibitem[{Snedden} and {Gaskell}(2007)]{sneddengaskell07}
{Snedden}, S.A.; {Gaskell}, C.M.
\newblock {The Case for Optically Thick High-Velocity Broad-Line Region Gas in
  Active Galactic Nuclei}.
\newblock {\em \apj} {\bf 2007}, {\em 669},~126--134.
\newblock {\url{https://doi.org/10.1086/521290}}.

\bibitem[{Marziani} \em{et~al.}(2006){Marziani}, {Dultzin-Hacyan}, and
  {Sulentic}]{marzianietal06}
{Marziani}, P.; {Dultzin-Hacyan}, D.; {Sulentic}, J.W.
\newblock {Accretion onto Supermassive Black Holes in Quasars: Learning from
  Optical/UV Observations}. In {\em New Developments in Black Hole Research};
  {Kreitler}, P.V., Ed.; Nova Press, New York,  2006; p. 123.

\bibitem[{Mineshige} \em{et~al.}(2000){Mineshige}, {Kawaguchi}, {Takeuchi}, and
  {Hayashida}]{mineshigeetal00}
{Mineshige}, S.; {Kawaguchi}, T.; {Takeuchi}, M.; {Hayashida}, K.
\newblock {Slim-Disk Model for Soft X-Ray Excess and Variability of Narrow-Line
  Seyfert 1 Galaxies}.
\newblock {\em \pasj} {\bf 2000}, {\em 52},~499--508,
  \href{http://xxx.lanl.gov/abs/arXiv:astro-ph/0003017}{{\normalfont
  [arXiv:astro-ph/0003017]}}.

\bibitem[{S{\k{a}}dowski}(2009)]{sadowski09}
{S{\k{a}}dowski}, A.
\newblock {Slim Disks Around Kerr Black Holes Revisited}.
\newblock {\em \apjs} {\bf 2009}, {\em 183},~171--178,
  \href{http://xxx.lanl.gov/abs/0906.0355}{{\normalfont
  [arXiv:astro-ph.HE/0906.0355]}}.
\newblock {\url{https://doi.org/10.1088/0067-0049/183/2/171}}.

\bibitem[{Dotan} and {Shaviv}(2011)]{dotanshaviv11}
{Dotan}, C.; {Shaviv}, N.J.
\newblock {Super-Eddington slim accretion discs with winds}.
\newblock {\em \mnras} {\bf 2011}, {\em 413},~1623--1632.
\newblock {\url{https://doi.org/10.1111/j.1365-2966.2011.18235.x}}.

\bibitem[{Abramowicz} and {Straub}(2014)]{abramowiczstaub14}
{Abramowicz}, M.A.; {Straub}, O.
\newblock {Accretion discs}.
\newblock {\em Scholarpedia} {\bf 2014}, {\em 9},~2408.
\newblock {\url{https://doi.org/10.4249/scholarpedia.2408}}.

\bibitem[Romero and Abraham(2000)]{romerozulema00}
Romero, G.E.; Abraham, Z.
\newblock Precession of relativistic jets in active galactic nuclei as a clue
  to binary supermassive black holes.
\newblock {\em International Journal of Modern Physics D} {\bf 2000}, {\em
  9},~173--184.
\newblock {\url{https://doi.org/10.1142/S0218271800000191}}.

\bibitem[{Krause} \em{et~al.}(2011){Krause}, {Burkert}, and
  {Schartmann}]{krauseetal11}
{Krause}, M.; {Burkert}, A.; {Schartmann}, M.
\newblock {Stability of cloud orbits in the broad-line region of active
  galactic nuclei}.
\newblock {\em \mnras} {\bf 2011}, {\em 411},~550--556,
  \href{http://xxx.lanl.gov/abs/1007.0112}{{\normalfont
  [arXiv:astro-ph.CO/1007.0112]}}.
\newblock {\url{https://doi.org/10.1111/j.1365-2966.2010.17698.x}}.

\bibitem[{Decarli} \em{et~al.}(2011){Decarli}, {Dotti}, and
  {Treves}]{decarlietal11}
{Decarli}, R.; {Dotti}, M.; {Treves}, A.
\newblock {Geometry and inclination of the broad-line region in blazars}.
\newblock {\em \mnras} {\bf 2011}, {\em 413},~39--46,
  \href{http://xxx.lanl.gov/abs/1011.5879}{{\normalfont [1011.5879]}}.
\newblock {\url{https://doi.org/10.1111/j.1365-2966.2010.18102.x}}.

\bibitem[{Afanasiev} \em{et~al.}(2019){Afanasiev}, {Popovi{\'c}}, and
  {Shapovalova}]{afanasievetal19}
{Afanasiev}, V.L.; {Popovi{\'c}}, L.{\v{C}}.; {Shapovalova}, A.I.
\newblock {Spectropolarimetry of Seyfert 1 galaxies with equatorial scattering:
  black hole masses and broad-line region characteristics}.
\newblock {\em \mnras} {\bf 2019}, {\em 482},~4985--4999,
  \href{http://xxx.lanl.gov/abs/1810.12164}{{\normalfont
  [arXiv:astro-ph.GA/1810.12164]}}.
\newblock {\url{https://doi.org/10.1093/mnras/sty2995}}.

\bibitem[{Wills} and {Browne}(1986)]{willsbrowne86}
{Wills}, B.J.; {Browne}, I.W.A.
\newblock {Relativistic beaming and quasar emission lines}.
\newblock {\em \apj} {\bf 1986}, {\em 302},~56--63.
\newblock {\url{https://doi.org/10.1086/163973}}.

\bibitem[{Marin}(2016)]{marin16}
{Marin}, F.
\newblock {Are there reliable methods to estimate the nuclear orientation of
  Seyfert galaxies?}
\newblock {\em \mnras} {\bf 2016}, {\em 460},~3679--3705,
  \href{http://xxx.lanl.gov/abs/1605.02904}{{\normalfont
  [arXiv:astro-ph.GA/1605.02904]}}.
\newblock {\url{https://doi.org/10.1093/mnras/stw1131}}.

\bibitem[Punsly \em{et~al.}()Punsly, Tramacere, Kharb, and
  Marziani]{punslyetal18}
Punsly, B.; Tramacere, A.; Kharb, P.; Marziani, P.
\newblock The Powerful Jet and Gamma-Ray Flare of the Quasar PKS 0438--436.

\bibitem[{Marziani} \em{et~al.}(2022){Marziani}, {Bon}, {Bon}, {D'Onofrio},
  {Punsly}, {{\'S}niegowska}, {Czerny}, {Panda}, {Mart\'{\i}nez Aldama}, {del
  Olmo}, {Deconto-Machado}, {Negrete}, {Dultzin}, {Buendia}, and
  {Garnica}]{marzianietal22a}
{Marziani}, P.; {Bon}, E.; {Bon}, N.; {D'Onofrio}, M.; {Punsly}, B.;
  {{\'S}niegowska}, M.; {Czerny}, B.; {Panda}, S.; {Mart\'{\i}nez Aldama},
  M.L.; {del Olmo}, A.;  et~al.
\newblock {The main sequence of quasars: The taming of the extremes}.
\newblock {\em Astronomische Nachrichten} {\bf 2022}, {\em 343},~e210082,
  \href{http://xxx.lanl.gov/abs/2111.04140}{{\normalfont
  [arXiv:astro-ph.GA/2111.04140]}}.
\newblock {\url{https://doi.org/10.1002/asna.20210082}}.

\bibitem[{Lynden-Bell}(1969)]{lynden-bell69}
{Lynden-Bell}, D.
\newblock {Galactic Nuclei as Collapsed Old Quasars}.
\newblock {\em \nat} {\bf 1969}, {\em 223},~690--694.
\newblock {\url{https://doi.org/10.1038/223690a0}}.

\bibitem[{Lin} and {Papaloizou}(1986)]{linpapaloizou86}
{Lin}, D.N.C.; {Papaloizou}, J.
\newblock {On the Tidal Interaction between Protoplanets and the Protoplanetary
  Disk. III. Orbital Migration of Protoplanets}.
\newblock {\em \apj} {\bf 1986}, {\em 309},~846.
\newblock {\url{https://doi.org/10.1086/164653}}.

\bibitem[{Artymowicz} and {Lubow}(1994)]{artymowiczlubow94}
{Artymowicz}, P.; {Lubow}, S.H.
\newblock {Dynamics of Binary-Disk Interaction. I. Resonances and Disk Gap
  Sizes}.
\newblock {\em \apj} {\bf 1994}, {\em 421},~651.
\newblock {\url{https://doi.org/10.1086/173679}}.

\bibitem[{Tiede} \em{et~al.}(2022){Tiede}, {Zrake}, {MacFadyen}, and
  {Haiman}]{tiedeetal22}
{Tiede}, C.; {Zrake}, J.; {MacFadyen}, A.; {Haiman}, Z.
\newblock {How Binaries Accrete: Hydrodynamic Simulations with Passive Tracer
  Particles}.
\newblock {\em \apj} {\bf 2022}, {\em 932},~24,
  \href{http://xxx.lanl.gov/abs/2111.04721}{{\normalfont
  [arXiv:astro-ph.GA/2111.04721]}}.
\newblock {\url{https://doi.org/10.3847/1538-4357/ac6c2b}}.

\bibitem[{Chen} \em{et~al.}(1989){Chen}, {Halpern}, and
  {Filippenko}]{chenetal89}
{Chen}, K.; {Halpern}, J.P.; {Filippenko}, A.V.
\newblock {Kinematic evidence for a relativistic Keplerian disk - ARP 102B}.
\newblock {\em \apj} {\bf 1989}, {\em 339},~742--751.
\newblock {\url{https://doi.org/10.1086/167332}}.

\bibitem[{Eracleous} and {Halpern}(2003)]{eracleoushalpern03}
{Eracleous}, M.; {Halpern}, J.P.
\newblock {Completion of a Survey and Detailed Study of Double-peaked Emission
  Lines in Radio-loud Active Galactic Nuclei}.
\newblock {\em \apj} {\bf 2003}, {\em 599},~886--908,
  \href{http://xxx.lanl.gov/abs/arXiv:astro-ph/0309149}{{\normalfont
  [arXiv:astro-ph/0309149]}}.
\newblock {\url{https://doi.org/10.1086/379540}}.

\bibitem[{Strateva} \em{et~al.}(2003){Strateva}, {Strauss}, {Hao}, {Schlegel},
  {Hall}, {Gunn}, {Li}, {Ivezi{\'c}}, {Richards}, {Zakamska}, {Voges},
  {Anderson}, {Lupton}, {Schneider}, {Brinkmann}, and {Nichol}]{stratevaetal03}
{Strateva}, I.V.; {Strauss}, M.A.; {Hao}, L.; {Schlegel}, D.J.; {Hall}, P.B.;
  {Gunn}, J.E.; {Li}, L.; {Ivezi{\'c}}, {\v Z}.; {Richards}, G.T.; {Zakamska},
  N.L.;  et~al.
\newblock {Double-peaked Low-Ionization Emission Lines in Active Galactic
  Nuclei}.
\newblock {\em AJ} {\bf 2003}, {\em 126},~1720--1749,
  \href{http://xxx.lanl.gov/abs/arXiv:astro-ph/0307357}{{\normalfont
  [arXiv:astro-ph/0307357]}}.
\newblock {\url{https://doi.org/10.1086/378367}}.

\bibitem[{Mengistue} \em{et~al.}(2023){Mengistue}, {Del Olmo}, {Marziani},
  {Povi{\'c}}, {Mart{\'\i}nez-Carballo}, {Perea}, and
  {M{\'a}rquez}]{terefemengistueetal23}
{Mengistue}, S.T.; {Del Olmo}, A.; {Marziani}, P.; {Povi{\'c}}, M.;
  {Mart{\'\i}nez-Carballo}, M.A.; {Perea}, J.; {M{\'a}rquez}, I.
\newblock {Optical and near-UV spectroscopic properties of low-redshift jetted
  quasars in the main sequence context}.
\newblock {\em \mnras} {\bf 2023}, {\em 525},~4474--4496,
  \href{http://xxx.lanl.gov/abs/2308.06080}{{\normalfont
  [arXiv:astro-ph.GA/2308.06080]}}.
\newblock {\url{https://doi.org/10.1093/mnras/stad2467}}.

\bibitem[{Peters}(1964)]{peters64}
{Peters}, P.C.
\newblock {Gravitational Radiation and the Motion of Two Point Masses}.
\newblock {\em Physical Review} {\bf 1964}, {\em 136},~1224--1232.
\newblock {\url{https://doi.org/10.1103/PhysRev.136.B1224}}.

\bibitem[{Deng}(2022)]{deng22}
{Deng}, H.
\newblock {Gravitational wave background from mergers of large primordial black
  holes}.
\newblock {\em \jcap} {\bf 2022}, {\em 2022},~037,
  \href{http://xxx.lanl.gov/abs/2110.02460}{{\normalfont
  [arXiv:astro-ph.CO/2110.02460]}}.
\newblock {\url{https://doi.org/10.1088/1475-7516/2022/03/037}}.

\bibitem[{Nguyen} and {Bogdanovi{\'c}}(2016)]{nguyenbogdanovic16}
{Nguyen}, K.; {Bogdanovi{\'c}}, T.
\newblock {Emission Signatures from Sub-parsec Binary Supermassive Black Holes.
  I. Diagnostic Power of Broad Emission Lines}.
\newblock {\em \apj} {\bf 2016}, {\em 828},~68,
  \href{http://xxx.lanl.gov/abs/1605.09389}{{\normalfont
  [arXiv:astro-ph.HE/1605.09389]}}.
\newblock {\url{https://doi.org/10.3847/0004-637X/828/2/68}}.

\bibitem[Hobbs \em{et~al.}(2010)Hobbs, Archibald, Arzoumanian, Backer, Bailes,
  Bhat, Burgay, Burke-Spolaor, Champion, Cognard, Coles, Cordes, Demorest,
  Desvignes, Ferdman, Finn, Freire, Gonzalez, Hessels, Hotan, Janssen, Jenet,
  Jessner, Jordan, Kaspi, Kramer, Kondratiev, Lazio, Lazaridis, Lee, Levin,
  Lommen, Lorimer, Lynch, Lyne, Manchester, McLaughlin, Nice, Oslowski, Pilia,
  Possenti, Purver, Ransom, Reynolds, Sanidas, Sarkissian, Sesana, Shannon,
  Siemens, Stairs, Stappers, Stinebring, Theureau, van Haasteren, van Straten,
  Verbiest, Yardley, and You]{hobbsetal10}
Hobbs, G.; Archibald, A.; Arzoumanian, Z.; Backer, D.; Bailes, M.; Bhat,
  N.D.R.; Burgay, M.; Burke-Spolaor, S.; Champion, D.; Cognard, I.;  et~al.
\newblock The International Pulsar Timing Array project: using pulsars as a
  gravitational wave detector.
\newblock {\em Classical and Quantum Gravity} {\bf 2010}, {\em 27},~084013.

\bibitem[Mingarelli \em{et~al.}(2017)Mingarelli, Lazio, Sesana, Greene, Ellis,
  Ma, Croft, Burke-Spolaor, and Taylor]{mingarellietal17}
Mingarelli, C.M.F.; Lazio, T.J.W.; Sesana, A.; Greene, J.E.; Ellis, J.A.; Ma,
  C.P.; Croft, S.; Burke-Spolaor, S.; Taylor, S.R.
\newblock The local nanohertz gravitational-wave landscape from supermassive
  black hole binaries.
\newblock {\em Nature Astronomy} {\bf 2017}, {\em 1},~886--892.
\newblock {\url{https://doi.org/10.1038/s41550-017-0299-6}}.

\bibitem[{Antonini} and {Perets}(2012)]{antoniniperets12}
{Antonini}, F.; {Perets}, H.B.
\newblock {Secular Evolution of Compact Binaries near Massive Black Holes:
  Gravitational Wave Sources and Other Exotica}.
\newblock {\em \apj} {\bf 2012}, {\em 757},~27,
  \href{http://xxx.lanl.gov/abs/1203.2938}{{\normalfont
  [arXiv:astro-ph.GA/1203.2938]}}.
\newblock {\url{https://doi.org/10.1088/0004-637X/757/1/27}}.

\bibitem[{Wang} \em{et~al.}(2009){Wang}, {Yan}, {Li}, {Chen}, {Xiang}, {Hu},
  {Ge}, and {Zhang}]{wangetal09sf}
{Wang}, J.M.; {Yan}, C.S.; {Li}, Y.R.; {Chen}, Y.M.; {Xiang}, F.; {Hu}, C.;
  {Ge}, J.Q.; {Zhang}, S.
\newblock {Evolution of Gaseous Disk Viscosity Driven by Supernova Explosions
  in Star-Forming Galaxies at High Redshift}.
\newblock {\em \apjl} {\bf 2009}, {\em 701},~L7--L11,
  \href{http://xxx.lanl.gov/abs/0907.4474}{{\normalfont
  [arXiv:astro-ph.CO/0907.4474]}}.
\newblock {\url{https://doi.org/10.1088/0004-637X/701/1/L7}}.

\bibitem[{Wang} \em{et~al.}(2010){Wang}, {Deng}, and {Wei}]{wangetal10sf}
{Wang}, J.; {Deng}, J.S.; {Wei}, J.Y.
\newblock {Ongoing star formation in AGN host galaxy discs: a view from
  core-collapse supernovae}.
\newblock {\em \mnras} {\bf 2010}, {\em 405},~2529--2533,
  \href{http://xxx.lanl.gov/abs/1003.1358}{{\normalfont
  [arXiv:astro-ph.CO/1003.1358]}}.
\newblock {\url{https://doi.org/10.1111/j.1365-2966.2010.16629.x}}.

\bibitem[{Wang} \em{et~al.}(2011){Wang}, {Ge}, {Hu}, {Baldwin}, {Li},
  {Ferland}, {Xiang}, {Yan}, and {Zhang}]{wangetal11sf}
{Wang}, J.M.; {Ge}, J.Q.; {Hu}, C.; {Baldwin}, J.A.; {Li}, Y.R.; {Ferland},
  G.J.; {Xiang}, F.; {Yan}, C.S.; {Zhang}, S.
\newblock {Star Formation in Self-gravitating Disks in Active Galactic Nuclei.
  I. Metallicity Gradients in Broad-line Regions}.
\newblock {\em \apj} {\bf 2011}, {\em 739},~3,
  \href{http://xxx.lanl.gov/abs/1107.3620}{{\normalfont
  [arXiv:astro-ph.GA/1107.3620]}}.
\newblock {\url{https://doi.org/10.1088/0004-637X/739/1/3}}.

\bibitem[{Wang} \em{et~al.}(2012){Wang}, {Du}, {Baldwin}, {Ge}, {Hu}, and
  {Ferland}]{wangetal12sf}
{Wang}, J.M.; {Du}, P.; {Baldwin}, J.A.; {Ge}, J.Q.; {Hu}, C.; {Ferland}, G.J.
\newblock {Star Formation in Self-gravitating Disks in Active Galactic Nuclei.
  II. Episodic Formation of Broad-line Regions}.
\newblock {\em \apj} {\bf 2012}, {\em 746},~137,
  \href{http://xxx.lanl.gov/abs/1202.0062}{{\normalfont
  [arXiv:astro-ph.CO/1202.0062]}}.
\newblock {\url{https://doi.org/10.1088/0004-637X/746/2/137}}.

\bibitem[{Artymowicz} \em{et~al.}(1993){Artymowicz}, {Lin}, and
  {Wampler}]{artymowiczetal93}
{Artymowicz}, P.; {Lin}, D.N.C.; {Wampler}, E.J.
\newblock {Star Trapping and Metallicity Enrichment in Quasars and Active
  Galactic Nuclei}.
\newblock {\em \apj} {\bf 1993}, {\em 409},~592.
\newblock {\url{https://doi.org/10.1086/172690}}.

\bibitem[{Lin}(1997)]{lin97}
{Lin}, D.N.C.
\newblock {Star/Disk Interaction in the Nuclei of Active Galaxies}.
\newblock In Proceedings of the IAU Colloq. 159: Emission Lines in Active
  Galaxies: New Methods and Techniques; {Peterson}, B.M.; {Cheng}, F.Z.;
  {Wilson}, A.S., Eds.,  1997, Vol. 113, {\em Astronomical Society of the
  Pacific Conference Series}, p.~64.

\bibitem[{Collin} and {Zahn}(1999)]{collinzahn99}
{Collin}, S.; {Zahn}, J.P.
\newblock {Star formation and evolution in accretion disks around massive black
  holes.}
\newblock {\em A\&Ap} {\bf 1999}, {\em 344},~433--449.

\bibitem[{Wang} \em{et~al.}(2023){Wang}, {Zhai}, {Li}, {Songsheng}, {Ho},
  {Chen}, {Liu}, {Du}, and {Yuan}]{wangetal23}
{Wang}, J.M.; {Zhai}, S.; {Li}, Y.R.; {Songsheng}, Y.Y.; {Ho}, L.C.; {Chen},
  Y.J.; {Liu}, J.R.; {Du}, P.; {Yuan}, Y.F.
\newblock {Star Formation in Self-gravitating Disks in Active Galactic Nuclei.
  III. Efficient Production of Iron and Infrared Spectral Energy
  Distributions}.
\newblock {\em \apj} {\bf 2023}, {\em 954},~84,
  \href{http://xxx.lanl.gov/abs/2311.06782}{{\normalfont
  [arXiv:astro-ph.GA/2311.06782]}}.
\newblock {\url{https://doi.org/10.3847/1538-4357/acdf48}}.

\bibitem[{Dittmann} and {Cantiello}(2024)]{dittmanncantiello24}
{Dittmann}, A.J.; {Cantiello}, M.
\newblock {A Semi-Analytical Model for Stellar Evolution in AGN Disks}.
\newblock {\em arXiv e-prints} {\bf 2024}, p. arXiv:2409.02981,
  \href{http://xxx.lanl.gov/abs/2409.02981}{{\normalfont
  [arXiv:astro-ph.GA/2409.02981]}}.
\newblock {\url{https://doi.org/10.48550/arXiv.2409.02981}}.

\bibitem[{Fabj} \em{et~al.}(2024){Fabj}, {Dittmann}, {Cantiello}, {Perna}, and
  {Samsing}]{fabjetal24}
{Fabj}, G.; {Dittmann}, A.J.; {Cantiello}, M.; {Perna}, R.; {Samsing}, J.
\newblock {Mapping the Outcomes of Stellar Evolution in the Disks of Active
  Galactic Nuclei}.
\newblock {\em arXiv e-prints} {\bf 2024}, p. arXiv:2408.16050,
  \href{http://xxx.lanl.gov/abs/2408.16050}{{\normalfont
  [arXiv:astro-ph.GA/2408.16050]}}.
\newblock {\url{https://doi.org/10.48550/arXiv.2408.16050}}.

\bibitem[{Liu} \em{et~al.}(2024){Liu}, {Wang}, and {Wang}]{liuetal24}
{Liu}, J.R.; {Wang}, Y.L.; {Wang}, J.M.
\newblock {Accretion-modified Stars in Accretion Disks of Active Galactic
  Nuclei: Observational Characteristics in Different Regions of the Disks}.
\newblock {\em \apj} {\bf 2024}, {\em 969},~37,
  \href{http://xxx.lanl.gov/abs/2405.02855}{{\normalfont
  [arXiv:astro-ph.HE/2405.02855]}}.
\newblock {\url{https://doi.org/10.3847/1538-4357/ad463a}}.

\bibitem[Fraix-Burnet \em{et~al.}(2017)Fraix-Burnet, Marziani, D'Onofrio, and
  Dultzin]{fraix-burnetetal17}
Fraix-Burnet, D.; Marziani, P.; D'Onofrio, M.; Dultzin, D.
\newblock The Phylogeny of Quasars and the Ontogeny of Their Central Black
  Holes.
\newblock {\em Frontiers in Astronomy and Space Sciences} {\bf 2017}, {\em
  4},~1.
\newblock {\url{https://doi.org/10.3389/fspas.2017.00001}}.

\bibitem[{Elitzur} and {Shlosman}(2006)]{elitzurshlosman06}
{Elitzur}, M.; {Shlosman}, I.
\newblock {The AGN-obscuring Torus: The End of the ``Doughnut'' Paradigm?}
\newblock {\em \apjl} {\bf 2006}, {\em 648},~L101--L104,
  \href{http://xxx.lanl.gov/abs/astro-ph/0605686}{{\normalfont
  [arXiv:astro-ph/astro-ph/0605686]}}.
\newblock {\url{https://doi.org/10.1086/508158}}.

\bibitem[{Portegies Zwart} and {McMillan}(2002)]{portegieszwartetal02}
{Portegies Zwart}, S.F.; {McMillan}, S.L.W.
\newblock {The Runaway Growth of Intermediate-Mass Black Holes in Dense Star
  Clusters}.
\newblock {\em \apj} {\bf 2002}, {\em 576},~899--907,
  \href{http://xxx.lanl.gov/abs/astro-ph/0201055}{{\normalfont
  [arXiv:astro-ph/astro-ph/0201055]}}.
\newblock {\url{https://doi.org/10.1086/341798}}.

\bibitem[{Portegies Zwart} \em{et~al.}(2004){Portegies Zwart}, {Baumgardt},
  {Hut}, {Makino}, and {McMillan}]{portegiesetal04}
{Portegies Zwart}, S.F.; {Baumgardt}, H.; {Hut}, P.; {Makino}, J.; {McMillan},
  S.L.W.
\newblock {Formation of massive black holes through runaway collisions in dense
  young star clusters}.
\newblock {\em \nat} {\bf 2004}, {\em 428},~724--726,
  \href{http://xxx.lanl.gov/abs/astro-ph/0402622}{{\normalfont
  [arXiv:astro-ph/astro-ph/0402622]}}.
\newblock {\url{https://doi.org/10.1038/nature02448}}.

\bibitem[{Gaete} \em{et~al.}(2024){Gaete}, {Schleicher}, {Lupi}, {Reinoso},
  {Fellhauer}, and {Vergara}]{gaeteetal24}
{Gaete}, B.; {Schleicher}, D.R.G.; {Lupi}, A.; {Reinoso}, B.; {Fellhauer}, M.;
  {Vergara}, M.C.
\newblock {Supermassive black hole formation via collisions in black hole
  clusters}.
\newblock {\em \aap} {\bf 2024}, {\em 690},~A378,
  \href{http://xxx.lanl.gov/abs/2406.13072}{{\normalfont
  [arXiv:astro-ph.HE/2406.13072]}}.
\newblock {\url{https://doi.org/10.1051/0004-6361/202450770}}.

\bibitem[{Chen} and {Lin}(2024)]{chenlin24}
{Chen}, Y.X.; {Lin}, D.N.C.
\newblock {The Population of Massive Stars in Active Galactic Nuclei Disks}.
\newblock {\em \apj} {\bf 2024}, {\em 967},~88,
  \href{http://xxx.lanl.gov/abs/2404.08780}{{\normalfont
  [arXiv:astro-ph.GA/2404.08780]}}.
\newblock {\url{https://doi.org/10.3847/1538-4357/ad3c3a}}.

\bibitem[{Neumayer} \em{et~al.}(2020){Neumayer}, {Seth}, and
  {B{\"o}ker}]{neumayeretal20}
{Neumayer}, N.; {Seth}, A.; {B{\"o}ker}, T.
\newblock {Nuclear star clusters}.
\newblock {\em \aapr} {\bf 2020}, {\em 28},~4,
  \href{http://xxx.lanl.gov/abs/2001.03626}{{\normalfont
  [arXiv:astro-ph.GA/2001.03626]}}.
\newblock {\url{https://doi.org/10.1007/s00159-020-00125-0}}.

\bibitem[{B{\"o}ker}(2010)]{boker09}
{B{\"o}ker}, T.
\newblock {Nuclear star clusters}.
\newblock In Proceedings of the Star Clusters: Basic Galactic Building Blocks
  Throughout Time and Space; {de Grijs}, R.; {L{\'e}pine}, J.R.D., Eds.,  2010,
  Vol. 266, {\em IAU Symposium}, pp. 58--63,
  \href{http://xxx.lanl.gov/abs/0910.4863}{{\normalfont
  [arXiv:astro-ph.CO/0910.4863]}}.
\newblock {\url{https://doi.org/10.1017/S1743921309990871}}.

\bibitem[{Tagawa} \em{et~al.}(2020){Tagawa}, {Haiman}, and
  {Kocsis}]{tagawaetal20}
{Tagawa}, H.; {Haiman}, Z.; {Kocsis}, B.
\newblock {Formation and Evolution of Compact-object Binaries in AGN Disks}.
\newblock {\em \apj} {\bf 2020}, {\em 898},~25,
  \href{http://xxx.lanl.gov/abs/1912.08218}{{\normalfont
  [arXiv:astro-ph.GA/1912.08218]}}.
\newblock {\url{https://doi.org/10.3847/1538-4357/ab9b8c}}.

\bibitem[{Barack} \em{et~al.}(2019){Barack}, {Cardoso}, {Nissanke}, {Sotiriou},
  {Askar}, {Belczynski}, {Bertone}, {Bon}, {Blas}, {Brito}, {Bulik}, {Burrage},
  {Byrnes}, {Caprini}, {Chernyakova}, {Chru{\'s}ciel}, {Colpi}, {Ferrari},
  {Gaggero}, {Gair}, {Garc{\'\i}a-Bellido}, {Hassan}, {Heisenberg}, {Hendry},
  {Heng}, {Herdeiro}, {Hinderer}, {Horesh}, {Kavanagh}, {Kocsis}, {Kramer}, {Le
  Tiec}, {Mingarelli}, {Nardini}, {Nelemans}, {Palenzuela}, {Pani}, {Perego},
  {Porter}, {Rossi}, {Schmidt}, {Sesana}, {Sperhake}, {Stamerra}, {Stein},
  {Tamanini}, {Tauris}, {Urena-L{\'o}pez}, {Vincent}, {Volonteri}, {Wardell},
  {Wex}, {Yagi}, {Abdelsalhin}, {Aloy}, {Amaro-Seoane}, {Annulli},
  {Arca-Sedda}, {Bah}, {Barausse}, {Barakovic}, {Benkel}, {Bennett}, {Bernard},
  {Bernuzzi}, {Berry}, {Berti}, {Bezares}, {Juan Blanco-Pillado},
  {Bl{\'a}zquez-Salcedo}, {Bonetti}, {Bo{\v{s}}kovi{\'c}}, {Bosnjak},
  {Bricman}, {Br{\"u}gmann}, {Capelo}, {Carloni}, {Cerd{\'a}-Dur{\'a}n},
  {Charmousis}, {Chaty}, {Clerici}, {Coates}, {Colleoni}, {Collodel},
  {Comp{\`e}re}, {Cook}, {Cordero-Carri{\'o}n}, {Correia}, {de la
  Cruz-Dombriz}, {Czinner}, {Destounis}, {Dialektopoulos}, {Doneva}, {Dotti},
  {Drew}, {Eckner}, {Edholm}, {Emparan}, {Erdem}, {Ferreira}, {Ferreira},
  {Finch}, {Font}, {Franchini}, {Fransen}, {Gal'tsov}, {Ganguly}, {Gerosa},
  {Glampedakis}, {Gomboc}, {Goobar}, {Gualtieri}, {Guendelman}, {Haardt},
  {Harmark}, {Hejda}, {Hertog}, {Hopper}, {Husa}, {Ihanec}, {Ikeda}, {Jaodand},
  {Jetzer}, {Jimenez-Forteza}, {Kamionkowski}, {Kaplan}, {Kazantzidis},
  {Kimura}, {Kobayashi}, {Kokkotas}, {Krolik}, {Kunz}, {L{\"a}mmerzahl},
  {Lasky}, {Lemos}, {Levi Said}, {Liberati}, {Lopes}, {Luna}, {Ma}, {Maggio},
  {Mangiagli}, {Martinez Montero}, {Maselli}, {Mayer}, {Mazumdar}, {Messenger},
  {M{\'e}nard}, {Minamitsuji}, {Moore}, {Mota}, {Nampalliwar}, {Nerozzi},
  {Nichols}, {Nissimov}, {Obergaulinger}, {Obers}, {Oliveri}, {Pappas},
  {Pasic}, {Peiris}, {Petrushevska}, {Pollney}, {Pratten}, {Rakic}, {Racz},
  {Radia}, {Ramazano{\u{g}}lu}, {Ramos-Buades}, {Raposo}, {Rogatko},
  {Rosca-Mead}, {Rosinska}, {Rosswog}, {Ruiz-Morales}, {Sakellariadou},
  {Sanchis-Gual}, {Sharan Salafia}, {Samajdar}, {Sintes}, {Smole}, {Sopuerta},
  {Souza-Lima}, {Stalevski}, {Stergioulas}, {Stevens}, {Tamfal},
  {Torres-Forn{\'e}}, {Tsygankov}, {{\"U}nl{\"u}t{\"u}rk}, {Valiante}, {van de
  Meent}, {Velhinho}, {Verbin}, {Vercnocke}, {Vernieri}, {Vicente},
  {Vitagliano}, {Weltman}, and {Whiting}]{baracketal19}
{Barack}, L.; {Cardoso}, V.; {Nissanke}, S.; {Sotiriou}, T.P.; {Askar}, A.;
  {Belczynski}, C.; {Bertone}, G.; {Bon}, E.; {Blas}, D.; {Brito}, R.;  et~al.
\newblock {Black holes, gravitational waves and fundamental physics: a
  roadmap}.
\newblock {\em Classical and Quantum Gravity} {\bf 2019}, {\em 36},~143001,
  \href{http://xxx.lanl.gov/abs/1806.05195}{{\normalfont
  [arXiv:gr-qc/1806.05195]}}.
\newblock {\url{https://doi.org/10.1088/1361-6382/ab0587}}.

\bibitem[Abbott \em{et~al.}(2016)Abbott, Abbott, Abbott, Abernathy, Acernese,
  Ackley, Adams, Adams, Addesso, Adhikari, Adya, Affeldt, Agathos, Agatsuma,
  Aggarwal, Aguiar, Aiello, Ain, Ajith, Allen, Allocca, Altin, Anderson,
  Anderson, Arai, Arain, Araya, Arceneaux, Areeda, Arnaud, Arun, Ascenzi,
  Ashton, Ast, Aston, Astone, Aufmuth, Aulbert, Babak, Bacon, Bader, Baker,
  Baldaccini, Ballardin, Ballmer, Barayoga, Barclay, Barish, Barker, Barone,
  Barr, Barsotti, Barsuglia, Barta, Bartlett, Barton, Bartos, Bassiri, Basti,
  Batch, Baune, Bavigadda, Bazzan, Behnke, Bejger, Belczynski, Bell, Bell,
  Berger, Bergman, Bergmann, Berry, Bersanetti, Bertolini, Betzwieser, Bhagwat,
  Bhandare, Bilenko, Billingsley, Birch, Birney, Birnholtz, Biscans, Bisht,
  Bitossi, Biwer, Bizouard, Blackburn, Blair, Blair, Blair, Bloemen, Bock,
  Bodiya, Boer, Bogaert, Bogan, Bohe, Bojtos, Bond, Bondu, Bonnand, Boom, Bork,
  Boschi, Bose, Bouffanais, Bozzi, Bradaschia, Brady, Braginsky, Branchesi,
  Brau, Briant, Brillet, Brinkmann, Brisson, Brockill, Brooks, Brown, Brown,
  Brown, Buchanan, Buikema, Bulik, Bulten, Buonanno, Buskulic, Buy, Byer,
  Cabero, Cadonati, Cagnoli, Cahillane, Bustillo, Callister, Calloni, Camp,
  Cannon, Cao, Capano, Capocasa, Carbognani, Caride, Diaz, Casentini, Caudill,
  Cavagli\`a, Cavalier, Cavalieri, Cella, Cepeda, Baiardi, Cerretani, Cesarini,
  Chakraborty, Chalermsongsak, Chamberlin, Chan, Chao, Charlton,
  Chassande-Mottin, Chen, Chen, Cheng, Chincarini, Chiummo, Cho, Cho, Chow,
  Christensen, Chu, Chua, Chung, Ciani, Clara, Clark, Cleva, Coccia, Cohadon,
  Colla, Collette, Cominsky, Constancio, Conte, Conti, Cook, Corbitt, Cornish,
  Corsi, Cortese, Costa, Coughlin, Coughlin, Coulon, Countryman, Couvares,
  Cowan, Coward, Cowart, Coyne, Coyne, Craig, Creighton, Creighton, Cripe,
  Crowder, Cruise, Cumming, Cunningham, Cuoco, Canton, Danilishin, D'Antonio,
  Danzmann, Darman, Da~Silva~Costa, Dattilo, Dave, Daveloza, Davier, Davies,
  Daw, Day, De, DeBra, Debreczeni, Degallaix, De~Laurentis, Del\'eglise,
  Del~Pozzo, Denker, Dent, Dereli, Dergachev, DeRosa, De~Rosa, DeSalvo,
  Dhurandhar, D\'{\i}az, Di~Fiore, Di~Giovanni, Di~Lieto, Di~Pace, Di~Palma,
  Di~Virgilio, Dojcinoski, Dolique, Donovan, Dooley, Doravari, Douglas, Downes,
  Drago, Drever, Driggers, Du, Ducrot, Dwyer, Edo, Edwards, Effler, Eggenstein,
  Ehrens, Eichholz, Eikenberry, Engels, Essick, Etzel, Evans, Evans, Everett,
  Factourovich, Fafone, Fair, Fairhurst, Fan, Fang, Farinon, Farr, Farr,
  Favata, Fays, Fehrmann, Fejer, Feldbaum, Ferrante, Ferreira, Ferrini,
  Fidecaro, Finn, Fiori, Fiorucci, Fisher, Flaminio, Fletcher, Fong, Fournier,
  Franco, Frasca, Frasconi, Frede, Frei, Freise, Frey, Frey, Fricke, Fritschel,
  Frolov, Fulda, Fyffe, Gabbard, Gair, Gammaitoni, Gaonkar, Garufi, Gatto,
  Gaur, Gehrels, Gemme, Gendre, Genin, Gennai, George, Gergely, Germain, Ghosh,
  Ghosh, Ghosh, Giaime, Giardina, Giazotto, Gill, Glaefke, Gleason, Goetz,
  Goetz, Gondan, Gonz\'alez, Castro, Gopakumar, Gordon, Gorodetsky, Gossan,
  Gosselin, Gouaty, Graef, Graff, Granata, Grant, Gras, Gray, Greco, Green,
  Greenhalgh, Groot, Grote, Grunewald, Guidi, Guo, Gupta, Gupta, Gushwa,
  Gustafson, Gustafson, Hacker, Hall, Hall, Hammond, Haney, Hanke, Hanks,
  Hanna, Hannam, Hanson, Hardwick, Harms, Harry, Harry, Hart, Hartman, Haster,
  Haughian, Healy, Heefner, Heidmann, Heintze, Heinzel, Heitmann, Hello,
  Hemming, Hendry, Heng, Hennig, Heptonstall, Heurs, Hild, Hoak, Hodge, Hofman,
  Hollitt, Holt, Holz, Hopkins, Hosken, Hough, Houston, Howell, Hu, Huang,
  Huerta, Huet, Hughey, Husa, Huttner, Huynh-Dinh, Idrisy, Indik, Ingram, Inta,
  Isa, Isac, Isi, Islas, Isogai, Iyer, Izumi, Jacobson, Jacqmin, Jang, Jani,
  Jaranowski, Jawahar, Jim\'enez-Forteza, Johnson, Johnson-McDaniel, Jones,
  Jones, Jonker, Ju, Haris, Kalaghatgi, Kalogera, Kandhasamy, Kang, Kanner,
  Karki, Kasprzack, Katsavounidis, Katzman, Kaufer, Kaur, Kawabe, Kawazoe,
  K\'ef\'elian, Kehl, Keitel, Kelley, Kells, Kennedy, Keppel, Key,
  Khalaidovski, Khalili, Khan, Khan, Khan, Khazanov, Kijbunchoo, Kim, Kim, Kim,
  Kim, Kim, Kim, King, King, Kinzel, Kissel, Kleybolte, Klimenko, Koehlenbeck,
  Kokeyama, Koley, Kondrashov, Kontos, Koranda, Korobko, Korth, Kowalska,
  Kozak, Kringel, Krishnan, Kr\'olak, Krueger, Kuehn, Kumar, Kumar, Kuo,
  Kutynia, Kwee, Lackey, Landry, Lange, Lantz, Lasky, Lazzarini, Lazzaro,
  Leaci, Leavey, Lebigot, Lee, Lee, Lee, Lee, Lenon, Leonardi, Leong, Leroy,
  Letendre, Levin, Levine, Li, Libson, Littenberg, Lockerbie, Logue, Lombardi,
  London, Lord, Lorenzini, Loriette, Lormand, Losurdo, Lough, Lousto, Lovelace,
  L\"uck, Lundgren, Luo, Lynch, Ma, MacDonald, Machenschalk, MacInnis, Macleod,
  Maga\~na Sandoval, Magee, Mageswaran, Majorana, Maksimovic, Malvezzi, Man,
  Mandel, Mandic, Mangano, Mansell, Manske, Mantovani, Marchesoni, Marion,
  M\'arka, M\'arka, Markosyan, Maros, Martelli, Martellini, Martin, Martin,
  Martynov, Marx, Mason, Masserot, Massinger, Masso-Reid, Matichard, Matone,
  Mavalvala, Mazumder, Mazzolo, McCarthy, McClelland, McCormick, McGuire,
  McIntyre, McIver, McManus, McWilliams, Meacher, Meadors, Meidam, Melatos,
  Mendell, Mendoza-Gandara, Mercer, Merilh, Merzougui, Meshkov, Messenger,
  Messick, Meyers, Mezzani, Miao, Michel, Middleton, Mikhailov, Milano, Miller,
  Millhouse, Minenkov, Ming, Mirshekari, Mishra, Mitra, Mitrofanov,
  Mitselmakher, Mittleman, Moggi, Mohan, Mohapatra, Montani, Moore, Moore,
  Moraru, Moreno, Morriss, Mossavi, Mours, Mow-Lowry, Mueller, Mueller, Muir,
  Mukherjee, Mukherjee, Mukherjee, Mukund, Mullavey, Munch, Murphy, Murray,
  Mytidis, Nardecchia, Naticchioni, Nayak, Necula, Nedkova, Nelemans, Neri,
  Neunzert, Newton, Nguyen, Nielsen, Nissanke, Nitz, Nocera, Nolting,
  Normandin, Nuttall, Oberling, Ochsner, O'Dell, Oelker, Ogin, Oh, Oh, Ohme,
  Oliver, Oppermann, Oram, O'Reilly, O'Shaughnessy, Ott, Ottaway, Ottens,
  Overmier, Owen, Pai, Pai, Palamos, Palashov, Palomba, Pal-Singh, Pan, Pan,
  Pankow, Pannarale, Pant, Paoletti, Paoli, Papa, Paris, Parker, Pascucci,
  Pasqualetti, Passaquieti, Passuello, Patricelli, Patrick, Pearlstone,
  Pedraza, Pedurand, Pekowsky, Pele, Penn, Perreca, Pfeiffer, Phelps, Piccinni,
  Pichot, Pickenpack, Piergiovanni, Pierro, Pillant, Pinard, Pinto, Pitkin,
  Poeld, Poggiani, Popolizio, Post, Powell, Prasad, Predoi, Premachandra,
  Prestegard, Price, Prijatelj, Principe, Privitera, Prix, Prodi, Prokhorov,
  Puncken, Punturo, Puppo, P\"urrer, Qi, Qin, Quetschke, Quintero,
  Quitzow-James, Raab, Rabeling, Radkins, Raffai, Raja, Rakhmanov, Ramet,
  Rapagnani, Raymond, Razzano, Re, Read, Reed, Regimbau, Rei, Reid, Reitze,
  Rew, Reyes, Ricci, Riles, Robertson, Robie, Robinet, Rocchi, Rolland,
  Rollins, Roma, Romano, Romano, Romanov, Romie, Rosi\ifmmode~\acute{n}\else
  \'{n}\fi{}ska, Rowan, R\"udiger, Ruggi, Ryan, Sachdev, Sadecki, Sadeghian,
  Salconi, Saleem, Salemi, Samajdar, Sammut, Sampson, Sanchez, Sandberg,
  Sandeen, Sanders, Sanders, Sassolas, Sathyaprakash, Saulson, Sauter, Savage,
  Sawadsky, Schale, Schilling, Schmidt, Schmidt, Schnabel, Schofield,
  Sch\"onbeck, Schreiber, Schuette, Schutz, Scott, Scott, Sellers, Sengupta,
  Sentenac, Sequino, Sergeev, Serna, Setyawati, Sevigny, Shaddock, Shaffer,
  Shah, Shahriar, Shaltev, Shao, Shapiro, Shawhan, Sheperd, Shoemaker,
  Shoemaker, Siellez, Siemens, Sigg, Silva, Simakov, Singer, Singer, Singh,
  Singh, Singhal, Sintes, Slagmolen, Smith, Smith, Smith, Smith, Son, Sorazu,
  Sorrentino, Souradeep, Srivastava, Staley, Steinke, Steinlechner,
  Steinlechner, Steinmeyer, Stephens, Stevenson, Stone, Strain, Straniero,
  Stratta, Strauss, Strigin, Sturani, Stuver, Summerscales, Sun, Sutton,
  Swinkels, Szczepa\ifmmode~\acute{n}\else \'{n}\fi{}czyk, Tacca, Talukder,
  Tanner, T\'apai, Tarabrin, Taracchini, Taylor, Theeg, Thirugnanasambandam,
  Thomas, Thomas, Thomas, Thorne, Thorne, Thrane, Tiwari, Tiwari, Tokmakov,
  Tomlinson, Tonelli, Torres, Torrie, T\"oyr\"a, Travasso, Traylor, Trifir\`o,
  Tringali, Trozzo, Tse, Turconi, Tuyenbayev, Ugolini, Unnikrishnan, Urban,
  Usman, Vahlbruch, Vajente, Valdes, Vallisneri, van Bakel, van Beuzekom,
  van~den Brand, Van Den~Broeck, Vander-Hyde, van~der Schaaf, van Heijningen,
  van Veggel, Vardaro, Vass, Vas\'uth, Vaulin, Vecchio, Vedovato, Veitch,
  Veitch, Venkateswara, Verkindt, Vetrano, Vicer\'e, Vinciguerra, Vine, Vinet,
  Vitale, Vo, Vocca, Vorvick, Voss, Vousden, Vyatchanin, Wade, Wade, Wade,
  Waldman, Walker, Wallace, Walsh, Wang, Wang, Wang, Wang, Wang, Ward, Ward,
  Warner, Was, Weaver, Wei, Weinert, Weinstein, Weiss, Welborn, Wen,
  We\ss{}els, Westphal, Wette, Whelan, Whitcomb, White, Whiting, Wiesner,
  Wilkinson, Willems, Williams, Williams, Williamson, Willis, Willke, Wimmer,
  Winkelmann, Winkler, Wipf, Wiseman, Wittel, Woan, Worden, Wright, Wu, Yablon,
  Yakushin, Yam, Yamamoto, Yancey, Yap, Yu, Yvert, Zadro\ifmmode~\dot{z}\else
  \.{z}\fi{}ny, Zangrando, Zanolin, Zendri, Zevin, Zhang, Zhang, Zhang, Zhang,
  Zhao, Zhou, Zhou, Zhu, Zucker, Zuraw, and Zweizig]{abbottetal15}
Abbott, B.P.; Abbott, R.; Abbott, T.D.; Abernathy, M.R.; Acernese, F.; Ackley,
  K.; Adams, C.; Adams, T.; Addesso, P.; Adhikari, R.X.;  et~al.
\newblock Observation of Gravitational Waves from a Binary Black Hole Merger.
\newblock {\em Phys. Rev. Lett.} {\bf 2016}, {\em 116},~061102.
\newblock {\url{https://doi.org/10.1103/PhysRevLett.116.061102}}.

\bibitem[{Ebisuzaki} \em{et~al.}(2001){Ebisuzaki}, {Makino}, {Tsuru}, {Funato},
  {Portegies Zwart}, {Hut}, {McMillan}, {Matsushita}, {Matsumoto}, and
  {Kawabe}]{ebisuzakietal01}
{Ebisuzaki}, T.; {Makino}, J.; {Tsuru}, T.G.; {Funato}, Y.; {Portegies Zwart},
  S.; {Hut}, P.; {McMillan}, S.; {Matsushita}, S.; {Matsumoto}, H.; {Kawabe},
  R.
\newblock {Missing Link Found? The ``Runaway'' Path to Supermassive Black
  Holes}.
\newblock {\em \apjl} {\bf 2001}, {\em 562},~L19--L22,
  \href{http://xxx.lanl.gov/abs/astro-ph/0106252}{{\normalfont
  [arXiv:astro-ph/astro-ph/0106252]}}.
\newblock {\url{https://doi.org/10.1086/338118}}.

\bibitem[{Mapelli}(2016)]{mapelli16}
{Mapelli}, M.
\newblock {Massive black hole binaries from runaway collisions: the impact of
  metallicity}.
\newblock {\em \mnras} {\bf 2016}, {\em 459},~3432--3446,
  \href{http://xxx.lanl.gov/abs/1604.03559}{{\normalfont
  [arXiv:astro-ph.GA/1604.03559]}}.
\newblock {\url{https://doi.org/10.1093/mnras/stw869}}.

\bibitem[{Belczynski} \em{et~al.}(2010){Belczynski}, {Bulik}, {Fryer},
  {Ruiter}, {Valsecchi}, {Vink}, and {Hurley}]{belczynskietal10}
{Belczynski}, K.; {Bulik}, T.; {Fryer}, C.L.; {Ruiter}, A.; {Valsecchi}, F.;
  {Vink}, J.S.; {Hurley}, J.R.
\newblock {On the Maximum Mass of Stellar Black Holes}.
\newblock {\em \apj} {\bf 2010}, {\em 714},~1217--1226,
  \href{http://xxx.lanl.gov/abs/0904.2784}{{\normalfont
  [arXiv:astro-ph.SR/0904.2784]}}.
\newblock {\url{https://doi.org/10.1088/0004-637X/714/2/1217}}.

\bibitem[{Giacobbo} and {Mapelli}(2018)]{giacobbomapelli18}
{Giacobbo}, N.; {Mapelli}, M.
\newblock {The progenitors of compact-object binaries: impact of metallicity,
  common envelope and natal kicks}.
\newblock {\em \mnras} {\bf 2018}, {\em 480},~2011--2030,
  \href{http://xxx.lanl.gov/abs/1806.00001}{{\normalfont
  [arXiv:astro-ph.HE/1806.00001]}}.
\newblock {\url{https://doi.org/10.1093/mnras/sty1999}}.

\bibitem[{Gr{\"o}bner} \em{et~al.}(2020){Gr{\"o}bner}, {Ishibashi}, {Tiwari},
  {Haney}, and {Jetzer}]{grobneretal20}
{Gr{\"o}bner}, M.; {Ishibashi}, W.; {Tiwari}, S.; {Haney}, M.; {Jetzer}, P.
\newblock {Binary black hole mergers in AGN accretion discs: gravitational wave
  rate density estimates}.
\newblock {\em \aap} {\bf 2020}, {\em 638},~A119,
  \href{http://xxx.lanl.gov/abs/2005.03571}{{\normalfont
  [arXiv:astro-ph.GA/2005.03571]}}.
\newblock {\url{https://doi.org/10.1051/0004-6361/202037681}}.

\bibitem[{Wang} \em{et~al.}(2021){Wang}, {Liu}, {Ho}, {Li}, and
  {Du}]{wangetal21a}
{Wang}, J.M.; {Liu}, J.R.; {Ho}, L.C.; {Li}, Y.R.; {Du}, P.
\newblock {Accretion-modified Stars in Accretion Disks of Active Galactic
  Nuclei: Gravitational-wave Bursts and Electromagnetic Counterparts from
  Merging Stellar Black Hole Binaries}.
\newblock {\em \apjl} {\bf 2021}, {\em 916},~L17,
  \href{http://xxx.lanl.gov/abs/2106.07334}{{\normalfont
  [arXiv:astro-ph.HE/2106.07334]}}.
\newblock {\url{https://doi.org/10.3847/2041-8213/ac0b46}}.

\bibitem[{Jermyn} \em{et~al.}(2022){Jermyn}, {Dittmann}, {McKernan}, {Ford},
  and {Cantiello}]{jermynetal22}
{Jermyn}, A.S.; {Dittmann}, A.J.; {McKernan}, B.; {Ford}, K.E.S.; {Cantiello},
  M.
\newblock {Effects of an Immortal Stellar Population in AGN Disks}.
\newblock {\em \apj} {\bf 2022}, {\em 929},~133,
  \href{http://xxx.lanl.gov/abs/2203.06187}{{\normalfont
  [arXiv:astro-ph.GA/2203.06187]}}.
\newblock {\url{https://doi.org/10.3847/1538-4357/ac5d40}}.

\bibitem[{Sicilia} \em{et~al.}(2022){Sicilia}, {Lapi}, {Boco}, {Spera}, {Di
  Carlo}, {Mapelli}, {Shankar}, {Alexander}, {Bressan}, and
  {Danese}]{siciliaetal22}
{Sicilia}, A.; {Lapi}, A.; {Boco}, L.; {Spera}, M.; {Di Carlo}, U.N.;
  {Mapelli}, M.; {Shankar}, F.; {Alexander}, D.M.; {Bressan}, A.; {Danese}, L.
\newblock {The Black Hole Mass Function Across Cosmic Times. I. Stellar Black
  Holes and Light Seed Distribution}.
\newblock {\em \apj} {\bf 2022}, {\em 924},~56,
  \href{http://xxx.lanl.gov/abs/2110.15607}{{\normalfont
  [arXiv:astro-ph.GA/2110.15607]}}.
\newblock {\url{https://doi.org/10.3847/1538-4357/ac34fb}}.

\bibitem[{LIGO Scientific Collaboration} \em{et~al.}(2015){LIGO Scientific
  Collaboration}, {Aasi}, {Abbott}, {Abbott}, {Abbott}, {Abernathy}, {Ackley},
  {Adams}, {Adams}, {Addesso}, {Adhikari}, {Adya}, {Affeldt}, {Aggarwal},
  {Aguiar}, {Ain}, {Ajith}, {Alemic}, {Allen}, {Amariutei}, {Anderson},
  {Anderson}, {Arai}, {Araya}, {Arceneaux}, {Areeda}, {Ashton}, {Ast}, {Aston},
  {Aufmuth}, {Aulbert}, {Aylott}, {Babak}, {Baker}, {Ballmer}, {Barayoga},
  {Barbet}, {Barclay}, {Barish}, {Barker}, {Barr}, {Barsotti}, {Bartlett},
  {Barton}, {Bartos}, {Bassiri}, {Batch}, {Baune}, {Behnke}, {Bell}, {Bell},
  {Benacquista}, {Bergman}, {Bergmann}, {Berry}, {Betzwieser}, {Bhagwat},
  {Bhandare}, {Bilenko}, {Billingsley}, {Birch}, {Biscans}, {Biwer},
  {Blackburn}, {Blackburn}, {Blair}, {Blair}, {Bock}, {Bodiya}, {Bojtos},
  {Bond}, {Bork}, {Born}, {Bose}, {Brady}, {Braginsky}, {Brau}, {Bridges},
  {Brinkmann}, {Brooks}, {Brown}, {Brown}, {Brown}, {Buchman}, {Buikema},
  {Buonanno}, {Cadonati}, {Calder{\'o}n Bustillo}, {Camp}, {Cannon}, {Cao},
  {Capano}, {Caride}, {Caudill}, {Cavagli{\`a}}, {Cepeda}, {Chakraborty},
  {Chalermsongsak}, {Chamberlin}, {Chao}, {Charlton}, {Chen}, {Cho}, {Cho},
  {Chow}, {Christensen}, {Chu}, {Chung}, {Ciani}, {Clara}, {Clark}, {Collette},
  {Cominsky}, {Constancio}, {Cook}, {Corbitt}, {Cornish}, {Corsi}, {Costa},
  {Coughlin}, {Countryman}, {Couvares}, {Coward}, {Cowart}, {Coyne}, {Coyne},
  {Craig}, {Creighton}, {Creighton}, {Cripe}, {Crowder}, {Cumming},
  {Cunningham}, {Cutler}, {Dahl}, {Dal Canton}, {Damjanic}, {Danilishin},
  {Danzmann}, {Dartez}, {Dave}, {Daveloza}, {Davies}, {Daw}, {DeBra}, {Del
  Pozzo}, {Denker}, {Dent}, {Dergachev}, {DeRosa}, {DeSalvo}, {Dhurandhar},
  {D{\textasciiacute}{\i}az}, {Di Palma}, {Dojcinoski}, {Dominguez}, {Donovan},
  {Dooley}, {Doravari}, {Douglas}, {Downes}, {Driggers}, {Du}, {Dwyer},
  {Eberle}, {Edo}, {Edwards}, {Edwards}, {Effler}, {Eggenstein}, {Ehrens},
  {Eichholz}, {Eikenberry}, {Essick}, {Etzel}, {Evans}, {Evans},
  {Factourovich}, {Fairhurst}, {Fan}, {Fang}, {Farr}, {Farr}, {Favata}, {Fays},
  {Fehrmann}, {Fejer}, {Feldbaum}, {Ferreira}, {Fisher}, {Frei}, {Freise},
  {Frey}, {Fricke}, {Fritschel}, {Frolov}, {Fuentes-Tapia}, {Fulda}, {Fyffe},
  and {Gair}]{LIGOcollaboration15}
{LIGO Scientific Collaboration}.; {Aasi}, J.; {Abbott}, B.P.; {Abbott}, R.;
  {Abbott}, T.; {Abernathy}, M.R.; {Ackley}, K.; {Adams}, C.; {Adams}, T.;
  {Addesso}, P.;  et~al.
\newblock {Advanced LIGO}.
\newblock {\em Classical and Quantum Gravity} {\bf 2015}, {\em 32},~074001,
  \href{http://xxx.lanl.gov/abs/1411.4547}{{\normalfont
  [arXiv:gr-qc/1411.4547]}}.
\newblock {\url{https://doi.org/10.1088/0264-9381/32/7/074001}}.

\bibitem[{Acernese} \em{et~al.}(2015){Acernese}, {Agathos}, {Agatsuma}, {Aisa},
  {Allemandou}, {Allocca}, {Amarni}, {Astone}, {Balestri}, {Ballardin},
  {Barone}, {Baronick}, {Barsuglia}, {Basti}, {Basti}, {Bauer}, {Bavigadda},
  {Bejger}, {Beker}, {Belczynski}, {Bersanetti}, {Bertolini}, {Bitossi},
  {Bizouard}, {Bloemen}, {Blom}, {Boer}, {Bogaert}, {Bondi}, {Bondu},
  {Bonelli}, {Bonnand}, {Boschi}, {Bosi}, {Bouedo}, {Bradaschia}, {Branchesi},
  {Briant}, {Brillet}, {Brisson}, {Bulik}, {Bulten}, {Buskulic}, {Buy},
  {Cagnoli}, {Calloni}, {Campeggi}, {Canuel}, {Carbognani}, {Cavalier},
  {Cavalieri}, {Cella}, {Cesarini}, {Mottin}, {Chincarini}, {Chiummo}, {Chua},
  {Cleva}, {Coccia}, {Cohadon}, {Colla}, {Colombini}, {Conte}, {Coulon},
  {Cuoco}, {Dalmaz}, {D'Antonio}, {Dattilo}, {Davier}, {Day}, {Debreczeni},
  {Degallaix}, {Del{\'e}glise}, {Pozzo}, {Dereli}, {Rosa}, {Fiore}, {Lieto},
  {Virgilio}, {Doets}, {Dolique}, {Drago}, {Ducrot}, {Endr{\H{o}}czi},
  {Fafone}, {Farinon}, {Ferrante}, {Ferrini}, {Fidecaro}, {Fiori}, {Flaminio},
  {Fournier}, {Franco}, {Frasca}, {Frasconi}, {Gammaitoni}, {Garufi},
  {Gaspard}, {Gatto}, {Gemme}, {Gendre}, {Genin}, {Gennai}, {Ghosh},
  {Giacobone}, {Giazotto}, {Gouaty}, {Granata}, {Greco}, {Groot}, {Guidi},
  {Harms}, {Heidmann}, {Heitmann}, {Hello}, {Hemming}, {Hennes}, {Hofman},
  {Jaranowski}, {Jonker}, {Kasprzack}, {K{\'e}f{\'e}lian}, {Kowalska}, {Kraan},
  {Kr{\'o}lak}, {Kutynia}, {Lazzaro}, {Leonardi}, {Leroy}, {Letendre}, {Li},
  {Lieunard}, {Lorenzini}, {Loriette}, {Losurdo}, {Magazz{\`u}}, {Majorana},
  {Maksimovic}, {Malvezzi}, {Man}, {Mangano}, {Mantovani}, {Marchesoni},
  {Marion}, {Marque}, {Martelli}, {Martellini}, {Masserot}, {Meacher},
  {Meidam}, {Mezzani}, {Michel}, {Milano}, {Minenkov}, {Moggi}, {Mohan},
  {Montani}, {Morgado}, {Mours}, {Mul}, {Nagy}, {Nardecchia}, {Naticchioni},
  {Nelemans}, {Neri}, {Neri}, {Nocera}, {Pacaud}, {Palomba}, {Paoletti},
  {Paoli}, {Pasqualetti}, {Passaquieti}, {Passuello}, {Perciballi}, {Petit},
  {Pichot}, {Piergiovanni}, {Pillant}, {Piluso}, {Pinard}, {Poggiani},
  {Prijatelj}, {Prodi}, {Punturo}, {Puppo}, {Rabeling}, {R{\'a}cz},
  {Rapagnani}, {Razzano}, {Re}, {Regimbau}, {Ricci}, {Robinet}, {Rocchi},
  {Rolland}, {Romano}, {Rosi{\'n}ska}, {Ruggi}, and {Saracco}]{acerneseetal15}
{Acernese}, F.; {Agathos}, M.; {Agatsuma}, K.; {Aisa}, D.; {Allemandou}, N.;
  {Allocca}, A.; {Amarni}, J.; {Astone}, P.; {Balestri}, G.; {Ballardin}, G.;
  et~al.
\newblock {Advanced Virgo: a second-generation interferometric gravitational
  wave detector}.
\newblock {\em Classical and Quantum Gravity} {\bf 2015}, {\em 32},~024001,
  \href{http://xxx.lanl.gov/abs/1408.3978}{{\normalfont
  [arXiv:gr-qc/1408.3978]}}.
\newblock {\url{https://doi.org/10.1088/0264-9381/32/2/024001}}.

\bibitem[{Punturo} \em{et~al.}(2010){Punturo}, {Abernathy}, {Acernese},
  {Allen}, {Andersson}, {Arun}, {Barone}, {Barr}, {Barsuglia}, {Beker},
  {Beveridge}, {Birindelli}, {Bose}, {Bosi}, {Braccini}, {Bradaschia}, {Bulik},
  {Calloni}, {Cella}, {Chassande Mottin}, {Chelkowski}, {Chincarini}, {Clark},
  {Coccia}, {Colacino}, {Colas}, {Cumming}, {Cunningham}, {Cuoco},
  {Danilishin}, {Danzmann}, {De Luca}, {De Salvo}, {Dent}, {De Rosa}, {Di
  Fiore}, {Di Virgilio}, {Doets}, {Fafone}, {Falferi}, {Flaminio}, {Franc},
  {Frasconi}, {Freise}, {Fulda}, {Gair}, {Gemme}, {Gennai}, {Giazotto},
  {Glampedakis}, {Granata}, {Grote}, {Guidi}, {Hammond}, {Hannam}, {Harms},
  {Heinert}, {Hendry}, {Heng}, {Hennes}, {Hild}, {Hough}, {Husa}, {Huttner},
  {Jones}, {Khalili}, {Kokeyama}, {Kokkotas}, {Krishnan}, {Lorenzini},
  {L{\"u}ck}, {Majorana}, {Mandel}, {Mandic}, {Martin}, {Michel}, {Minenkov},
  {Morgado}, {Mosca}, {Mours}, {M{\"u}ller{\textendash}Ebhardt}, {Murray},
  {Nawrodt}, {Nelson}, {Oshaughnessy}, {Ott}, {Palomba}, {Paoli}, {Parguez},
  {Pasqualetti}, {Passaquieti}, {Passuello}, {Pinard}, {Poggiani}, {Popolizio},
  {Prato}, {Puppo}, {Rabeling}, {Rapagnani}, {Read}, {Regimbau}, {Rehbein},
  {Reid}, {Rezzolla}, {Ricci}, {Richard}, {Rocchi}, {Rowan}, {R{\"u}diger},
  {Sassolas}, {Sathyaprakash}, {Schnabel}, {Schwarz}, {Seidel}, {Sintes},
  {Somiya}, {Speirits}, {Strain}, {Strigin}, {Sutton}, {Tarabrin},
  {Th{\"u}ring}, {van den Brand}, {van Leewen}, {van Veggel}, {van den Broeck},
  {Vecchio}, {Veitch}, {Vetrano}, {Vicere}, {Vyatchanin}, {Willke}, {Woan},
  {Wolfango}, and {Yamamoto}]{punturoetal10}
{Punturo}, M.; {Abernathy}, M.; {Acernese}, F.; {Allen}, B.; {Andersson}, N.;
  {Arun}, K.; {Barone}, F.; {Barr}, B.; {Barsuglia}, M.; {Beker}, M.;  et~al.
\newblock {The Einstein Telescope: a third-generation gravitational wave
  observatory}.
\newblock {\em Classical and Quantum Gravity} {\bf 2010}, {\em 27},~194002.
\newblock {\url{https://doi.org/10.1088/0264-9381/27/19/194002}}.

\bibitem[{Maggiore} \em{et~al.}(2020){Maggiore}, {Van Den Broeck}, {Bartolo},
  {Belgacem}, {Bertacca}, {Bizouard}, {Branchesi}, {Clesse}, {Foffa},
  {Garc{\'\i}a-Bellido}, {Grimm}, {Harms}, {Hinderer}, {Matarrese}, {Palomba},
  {Peloso}, {Ricciardone}, and {Sakellariadou}]{maggioreetal20}
{Maggiore}, M.; {Van Den Broeck}, C.; {Bartolo}, N.; {Belgacem}, E.;
  {Bertacca}, D.; {Bizouard}, M.A.; {Branchesi}, M.; {Clesse}, S.; {Foffa}, S.;
  {Garc{\'\i}a-Bellido}, J.;  et~al.
\newblock {Science case for the Einstein telescope}.
\newblock {\em \jcap} {\bf 2020}, {\em 2020},~050,
  \href{http://xxx.lanl.gov/abs/1912.02622}{{\normalfont
  [arXiv:astro-ph.CO/1912.02622]}}.
\newblock {\url{https://doi.org/10.1088/1475-7516/2020/03/050}}.

\end{thebibliography}

\PublishersNote{}
\end{adjustwidth}
\end{document}